\documentclass[12pt]{elsarticle}
\usepackage{graphicx}                         
\usepackage{axodraw4j}
\usepackage{hyperref}
\usepackage{epstopdf}                   
\usepackage{subcaption}
\usepackage{a41}
\usepackage{booktabs}                                                                
\usepackage{xcolor}                     
\usepackage{slashed}
\usepackage[rflt]{floatflt}
\usepackage{float}
\usepackage{hyperref}
\hypersetup{colorlinks=true, 
	linkcolor=blue, filecolor=blue, urlcolor=mygreen}
\usepackage{times}    
\usepackage{array}
\usepackage[english]{babel}
\usepackage[T1]{fontenc}
\usepackage{ae}
\usepackage{url}
\usepackage{amsmath, amsthm, amssymb}
\usepackage{slashed}

\usepackage{rotating}
\usepackage{graphicx}
\usepackage{comment}
\usepackage{xcolor}
\usepackage{array}
\usepackage{hyperref} 
\usepackage{cleveref}
\newcounter{mmacnt}
\newcolumntype{C}[1]{>{\centering\arraybackslash}p{#1}}
\def\restartmma{\setcounter{mmacnt}{0}}
\restartmma \catcode`|=\active
\def|#1|{\mathrm{#1}}
\catcode`|=12
\newenvironment{mma}{
\par\smallskip
\catcode`|=\active
\parskip=0pt\parindent=0pt 
\small
\def\In##1\\{%
\def\linebreak{\hfill\break\null\qquad}%
\refstepcounter{mmacnt}
\hangindent=2.5em\hangafter=0
\leavevmode
\llap{\tiny\sffamily In[\arabic{mmacnt}]:=\kern.5em}%
\mathversion{bold}\footnotesize$
\displaystyle##1$\normalsize
\mathversion{normal}\par
 }%
\def\Print##1\\{%
\def\linebreak{\hfill\break}%
\hangindent=2.5em\hangafter=0
\leavevmode ##1\par}%
\def\Out##1\\{%
\def\linebreak{$\hfill\break\null\hfill$}%
\kern\abovedisplayskip\par
\hangindent=2.5em\hangafter=0
\leavevmode
\llap{\tiny\sffamily Out[\arabic{mmacnt}]=\kern.5em}
\footnotesize$\displaystyle##1$
\normalsize\hfill\null\par
\kern\belowdisplayskip
}%
\def\Warning##1##2\\{%
\def\linebreak{\hfill\break}%
\hangindent=2.5em\hangafter=0
\leavevmode
{\scriptsize##1 : ##2}\par}%
}{%
\par\smallskip
}

\usepackage{color}
\newenvironment{fshaded}{%
\MakeFramed {\FrameRestore}
}%
{\endMakeFramed}

\makeatletter
\def\ps@pprintTitle{%
\let\@oddhead\@empty
\let\@evenhead\@empty
\def\@oddfoot{\reset@font\hfil\thepage\hfil}
\let\@evenfoot\@oddfoot
}
\makeatother
\usepackage{tikz}
\usetikzlibrary{matrix}
\allowdisplaybreaks[4]

\begin{document}
\begin{frontmatter}
\title{\Large
\textbf{
One-loop contributing
for $A \rightarrow Z Z \gamma$
in the Two-Higgs-Doublet Model
with Vector-Like Quarks}}
\author[1,2,3,4]{Dzung Tri Tran}
\author[1,2]{Vo Quoc Phong}
\author[5]{Khiem Hong Phan}
\ead{khiem.phan@eiu.edu.vn}
\address[1]{\it Department of Theoretical
Physics, Faculty of Physics and Engineering
Physics, University of Science, Ho Chi Minh
City, Vietnam}
\address[2]{\it Vietnam National University,
Ho Chi Minh City, Vietnam}
\address[3]{\it Institute of Fundamental
and Applied Sciences, Duy Tan University,
Ho Chi Minh City $70000$, Vietnam}
\address[4]{\it Faculty of Natural Sciences,
Duy Tan University, Da Nang City $50000$,
Vietnam}
\address[5]{\it Eastern International
University, 81 Nam Ky Khoi Nghia Street,
Binh Duong Ward, Ho Chi Minh City,
Vietnam}
\pagestyle{myheadings}
\markright{}
\begin{abstract} 
In this paper, we present the first
results for the one-loop decay
$A \rightarrow ZZ\gamma$ in the Two-Higgs-Doublet
Model and its extension with Vector-Like
Quarks. We calculate
the one-loop partial decay widths and branching
ratios for both unpolarized and polarized
final-state $Z$ bosons. The effects of
Vector-Like Quarks propagating in the loop
are also examined in our phenomenological
analysis. We find that the branching ratios can reach
$10^{-5}-10^{-4}$ for several benchmark points
within the viable parameter space of the considered models.
The contributions from Vector-Like Quarks can modify
the total rates by approximately $5\%$. With the
high luminosities expected at future multi-TeV
muon colliders, these effects may be experimentally
accessible.
\end{abstract}
\begin{keyword}
Higgs phenomenology, one-loop corrections,
analytic methods for quantum field theory,
and dimensional regularization.
\end{keyword}
\end{frontmatter}
\section{Introduction}
It is well known that the nature of the Higgs sector
remains an open question. Many extensions of the
Standard Model (SM) enlarge the scalar sector by introducing new scalar fields, thereby predicting additional Higgs bosons beyond the SM (BSM). Future colliders, such as the High-Luminosity Large Hadron Collider (HL-LHC) and future lepton colliders (LC), are expected to probe additional scalar bosons in BSM. Among the possible scalar-boson productions, searches for CP-odd Higgs bosons are of particular interest at future colliders. The direct production of a light CP-odd Higgs boson at the Tevatron and the LHC has been investigated in Ref.~\cite{Dermisek:2009fd}. The decay of a CP-odd Higgs boson into a $Zh$ final state has been searched for in $pp$ collisions at $\sqrt{s}=8$ TeV using the ATLAS detector, as reported in Ref.~\cite{ATLAS:2015kpj}. In addition, searches for a light pseudoscalar Higgs boson have been performed in several decay channels, including the $\mu\mu\tau\tau$ final state in $pp$ collisions at $\sqrt{s}=13$ TeV~\cite{CMS:2020ffa}, the $\tau^+\tau^-$ final state~\cite{ATLAS:2024rzd}, and the dimuon final state~\cite{CMS:2012fgd}. The final state consisting of a heavy CP-even Higgs boson and a $Z$ boson produced via the decay of a pseudoscalar Higgs boson has also been studied in Ref.~\cite{ATLAS:2023zkt}.

From a phenomenological perspective, the impact of radiative corrections on the decay branching ratios of the CP-odd Higgs boson in two-Higgs-doublet models (2HDM) has examined in Refs.~\cite{Aiko:2022gmz,Kanemura:2017gbi,Kanemura:2019slf,Aiko:2023xui}. The decay $A\to h Z^{(*)}$ has been studied within the 2HDM under the inverted-hierarchy scenario at the LHC~\cite{Akeroyd:2023kek}. A complete one-loop treatment of the decay processes $A\to \tilde f_1\tilde f_2$ and $\tilde f_2\to \tilde f_1 A$ has been performed in various BSM scenarios~\cite{Weber:2003tw}. A Padé-improved prediction for the decay rate of the CP-odd Higgs boson into two gluons has been presented in Ref.~\cite{Chishtie:1999dd}. Furthermore, the phenomenology of pseudoscalar boson decays into gauge bosons has been investigated at the LHC and at a future
$100$ TeV collider~\cite{Arhrib:2018pdi}. In Ref.~\cite{Arco:2025ydq}, the use of top-quark spin correlations to distinguish the pseudoscalar $A\to ZH$ and scalar $H\to ZA$ signatures in the $Zt\bar{t}$ final state has been explored at the LHC. The interplay between the $A\to\gamma\gamma$ decay, electric dipole moments, and the diphoton excesses at $95$ GeV and $152$ GeV has been examined within the 2HDM in Ref.~\cite{Banik:2024ugs}. One-loop contributions to the decays $A\to\ell\bar{\ell}V$, with $\ell=e,\mu$ and $V=\gamma,Z$, in extensions of the SM have been calculated in Ref.~\cite{Phan:2024zus}, while the decays $A\to Z\gamma\gamma$ and $\phi\to Z\gamma\gamma$, with $\phi=h,H$, have been investigated in Refs.~\cite{Sanchez-Velez:2018xdj,TranDT:2025EPJC}. The associated production of $AZ$ has been studied at the LHC~\cite{Yin:2002sq}, in high-energy $e^+e^-$ collisions~\cite{Akeroyd:1999gu,Akeroyd:2001aka}, and in $e\gamma$ collisions~\cite{Sasaki:2017fvk}. Other production mechanisms for the $A$ boson at high-energy $e^+e^-$ colliders have also been investigated~\cite{Farris:2003pn,Arhrib:2002ti}.

In this work, we present the first systematic
study of the one-loop decay $A \to ZZ\gamma$
in the 2HDM and its extension with Vector-Like
Quarks. We calculate the
corresponding one-loop partial decay widths
and branching ratios for both unpolarized
and polarized final-state $Z$ bosons.
The effects of Vector-Like Quarks in the
loop are systematically analyzed and
quantified. Our results show that the
branching ratios can reach
$10^{-5}-10^{-4}$ for several benchmark
scenarios within the phenomenologically
viable parameter space. Furthermore,
the Vector-Like Quark contributions can
modify the total decay rates by up to
approximately $5\%$. These results suggest
that the effects of Vector-Like Quarks
could be probed at future multi-TeV muon
colliders, where the high integrated
luminosities may provide sufficient
sensitivity to these loop-induced
contributions.

The remainder of this paper is organized as follows.
In Section~2, we introduce the 2HDM
and its extension with Vector-Like Quarks.
Section~3 is devoted to the calculation of the one-loop
contributions to the decay $A \to ZZ\gamma$ in the
2HDM-VLQ, with details of the analytical calculations
presented therein. In Section~4, we present the
corresponding phenomenological results and discuss
their implications. Finally, Section~5 summarizes our
main conclusions and provides an outlook for future
studies. The polarization vectors of the final-state
$Z$ bosons are collected in the Appendix.
\section{Two-Higgs-Doublet Model 
with Vector-Like Quarks}
In this section, we briefly review the
Two-Higgs-Doublet Model (2HDM) extended with
Vector-Like Quarks (2HDM-VLQ). For a detailed
discussion of the theoretical framework and
phenomenological implications of the 2HDM,
we refer the interested reader to
Ref.~\cite{Branco:2011iw}. The scalar Higgs
potential in the 2HDM  with
respect to the gauge-invariant
renormalizable form,
is given as follows:
\begin{eqnarray}
\label{Eq:2HDMpotential}
\mathcal{V}_{\text{2HDM}} \,
(\Phi_1, \Phi_2) 
&=& 
m_{11}^2 \Phi_1^\dagger \Phi_1 
+ 
m_{22}^2 \Phi_2^\dagger \Phi_2 
- 
\Big[ 
m_{12}^2 \Phi_1^\dagger \Phi_2 
+ 
\text{h.c.} 
\Big] 
+ 
\dfrac{\lambda_1}{2}
(\Phi_1^\dagger \Phi_1)^2 
+ 
\dfrac{\lambda_2}{2}
(\Phi_2^\dagger \Phi_2)^2
\nonumber\\
&&
+
\lambda_3
(\Phi_1^\dagger \Phi_1)
(\Phi_2^\dagger \Phi_2)
+ 
\lambda_4 
(\Phi_1^\dagger \Phi_2)
(\Phi_2^\dagger \Phi_1) 
+ 
\dfrac{1}{2}
\Big[
\lambda_5 
(\Phi_1^\dagger \Phi_2)^2 
+ 
\text{h.c.}
\Big].
\end{eqnarray}
In this work, we focus on the CP-conserving version of the 2HDM. Accordingly, all parameters in the scalar potential are assumed to be real. The model is subject to a discrete $Z_2$ symmetry, under which the Higgs doublets transform as $\Phi_1 \rightarrow \Phi_1$ and $\Phi_2 \rightarrow -\Phi_2$, while allowing for a softly broken $Z_2$ term of the form
$m_{12}^2 \Phi_1^\dagger \Phi_2 + \text{h.c.}$.
After electroweak symmetry breaking (EWSB),
the two Higgs doublet fields are parameterized as follows:
\begin{eqnarray}
\Phi_i
&=&
\begin{bmatrix}
	\omega^{+}_{i}
	\\
	\dfrac{1}{\sqrt{2}}
	\big(
	v_{i}
	+
	h_{i}
	+
	i \, 
	z_{i}
	\big)
\end{bmatrix}
,
\end{eqnarray}
where $v_{1}$ and $v_{2}$ are real 
vacuum expectation values (VEVs).
Their combined value
$v^2 = v_{1}^2 + v_{2}^2$
tend to $246$ GeV.
As consided as model parameter
which is used in phenomenological
analys, the VEVs ratio is used as
$t_{\beta} \equiv \tan \beta
= v_{2} / v_{1}$.
As a result, we acquire that
$v_{1} = v \cos \beta
\equiv v \, c_{\beta}$ as well as
$v_{2} = v \sin \beta
\equiv v \, s_{\beta}$.
To obtain the scalar mass
eigenstates, we perform the
following rotations:
\begin{eqnarray}
\begin{pmatrix}
	h_{1}
	\\[0.125cm]
	h_{2}
\end{pmatrix}
=
R (\alpha)
\begin{pmatrix}
	H
	\\[0.125cm]
	h
\end{pmatrix}
,
\qquad
\begin{pmatrix}
	z_{1}
	\\[0.125cm]
	z_{2}
\end{pmatrix}
=
R (\beta)
\begin{pmatrix}
	G^0
	\\[0.125cm]
	A
\end{pmatrix}
,
\qquad
\begin{pmatrix}
	\omega^{\pm}_{1}
	\\[0.125cm]
	\omega^{\pm}_{2}
\end{pmatrix}
=
R (\beta)
\begin{pmatrix}
	G^\pm
	\\[0.125cm]
	H^\pm 
\end{pmatrix}
.
\end{eqnarray}
Where the rotation matrix is given by
\begin{eqnarray}
R (\theta)
&=&
\begin{pmatrix}
	c_{\theta}
	&
	-s_{\theta}
	\\[0.125cm]
	s_{\theta}
	&
	c_{\theta}
\end{pmatrix}
.
\end{eqnarray}
The 2HDM is characterized by
seven independent parameters, which can
be chosen as
\begin{eqnarray}
\mathcal{P}_{\text{2HDM}}
&=&
\big\{
M_{h},
M_{H},
M_{A},
M_{H^\pm},
t_{\beta},
s_{\beta - \alpha},
m_{12}^2
\big\}
,
\end{eqnarray}
where the shorthand notation
$s_{\beta-\alpha}
=\sin(\beta-\alpha)$ is
used for the mixing term.

As mentioned above, in order to suppress
flavor-changing neutral currents (FCNCs)
at tree level, a $Z_2$ symmetry is imposed
on the fermion sector. Consequently, the
CP-conserving 2HDM can be classified into
four distinct types of Yukawa
interactions~\cite{Aoki:2009PRD},
namely Type-I, Type-II, Type-X
(or lepton-specific), and Type-Y
(or flipped) 2HDMs. The Yukawa
Lagrangian can generally be written as
\begin{eqnarray}
\label{Eq:YukLag}
-
\mathcal{L}_{Y}
&=&
\underbrace{
	\;
	Y^{u} 
	\bar{Q}^{0}_{L}
	\tilde{\Phi}_{i}
	u^{0}_{R}
	+
	Y^{d} 
	\bar{Q}^{0}_{L}
	\Phi_{j}
	d^{0}_{R}
	\;}_{\text{quarks}}
+
\underbrace{
	\;
	Y^{l}
	\bar{L}^{0}_{L}
	\Phi_{k}
	l^{0}_{R}
	\;}_{\text{leptons}}
+
\text{h.c.}.
\end{eqnarray}
where $\tilde{\Phi}_{i}
= i\sigma_2 \Phi_{i}^{*}$,
and $\Phi_{i,j,k}$ denote either
$\Phi_1$ or $\Phi_2$,
depending on the type of the 2HDM.

In this framework, we will be
discussing only 2HDM Type-II,
where $\Phi_{2}$ only couples to
up-type quarks and $\Phi_{1}$
interacted with both charged leptons
and down-type quarks,
also including its supersymmetric
incarnation like the minimal
supersymmetric Standard Model (MSSM).
The purpose of this paper is to
investigate whether a similar
phenomenon can be triggered and
enlarged by the presence of
Vector-Like Quarks (VLQs)
propagating in loop diagrams
alongside the particle states of
the 2HDM Type-II.

We now move on to discuss the
VLQs side of the model and
we start by listing the
representations of all gauge-covariant
multiplets. First of all, VLQs are (heavy)
spin-$1/2$ states that transform as 
triplets under color, but differently 
from the Standard Model (SM) quarks, 
their left- and right-handed 
couplings have the same 
electroweak (EW) quantum numbers.
Consequently, if the scalar sector 
only includes $SU(2)_L$ doublets, 
as is the case of the SM 
as well as 2HDM, some new 
vector-like quarks coupling to 
these ones with renormalizable couplings 
can only appear in seven scenarios of 
gauge-covariant multiplets with 
two $SU(2)_L$ doublet scalar 
fields $\Phi_1$ and $\Phi_2$, 
in which their identical hypercharge 
$Y = 1/2$.
The canonical representation of 
vector-like 
quarks~\cite{Arhrib:2024EPJC,
Arhrib:2025EPJC}
includes one of the following seven 
multiplet scenarios 
(isospin, hypercharge): two singlets, 
three doublets, and two triplets 
presented as detailed below
\begin{itemize}
\item Singlets: 
$T^0_{L,R} \;\; \Big(0,\dfrac{2}{3}\Big)$, 
\hspace{0.5cm}
$B^0_{L,R} \;\; \Big(0,-\dfrac{1}{3}\Big)$
\item Doublets: 
$(X \,\, T^0)_{L,R} \;\; \Big(\dfrac{1}{2},\dfrac{7}{6}\Big)$, 
\hspace{0.5cm}
$(T^0 \, B^0)_{L,R} \;\; \Big(\dfrac{1}{2},\dfrac{1}{6}\Big)$, 
\hspace{0.5cm}
$(B^0 \,\, Y)_{L,R} \;\; \Big(\dfrac{1}{2},-\dfrac{5}{6}\Big)$
\item Triplets: 
$(X \,\, T^0 \, B^0)_{L,R} \;\; \Big(1,\dfrac{2}{3}\Big)$, 
\hspace{0.5cm}
$(T^0 \, B^0 \,\, Y)_{L,R} \;\; \Big(1,-\dfrac{1}{3}\Big)$
\end{itemize}
in which we apply a zero superscript 
on weak-eigenstates $T^0$ and $B^0$ to 
distinguish them from mass-eigenstates 
$T$ and $B$. 
Additionally, these new quark fields 
$T^0$ and $B^0$ have electric charges 
$2/3$ and $-1/3$, respectively. 
It means that they carry the same 
electric charge as the corresponding 
SM top $t$ and bottom $b$ quarks. 
In most of these scenarios, such VLQs 
appear as partners of the third 
generation of quarks, as well as 
mixing with top and bottom quarks.
Further, if their mass is not 
exceedingly large for the LHC 
kinematic reach, they can be 
accessible in its detectors 
in a variety of final states.
Note that some of the multiplets include 
quarks $X$ of electric charge $5/3$, and 
$Y$ with charge $-4/3$, in which case 
the weak- and mass-eigenstates coincide 
as long as only one such multiplet 
is present.

In the fact that these VLQs $T$ and 
$B$ have the same electric charges 
corresponding to the SM top $t$ and 
bottom $b$ quarks, this situation leads 
to a deviation in their couplings to 
the Z boson.
Atomic parity violation experiments and 
the measurements of both $R_c$ and 
$R_b$ at LEP impose
constraints~\cite{Aguilar:2003PRD}
on these deviations for the up $u$ and 
charm $c$ quarks as well as for mixing 
with the new fields coupling to 
the lighter quarks $d$ and $s$, 
significantly stronger than those 
for these top and bottom quarks,
respectively.
Consequently, we now turn to 
the mixing of the new partners to 
the third generation of SM quarks.
Furthermore, when only the top quark 
$t$ ``mixes'' with $T$, the relation 
between mass-eigenstates $T_{L,R}$ 
$(u, c, t, T)$ and weak-eigenstates 
$T^{0}_{L,R}$ can be factored into 
two $2\times2$ unitary matrices 
$U^{u}_{L,R}$, which are given by
\begin{eqnarray}
\label{Eq:tTmixing}
\begin{pmatrix}
	t_{L,R}
	\\[0.125cm]
	T_{L,R}
\end{pmatrix}
&=&
U^{u}_{L,R}
\begin{pmatrix}
	t^{0}_{L,R}
	\\[0.125cm]
	T^{0}_{L,R}
\end{pmatrix}
.
\end{eqnarray}
In contrast to the up-type quark sector, 
the addition of new weak-eigenstates 
$B^{0}_{L,R}$ of charge $-1/3$ in the 
down-type quark sector results in four 
mass-eigenstates $B_{L,R}$ $(d, s, b, B)$.
Therefore, the $2\times2$ unitary matrices 
$U^{d}_{L,R}$ define the dominant $b - B$ 
mixing as follows
\begin{eqnarray}
\label{Eq:bBmixing}
\begin{pmatrix}
	b_{L,R}
	\\[0.125cm]
	B_{L,R}
\end{pmatrix}
&=&
U^{d}_{L,R}
\begin{pmatrix}
	b^{0}_{L,R}
	\\[0.125cm]
	B^{0}_{L,R}
\end{pmatrix}
.
\end{eqnarray}
In additional, we occasionally 
apply that the shorthand notations 
$s^{u,d}_{L,R} \equiv 
\sin \theta^{u,d}_{L,R}$ and 
$c^{u,d}_{L,R} 
\equiv \cos \theta^{u,d}_{L,R}$ with 
$\theta$ is the mixing angle between 
mass- and weak-eigenstates.
These mentioned mixing matrices
$U^{u,d}_{L,R}$ with $\phi_{u,d}$
are two possible CP-violating phases 
which will be ignored in this work, 
are given by 
\begin{eqnarray}
U^{u,d}_{L,R}
&=&
\begin{pmatrix}
	c^{u,d}_{L,R}
	&
	- s^{u,d}_{L,R} \,
	e^{i \phi_{u,d}}
	\\[0.25cm]
	s^{u,d}_{L,R} \,
	e^{-i \phi_{u,d}}
	&
	c^{u,d}_{L,R}
\end{pmatrix}
.
\end{eqnarray}
Note that to ease the notation, we have 
dropped the superscripts $u(d)$ whenever 
the mixing occurs only in 
the up(down)-type quark sector.
In the weak-eigenstates basis, the 
diagonalization of the mass matrices 
makes the Lagrangian as in 
Eq.~(\ref{Eq:YukLag}) of 
the third generation and 
heavy quark mass terms such that
\begin{eqnarray}
\mathcal{L}_{Y,\text{mass}}
&=&
-
\begin{pmatrix}
	\bar{t}^{0}_{L}
	&\hspace{-0.125cm}
	\bar{T}^{0}_{L}
\end{pmatrix}
\underbrace{
	\;
	\begin{pmatrix}
		Y^{u}_{33}
		\frac{v}{\sqrt{2}}
		&
		Y^{u}_{34}
		\frac{v}{\sqrt{2}}
		\\[0.35cm]
		Y^{u}_{43}
		\frac{v}{\sqrt{2}}
		&
		M^{0}_{u}
	\end{pmatrix}
	\;}_{\mathcal{M}^{u}}
\begin{pmatrix}
	t^{0}_{R}
	\\[0.12cm]
	T^{0}_{R}
\end{pmatrix}
-
\begin{pmatrix}
	\bar{b}^{0}_{L}
	&\hspace{-0.125cm}
	\bar{B}^{0}_{L}
\end{pmatrix}
\underbrace{
	\;
	\begin{pmatrix}
		Y^{d}_{33}
		\frac{v}{\sqrt{2}}
		&
		Y^{d}_{34}
		\frac{v}{\sqrt{2}}
		\\[0.35cm]
		Y^{d}_{43}
		\frac{v}{\sqrt{2}}
		&
		M^{0}_{d}
	\end{pmatrix}
	\;}_{\mathcal{M}^{d}}
\begin{pmatrix}
	b^{0}_{R}
	\\[0.12cm]
	B^{0}_{R}
\end{pmatrix}
+
\text{h.c.}
\nonumber\\
\end{eqnarray}
in which two bare masses $M^{0}_{u,d}$ 
and the $3\times4$ Yukawa coupling 
matrices $Y^{u,d}_{ij}$ are denoted 
in the mixing matrices $\mathcal{M}^{u,d}$.
Note that these bare mass terms are not 
related to the Higgs mechanism: they are
gauge invariant terms and can appear 
as such in the Yukawa Lagrangian, or 
they can be generated by another Yukawa 
coupling to a scalar multiplet that 
acquires the other VEV $v' \gg v = 246$ 
GeV.

In the next paragraph, 
by using standard techniques of 
diagonalization, the mixing matrices 
$\mathcal{M}^{q}$ for $q = u,d$ are 
presented as
\begin{eqnarray}
U^{q}_{L}
\mathcal{M}^{q}
(U^{q}_{R})^\dagger
&=&
\mathcal{M}^{q}_{\text{diag}}
\end{eqnarray}
where the 2HDM Type-II is considered 
here, these diagonal ones $\mathcal{M}^{q}_{\text{diag}}$ 
acquire the form of that,
\begin{eqnarray}
\mathcal{M}^{u}_{\text{diag}}
=
\dfrac{1}{s_{\beta}}
\begin{pmatrix}
	m_{t}
	&
	0
	\\
	0
	&
	m_{T}
\end{pmatrix}
,
\qquad && \qquad
\mathcal{M}^{d}_{\text{diag}}
=
\dfrac{1}{c_{\beta}}
\begin{pmatrix}
	m_{b}
	&
	0
	\\
	0
	&
	m_{B}
\end{pmatrix}
.
\end{eqnarray}
Regarding to the mixing matrices 
$\mathcal{M}^{q}$, it should be noted 
that when multiplets contain both $T$
and $B$ quarks, the bare mass value 
$M^{0}$ remains identical for both
the up- and down-type quark sectors. 
For the singlet and triplet cases 
$Y^{q}_{43} = 0$, while for the doublet 
one has $Y^{q}_{34} = 0$. 
Moreover, for the $(X \,\, T^0 \, B^0)$ 
triplet, one has $Y^{d}_{34} = \sqrt{2} 
\, Y^{u}_{34}$, and for the 
$(T^0 \, B^0 \,\, Y)$ triplet one has 
$Y^{u}_{34} = \sqrt{2} \, Y^{d}_{34}$.
In addition, these mixing angles in 
the left- and right-handed sectors 
\big[$U^{q}_{L}$ and 
$(U^{q}_{R})^\dagger$\big] 
are not independent parameters. 
By utilizing the mass diagonalization 
and depending on the VLQs representation, 
one can find that
\begin{alignat}{2}
\tan \theta^{q}_{R}
&=
\dfrac{m_{q}}{m_{Q}}
\tan \theta^{q}_{L}
\qquad
&
&(\text{singlets, triplets}),
\\
\tan \theta^{q}_{L}
&=
\dfrac{m_{q}}{m_{Q}}
\tan \theta^{q}_{R}
\qquad
&
&(\text{doublets})
\label{Eq:doubletrelation}
\end{alignat}
in which $(q, m_{q}, m_{Q}) \equiv 
(u, m_{t}, m_{T})$ and 
$(d, m_{b}, m_{B})$ is denoted 
for convenience, thus one of 
the mixing angles is always dominant, 
especially in the down quark sector,
because of the fact that mass value 
of $m_{b} \ll m_{t}$.
It can be easily argued that all 
representations of the VLQs scenarios 
can be characterized by only two 
parameters, except for the 
$(T^0 \, B^0)$ doublet model 
which requires three parameters. 
In our numerical examination, 
the following input parameters 
in the mass-eigenstates 
(dropped the zero superscript) of 
these corresponding VLQs couplings 
for all scenarios will be used 
as in Table~\ref{VLQsParameters}.
Furthermore, some other parameters 
such as these mixing angles 
$\theta^{u,d}_{L}$ for doublet 
scenarios can be determined by 
using Eq.~(\ref{Eq:doubletrelation}),
as well as these masses $m_{X,Y}$ of 
the VLQs with corresponding exotic 
electric charges $5/3$ and $-4/3$ 
could be found in 
Ref.~\cite{Aguilar:2013PRD}.
\begin{center}
\begin{table}[ht]
\centering
\begin{tabular}{lC{1.75cm}
C{1.75cm} C{2.5cm} C{2.0cm} C{2.0cm}}
\toprule
& 
\multicolumn{2}{c}{Singlet scenarios} 
& 
\multicolumn{3}{c}{Doublet scenarios}  
\\
\raisebox{-0.5ex}[0pt]{Representations}
& 
(T) 
& 
(B) 
& 
(TB) 
& 
(XT) 
& 
(BY) 
\\
\midrule
Input Parameters 
& 
$m_T, s_L^u$ 
& 
$m_B, s_L^d$ 
& 
$m_T, s_R^u, s_R^d$ 
& 
$m_T, s_R^u$ 
& 
$m_B, s_R^d$ 
\\
\bottomrule
\end{tabular}
\caption{
\label{VLQsParameters}
All input representations of 
the VLQs scenarios 
in the mass-eigenstates 
and mixing angle factors.
}
\end{table}
\end{center}
Note that these (XTB) or (TBY) 
triplet scenarios are not listed 
in Table~\ref{VLQsParameters}, which 
consistently shares similarities with 
the (T) or the (B) singlet going along 
with (TB) doublet cases 
within the 2HDM Type-II framework, 
respectively, are temporarily 
neglected in this work.

At the next paragraphs, we consider 
in details the interaction Lagrangian 
for these couplings of charged Higgs 
$H^\pm$, scalar CP-even $h,H$ and 
CP-odd $A$ Higgs particles, 
as well as these two massive vector 
bosons $W^\pm$ and $Z$ in a 
representative scenario 
for our numerical phenomenology.
In particular, the 2HDM Type II 
supplemented by vector-like quarks in 
the representation of doublet XT 
is predominantly considered 
as an investigated example. 
The most general Yukawa Lagrangian for 
the doublet VLQs XT scenario, 
which is augmented from the quark 
sector in Eq.~(\ref{Eq:YukLag}) for 
the 2HDM Type II, with 
the new weak iso-spin quark doublet 
$\bar{Q}^{0}_{L4} = (\bar{u}^0_{L4} \,\, 
\bar{d}^0_{L4}) \equiv (\bar{X}_{L} \,\, 
\bar{T}^0_{L})$, in which only $T^0$ 
primarily mix with the third quark 
generation $(i = j = 3)$ such as 
SM top quark $t$, that can be 
presented as detailed below
\begin{eqnarray}
-
\mathcal{L}_{Y}
&=&
Y^{u}_{ij} \,
\bar{Q}^{0}_{L i}
\tilde{\Phi}_{2}
u^{0}_{R j}
+
Y^{d}_{ij} \,
\bar{Q}^{0}_{L i}
\Phi_{1}
d^{0}_{R j}
+
\underbrace{
	\;
	Y^{u}_{4j} \,
	\bar{Q}^{0}_{L4}
	\Phi_{2}
	u^{0}_{R j}
	\;}_{\text{doublet XT}}
+
\text{h.c.}
\end{eqnarray}
Furthermore, due to the new doublet 
$\bar{Q}^{0}_{L4}$ has a hyper-charge 
value of $7/6$, we then extend 
the Lagrangian by exchanging 
$\tilde{\Phi}_{2} \leftrightarrow 
\Phi_{2}$ for the case of $\Phi_{2}$, 
that only couples to up-type quarks 
in 2HDM--II model, such that
\begin{eqnarray}
\mathcal{L}_{Y}
&=&
-  
\dfrac{Y^{u}_{33}}{\sqrt{2}}
\big(
v_{2}
+
h_{2}
-
i \, 
z_{2}
\big)
\bar{t}^{0}_{L}
t^{0}_{R}
- 
\dfrac{Y^{d}_{33}}{\sqrt{2}}
\big(
v_{1}
+
h_{1}
+
i \, 
z_{1}
\big)
\bar{b}^{0}_{L}
b^{0}_{R}
-  
\dfrac{Y^{u}_{43}}{\sqrt{2}}
\big(
v_{2}
+
h_{2}
+
i \, 
z_{2}
\big)
\bar{T}^{0}_{L}
t^{0}_{R}
\nonumber 
\\[0.25cm]
&&
+
Y^{u}_{33} \,
\omega^{-}_{2}
\bar{b}^{0}_{L}
t^{0}_{R}
- 
Y^{d}_{33} \,
\omega^{+}_{1}
\bar{t}^{0}_{L}
b^{0}_{R}
- 
Y^{u}_{43} \,
\omega^{+}_{2}
\bar{X}_{L}
t^{0}_{R}
+
\text{h.c.}
,
\label{Eq:LagYukDoubletXT}
\end{eqnarray}
where the charge $5/3$ quark $X$ 
only obviously has charged current 
interactions with charge $2/3$ quarks 
such as $t$ and $T$ as depicted 
in the last term. 
By expanding VEVs for 
the mixing mass matrices of 
up-type quarks as well as 
the down-type quark sector is still
systematically diagonalized 
in this case, we then obtain 
that all relevant 
couplings~\cite{Benbrik:2025PRD} 
for these scalar charged and 
neutral Higgs bosons are acquired 
in detail as follows:
\begin{eqnarray}
\mathcal{L}_{Y}^{H^\pm}
&=&
\Big[
c_{\beta}
Y^{u}_{33} \,
\big(
c^{u}_{R}
\,
\bar{t}
+
s^{u}_{R}
\,
\bar{T}
\big)
P_{L}
\,
b 
+
s_{\beta}  
Y^{d}_{33} \,
\big(
c^{u}_{L}
\,
\bar{t}
+
s^{u}_{L}
\,
\bar{T}
\big)
P_{R}
\,
b
\\
&&
\hspace{0.5cm}
- 
c_{\beta}
Y^{u}_{43} \,
\bar{X}
P_{R}
\big(
c^{u}_{R}
\,
t
+
s^{u}_{R}
\,
T
\big)
\Big]
H^{+}
+
\text{h.c.}
,
\nonumber
\\
\mathcal{L}_{Y}^{h,H}
&=&
-  
\dfrac{1}{\sqrt{2}}
\Big[
Y^{u}_{33}
\big(
c^{u}_{L}
\,
\bar{t}
+
s^{u}_{L}
\,
\bar{T}
\big)
\big(
c^{u}_{R}
\,
t
+
s^{u}_{R}
\,
T
\big)
P_{R}
\big(
s_{\alpha} 
H 
+ 
c_{\alpha} 
h
\big)
\\
&&\hspace{1.25cm}
+
Y^{u}_{43}
\big(
-
s^{u}_{L}
\,
\bar{t}
+
c^{u}_{L}
\,
\bar{T}
\big)
\big(
c^{u}_{R}
\,
t
+
s^{u}_{R}
\,
T
\big)
P_{R}
\big(
s_{\alpha} 
H 
+ 
c_{\alpha} 
h
\big)
\nonumber
\\
&&\hspace{1.25cm}
+
Y^{d}_{33}
\,
\bar{b}
\,
b
P_{R}
\big(
c_{\alpha} 
H 
- 
s_{\alpha} 
h
\big)
\Big]
+
\text{h.c.}
,
\nonumber
\\
\mathcal{L}_{Y}^{A}
&=&
-
\dfrac{i}{\sqrt{2}}
\Big[
-
c_{\beta} 
Y^{u}_{33}
\big(
c^{u}_{L}
\,
\bar{t}
+
s^{u}_{L}
\,
\bar{T}
\big)
\big(
c^{u}_{R}
\,
t
+
s^{u}_{R}
\,
T
\big)
P_{R}
\\
&&\hspace{1.5cm}
+
c_{\beta} 
Y^{u}_{43}
\big(
-
s^{u}_{L}
\,
\bar{t}
+
c^{u}_{L}
\,
\bar{T}
\big)
\big(
c^{u}_{R}
\,
t
+
s^{u}_{R}
\,
T
\big)
P_{R}
-
s_{\beta} 
Y^{d}_{33}
\,
\bar{b}
\,
b
P_{R}
\Big]
A
+
\text{h.c.}
,
\nonumber
\end{eqnarray}
in which these above mentioned Yukawa 
factors from solving of the diagonal 
mixing matrices are reading as below
\begin{eqnarray}
Y^{u}_{33}
=
\dfrac{\sqrt{2}}{v s_{\beta}}
\,
\dfrac{c^{u}_{L}}{c^{u}_{R}}
m_{t}
,
\qquad \quad
Y^{u}_{43}
=
\dfrac{\sqrt{2}}{v s_{\beta}}
\,
\dfrac{
	{s^{u}_{R}}^2
	-
	{s^{u}_{L}}^2
}{
	s^{u}_{L}
	c^{u}_{R}
}
m_{t}
,
\qquad \quad
Y^{d}_{33}
=
\dfrac{\sqrt{2}}{v c_{\beta}}
\,
m_{b}
.
\end{eqnarray}
Additionally, as considered
in Ref.~\cite{TranDT:2025EPJC},
we also utilize these useful relations 
between two mixing angles $\alpha$ 
and $\beta$ in terms of 
$s_{\beta - \alpha}$ and 
$t_{\beta}$ in order to express 
all viable mixing angle ratios 
for these above couplings. 
In the 2HDM Type II with XT doublet 
scenario, there are both new 
left- and right-handed 
VLQ doublets under one group 
$SU(2)_{L}$ that have in addition to 
the standard model fields, the quark 
interactions with $W^{\pm}$ and $Z$ 
bosons~\cite{Arhrib:2025EPJC} 
are given in a term-by-term manner 
as detailed below
\begin{eqnarray}
\mathcal{L}_{W^\pm}
&=&
-
\dfrac{g}{\sqrt{2}}
\Big[
\bar{t}^{0}_{L}
\gamma^{\mu}
b_{L}
W^{+}_\mu
+
\bar{X}_{L}
\gamma^{\mu}
\,
T^{0}_{L}
W^{+}_\mu
+
\bar{X}_{R}
\gamma^{\mu}
\,
T^{0}_{R}
W^{+}_\mu
\Big]
+
\text{h.c.}
,
\\
\mathcal{L}_{Z}
&=&
-
\dfrac{g}{2 c_{W}}
\Big[
Z_{\mu}
\bar{t}^{0}_{L}
\gamma^{\mu}
t^{0}_{L}
-
2 Q_{t} s_{W}^2
Z_{\mu}
\bar{t}
\gamma^{\mu}
t
-
Z_{\mu}
\bar{b}_{L}
\gamma^{\mu}
b_{L}
-
2 Q_{b} s_{W}^2
Z_{\mu}
\bar{b}
\gamma^{\mu}
b
\nonumber
\\
&&\hspace{1.25cm}
-
Z_{\mu}
\bar{T}^{0}_{L}
\gamma^{\mu}
T^{0}_{L}
-
Z_{\mu}
\bar{T}^{0}_{R}
\gamma^{\mu}
T^{0}_{R}
-
2 Q_{T} s_{W}^2
Z_{\mu}
\bar{T}
\gamma^{\mu}
T
\\
&&\hspace{1.25cm}
+
Z_{\mu}
\bar{X}_{L}
\gamma^{\mu}
X_{L}
+
Z_{\mu}
\bar{X}_{R}
\gamma^{\mu}
X_{R}
-
2 Q_{X} s_{W}^2
Z_{\mu}
\bar{X}
\gamma^{\mu}
X
\Big]
,
\nonumber
\end{eqnarray}
then the rotation between mass- 
and weak-eigenstates is fully 
accomplished as in both 
Eqs.~(\ref{Eq:tTmixing}) and
(\ref{Eq:bBmixing}) in order to
acquire these neutral and 
charged current interactions.

In conclusion, we summarize all 
relevant coupling representations 
in the following terms of 
three interaction types 
for the VLQs XT doublet scenario 
in our numerical investigations 
afterwards. 
Furthermore, these appeared 
coupling factors such as 
$\mathcal{X}^{L,R}_{qq,\ldots}$, 
$\mathcal{Y}^{\,h,H,A}_{qq,\ldots}$ 
and so forth are obviously 
expressed in terms of 
these mixing angle factors
$s^{u,d}_{L,R}$ and $c^{u,d}_{L,R}$, 
the corresponding SM quark $m_{q}$ 
as well as VLQ $m_{Q}$ masses, 
two 2HDM parameters such as 
$s_{\beta - \alpha}$ and $t_{\beta}$.
All of the above could also be found 
consistently in 
Ref.~\cite{Arhrib:2025EPJC} with 
our excellent agreements.
In addition, within the scope of 
this paper, we do not temporarily 
consider these light-heavy interactions.
All representative types of 
the relevant interaction Lagrangian 
with the chiral projectors 
$P_{L,R} = (1 \mp \gamma^5)/2$ 
as well as the same coupling structure 
for these light and heavy scalar 
CP-even Higgs bosons correspondingly 
denoted as $\phi \equiv h, H$
are listed in Refs.~\cite{Arhrib:2024EPJC, 
Arhrib:2025EPJC, Benbrik:2025PRD}, 
as detailed below:
\subsection{Light-light interactions}
The Lagrangian for a light quark $q \equiv t,b$
modified by mixing with vector-like quarks,
reads as follows:
\begin{eqnarray}
\mathcal{L}_{Z}
&=&
-
\dfrac{g}{2 c_{W}}
\,
\bar{q}
\,
\gamma^{\mu}
\Big[
\pm
\mathcal{X}^{L}_{qq}
P_{L}
\pm
\mathcal{X}^{R}_{qq}
P_{R}
-
2 Q_{q} s_{W}^2
\Big]
q
\,
Z_\mu,
\\
\mathcal{L}_{\phi}
&=&
-
\dfrac{g \, m_{t}}{2 M_{W}}
\mathcal{Y}^{\,\phi}_{tt}
\,\,
\bar{t} t \phi
-
\dfrac{g \, m_{b}}{2 M_{W}}
\mathcal{Y}^{\,\phi}_{bb}
\,\,
\bar{b} b \phi,
\\
\mathcal{L}_{A}
&=&
-
i
\dfrac{g \, m_{t}}{2 M_{W}}
\mathcal{Y}^{A}_{tt}
\,
\bar{t}
\,
\gamma^5
t
A
+
i
\dfrac{g \, m_{b}}{2 M_{W}}
\mathcal{Y}^{A}_{bb}
\,
\bar{b} 
\,
\gamma^5
b
A
\end{eqnarray}
where the plus (minus) sign
in the $Z$ coupling term for $t \, (b)$.
Note that all coupling coefficients
can be found
in Refs.~\cite{Arhrib:2024EPJC,
Arhrib:2025EPJC,Benbrik:2025PRD}.
\subsection{Heavy-heavy interactions}
The Lagrangian for these heavy 
vector-like quarks $Q \equiv X, T, 
B, Y$ follows a similar notation,
\begin{eqnarray}
\mathcal{L}_{Z}
&=&
-
\dfrac{g}{2 c_{W}}
\,
\bar{Q}
\,
\gamma^{\mu}
\Big[
\pm
\mathcal{X}^{L}_{QQ}
P_{L}
\pm
\mathcal{X}^{R}_{QQ}
P_{R}
-
2 Q_{Q} s_{W}^2
\Big]
Q
\,
Z_\mu,
\\
\mathcal{L}_{\phi}
&=&
-
\dfrac{g \, m_{T}}{2 M_{W}}
\mathcal{Y}^{\,\phi}_{TT}
\,
\bar{T} T \phi
-
\dfrac{g \, m_{B}}{2 M_{W}}
\mathcal{Y}^{\,\phi}_{BB}
\,
\bar{B} B \phi,
\\
\mathcal{L}_{A}
&=&
-
i
\dfrac{g \, m_{T}}{2 M_{W}}
\mathcal{Y}^{A}_{TT}
\,
\bar{T}
\gamma^5
T
A
+
i
\dfrac{g \, m_{B}}{2 M_{W}}
\mathcal{Y}^{A}_{BB}
\,
\bar{B}
\gamma^5
B
A
\end{eqnarray}
in which  the plus (minus) sign 
in the $Z$ coupling term for each 
$X, T \, (B, Y)$.
All coupling coefficients
can be found
in Refs.~\cite{Arhrib:2024EPJC,
Arhrib:2025EPJC,Benbrik:2025PRD}.
\subsection{Light-heavy interactions}
Subsequently, these coupling terms 
involving a heavy $Q \equiv T, B$ and
a light $q \equiv t, b$ quark are reading
as follows:
\begin{eqnarray}
\mathcal{L}_{Z}
&=&
-
\dfrac{g}{2 c_{W}}
\,
\bar{q}
\,
\gamma^{\mu}
\Big[
\pm
\mathcal{X}^{L}_{qQ}
P_{L}
\pm
\mathcal{X}^{R}_{qQ}
P_{R}
\Big]
Q
Z_\mu
+
\text{h.c.},
\\
\mathcal{L}_{\phi}
&=&
-
\dfrac{g \, m_{T}}{2 M_{W}}
\bar{t}
\Big[
\mathcal{Y}^{L}_{\phi\,tT}
P_{L}
+
\mathcal{Y}^{R}_{\phi\,tT}
P_{R}
\Big]
T
\phi
-
\dfrac{g \, m_{B}}{2 M_{W}}
\bar{b}
\Big[
\mathcal{Y}^{L}_{\phi\,bB}
P_{L}
+
\mathcal{Y}^{R}_{\phi\,bB}
P_{R}
\Big]
B
\phi
+
\text{h.c.},
\\
\mathcal{L}_{A}
&=&
i
\dfrac{g \, m_{T}}{2 M_{W}}
\,
\bar{t}
\Big[
\mathcal{Y}^{L}_{AtT}
\,
P_{L}
-
\mathcal{Y}^{R}_{AtT}
\,
P_{R}
\Big]
T
A
-
i
\dfrac{g \, m_{B}}{2 M_{W}}
\,
\bar{b}
\Big[
\mathcal{Y}^{L}_{AbB}
\,
P_{L}
-
\mathcal{Y}^{R}_{AbB}
\,
P_{R}
\Big]
B
A
+
\text{h.c.}
\end{eqnarray}
in which  the plus and minus sign 
in the $Z$ coupling term are denoted 
for the each $(t, T)$ and $(b, B)$ 
pair, respectively. We note that
all coupling coefficients can be taken
in Refs.~\cite{Arhrib:2024EPJC,
Arhrib:2025EPJC,Benbrik:2025PRD}.
\section{One--loop induced expressions
for $A \rightarrow Z Z \gamma$}
\begin{figure}[ht]
\centering
\includegraphics[width=15cm,height=9cm]
{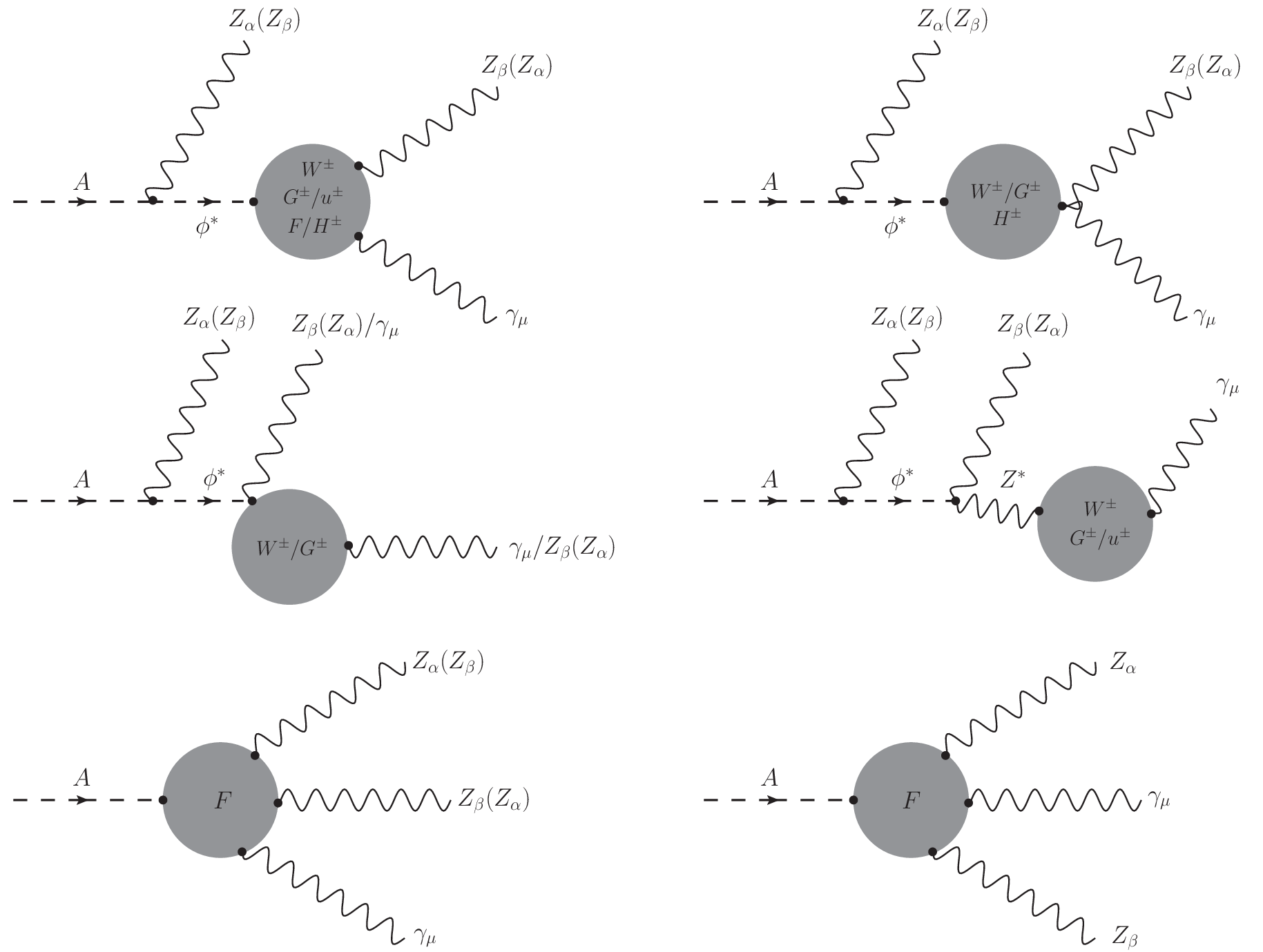}
\caption{
\label{Fig:diags}
All one-loop Feynman diagrams contributing
to the decay process $A \rightarrow ZZ\gamma$
are shown in the general $R_{\xi}$ gauge.
These diagrams include contributions from
fermions $F$ (including both SM quarks and
VLQs), $W^\pm$ gauge bosons, charged
Goldstone bosons $G^\pm$, charged ghosts
$u^\pm$, and charged Higgs bosons $H^\pm$.
Additional diagrams are obtained by
exchanging the two final-state $Z$ bosons.
}
\end{figure}

We now turn to the most relevant 
aspects of our contributing 
calculation. To begin with,
we will analyze
all detailed examinations about 
the participative decaying diagrams 
of one-loop induced process 
$A \rightarrow Z Z \gamma$ with 
the corresponding Lorentz indices 
to three vector bosons in final state 
are denoted as $Z_\alpha, Z_\beta$ 
and $\gamma_\mu$ in this section.
Within the 't Hooft--Feynman gauge
(HF), these complete induced
one-loop Feynman diagrams with 
getting primary contributions 
and obeying totally 
ultraviolet (UV) divergence 
finiteness related to 
this process are topologically 
shown in Fig.~\ref{Fig:diags}.
Furthermore, the Feynman rules 
for the 2HDM in our calculations 
can be acquired once the relevant 
2HDM Lagrangian is expanded and 
developed in terms of mass-eigenstates 
and could be found, for instance, 
in Refs.~\cite{Branco:2011iw}
for scalar Higgs sectors, 
as well as in Ref.~\cite{Aoki:2009PRD}
for specially studying about all types 
of Yukawa couplings in fermion sector.
Besides that since there is no new 
vector boson appearing in this model, 
the Lagrangian for all relevant 
interactions for vector gauge boson 
and ghost particles is also considered 
as in the SM, but with a tiny modification
in its couplings interacting with 
these Higgs bosons.
In the described Fig.~\ref{Fig:diags}, 
there are three sets of one-loop 
Feynman diagrams under our investigation 
that induce this decay process such as 
triangle, self-energy and box diagrams 
as can be seen as three distinct topologies.
The first topology is mainly considered 
as reducible diagrams and arises from 
the $A Z \phi$ coupling, which is 
the so-called pure two-body decay 
at tree-level, followed by 
the off-shell one-loop decay process 
$\phi^* \rightarrow Z \gamma$, in which 
the virtual propagating particle 
$\phi$ participated by the both scalar 
CP-even $h,H$ Higgs bosons.
From the initial three general 
diagrams corresponding to this topology, 
their contributions are consistently 
emerging from the intermediate 
scalar boson via loops carrying 
charged fermions $F$ 
(including both SM quarks and VLQs), 
the $W^\pm$ gauge boson accompanied by 
their associated Goldstone bosons $G^\pm$
and ghost $u^\pm$ particles in 
the context of the HF gauge, and 
in addition to the charged scalar 
boson $H^\pm$ appearing in the 2HDM model.
A next diagram for the second category 
has also the same topological structure 
in terms of both Higgs poles 
$\phi \equiv h,H$ with the former one, 
but the $\phi Z Z$ coupling combined 
with the off-shell $Z-\gamma$ mixing 
diagrams by a virtual massive 
vector boson $Z^*$ is considered here. 
It should be noted that 
all charged fermions $F$ 
and the charged Higgs particle $H^\pm$ 
have no contributions in this type of 
mixing diagram. 
Furthermore, other analogous types for 
mixture contributions also listed in 
this model such as these mixing diagrams 
of the off-shell $G^0 - \gamma$ for 
the neutral Goldstone bosons as well as 
the mixing $A - \gamma$ for the CP-odd 
Higgs boson, are considered for 
completeness and neglected by 
the analytical cancellations of 
their contributions.
Moreover, for the triangle and 
self-energy diagrams, there are such 
additional diagrams that are obtained 
by exchanging the $Z$ bosons 
in the final state.

All these one-loop calculations in 
the 2HDM are performed in details 
within the on-shell renormalization 
scheme for both gauge and 
fermion sector as outlined in 
Ref.~\cite{Aoki:1982PTPS}, 
as well as for scalar sector employed 
by the improved on-shell renormalization 
scheme as carried out in Ref.~\cite{Kanemura:2017PRD}.
Moreover, in order to regularize 
all one-loop Lorentz-covariant 
tensor integrals, the dimensional 
regularization is utilized in the
space-time dimension $D = 4 - 2 \epsilon$, 
as described in Ref.~\cite{Hahn:1999CPC}.
Therefore, sum of these one-loop 
contributions from both triangle and 
self-energy diagrams relevant to 
the sub-process 
$\phi^* \rightarrow Z \gamma$, excluding 
charged fermions $F$ and Higgs bosons 
$H^\pm$, is also employed 
in our computing. 
After the reduction scheme which 
are reduced down to a combination of 
one-, two-, three-, and four-point 
Passarino-Veltman (PV) scalar functions 
such as $A_0, B_0, C_0$ and $D_0$ 
in order to solve the loop integrals 
is analytically conducted.
Then we confirm that the result has 
an UV-divergent cancellation of 
the terms proportional to the 
$\epsilon^{-1}$-pole, which totally 
vanished by expanding in the general 
dimension $D$ for the sum of 
these various terms involving 
the Passarino-Veltman 
decomposition~\cite{Denner:1993FP} 
of the tensor integrals. 
It means that our final result has 
ultraviolet divergences canceled out 
when the PV reduction scheme was applied 
for those contributions.

The last topology consists of 
these one-loop diagrams with 
four external legs is depicted in the 
two remaining ones. 
There are three additional diagrams 
that are acquired by exchanging 
the $Z$ bosons in total.
Due to CP invariance in the scalar 
Higgs sector of 2HDM, the decay process 
cannot proceed via box diagrams with 
both gauge $W^\pm$ and charged Higgs 
$H^\pm$ bosons in loops. 
In more details, this could be explained 
by the CP-odd scalar $A$ does not directly 
interact with a pair of these charged 
vector gauge or scalar Higgs bosons, 
as well as even though a pair of 
$A \, W^\pm H^\mp$ vertex can exist 
in the 2HDM, the absence of $V W^\pm H^\mp$ 
coupling for $V \equiv Z, \gamma$ 
at the tree level is always consulted 
in our current results.
The main contributions of box diagrams 
are thus expected to arise from 
the heaviest fermions.
As a result, the dynamical content for 
these box diagrams is rather simple 
in the sense that there is only one kind 
of particle circulating into the loop, 
namely, SM charged fermions with 
top quark $t$ taken as an example, and 
then new charged states like VLQs are 
fully emerged exchanging in these loops.
This analogous expanding is also performed 
in one-loop triangle diagrams for 
fermion contributions in the next 
phenomenological calculations.
Note that, in the trace of a chain 
of Dirac matrices, we have used 
the 't Hooft-Veltman 
scheme~\cite{Hooft:1972NPB} for 
the treatment of $\gamma^5$ 
matrices appears in general $D$ 
dimensions related to all fermions 
exchanging in these mentioned 
one-loop diagrams.

We now move on to briefly describe 
the all tools used in our next calculations.
These above Feynman diagrams are 
fully generated with the help of 
\textsf{FeynArts} 
package~\cite{Hahn:2001CPC} with 
extending in general new couplings of 
the VLQs representations 
investigated in this study.
To completely evaluate the loop integrals 
and absolutely perform the PV reduction to 
scalar basis functions, 
we have analytically utilized 
the advantage of 
\textsf{FeynCalc}~\cite{Mertig:1991CPC, Mertig:2016CPC} and have clarified 
the cancellation of UV-divergent poles 
in the final result by using two 
interface tools such as 
\textsf{Package-X}~\cite{Patel:2015CPC} 
and \textsf{LoopTools}~\cite{Hahn:1999CPC}
packages with
more useful options in our analysis, 
which are connected to \textsf{FeynCalc} 
via the \textsf{FeynHelpers} 
package~\cite{Vladyslav:2017CPC}.
Subsequently, these loop functions are 
numerically employed by utilizing 
\textsf{CollierLink}~\cite{Patel:2015CPC} 
that provides a \textsf{Mathematica}'s 
calculating platform
linked to the \textsf{Collier} 
library~\cite{Denner:2006NPB, 
Denner:2011NPB, Denner:2017CPC}. 
In the end, for the multidimensional 
numerical integrating over 
the phase-space variables, 
we apply the tool 
\textsf{Vegas}~\cite{Peter:1978JCP} 
from the \textsf{Cuba} 
library~\cite{Hahn:2005CPC}.
In summary, this methodology allows 
for the numerical calculation of 
partial decay rates and their 
invariant-mass distributions through 
these mentioned computational programs. 
In our phenomenological investigation, 
we examine the branching ratios (Br) 
of these considered decay channels 
across the specific model's parameter 
spaces.

Regarding to these subsequent 
discussions, we will supply a detailed 
derivation of the computations of 
such process $A \rightarrow Z Z \gamma$ 
in progressive.
The one-loop induced amplitude 
for the process 
$A \, (p) \rightarrow Z_{\alpha} (k_1) \, Z_{\beta} (k_2) \, \gamma_{\mu} (k_3)$ 
could, in general, be presented 
through the following gauge-invariant
Lorentz structures as below
\begin{eqnarray}
\mathcal{M}_{A \rightarrow ZZ\gamma}
&=&
\mathcal{M}^{\alpha \beta \mu} \,
\varepsilon^*_\alpha (k_1) \,
\varepsilon^*_\beta (k_2) \,
\varepsilon^*_\mu (k_3)
,
\label{Eq:Amplitude}
\end{eqnarray}
where $\varepsilon^*_\nu (k_l)$ for 
$l = 1,2,3$ along with 
the relevant tensor index 
$\nu \equiv \alpha, \beta, \mu$ 
represents for the polarization 
vectors corresponding to 
the neutral gauge bosons 
in the final state. 
Besides that the invariant masses, 
due to the on-shell massive conditions, 
are listed such as $p^2 = M_{A}^2, \; 
k_1^2 = k_2^2 = M_Z^2, \; k_3^2 = 0$.
In addition, the following Lorentz 
invariant kinematic variables $
k_{ij}
=
(k_i + k_j)^2
$
within $i < j$ for $i,j = 1,2,3$ 
are consistently expressed in 
whole subsequent calculations 
regarding to all the scalar products 
between the four-momenta terms.
These variables, which are not 
all independent, satisfy the 
Mandelstam relation as 
$k_{12} + k_{13} + k_{23}
=
M_{A}^2 + 2 M_Z^2$ 
by four-momentum conservation.
In general, the Lorentz structure 
of $\mathcal{M}^{\alpha \beta \mu}$, 
parameterized in terms of totally 
36 basic form factors, including 
these terms $\mathcal{F}_{ij}$ and 
$\mathcal{F}_{ijk}$ for 
$i,j,k = 1,2,3$ are the coefficients 
of metric tensor and four-momenta, 
is provided in detail below
\begin{eqnarray}
\mathcal{M}^{\alpha \beta \mu}
&=&
\sum \limits_{i = 1}^3
\Big[
\mathcal{F}_{1i} \,
k_i^\alpha 
g^{\beta \mu}
+
\mathcal{F}_{2i} \,
k_i^\beta
g^{\alpha \mu}
+
\mathcal{F}_{3i} \,
k_i^\mu 
g^{\alpha \beta}
\Big]
+
\sum \limits_{i, j, l = 1}^3
\mathcal{F}_{i j l} \,
k_{i}^\alpha
k_{j}^\beta
k_{l}^\mu
.
\end{eqnarray}
Within the context of this paper, 
furthermore, since these VLQs 
light-heavy interactions are not 
temporarily scrutinized in all 
one-loop contributions of 
the charged fermions $F$, the arising 
term involving the $\gamma^5$ matrix 
in the trace of a chain of Dirac 
matrices consistently disappears.
This would eliminate the formation 
of all necessary Levi-Civita tensors 
in the invariant amplitude of 
such fermion loops.  

In addition, because of the 
transversality condition for 
the external vector bosons, 
the following relation holds as 
$k_1^\alpha \,
\varepsilon^*_\alpha (k_1)
=
k_2^\beta \,
\varepsilon^*_\beta (k_2)
=
k_3^\mu \,
\varepsilon^*_\mu (k_3)
=
0$.
As a result, we are then likely to 
omit all the terms proportional 
to $k_1^\alpha, k_2^\beta$ and $k_3^\mu$ 
in the invariant amplitude 
in order to rationalize enormously 
our current computing which their 
resulting expression only consists of 
14 form factors as follows:
\begin{eqnarray}
\label{Eq:Amp14terms}
\mathcal{M}^{\alpha \beta \mu}
&=&
\Big[
\mathcal{F}_{12} \,
k_2^\alpha
+
\mathcal{F}_{13} \,
k_3^\alpha  
\Big]
g^{\beta \mu}
+
\Big[
\mathcal{F}_{21} \,
k_1^\beta
+
\mathcal{F}_{23} \,
k_3^\beta
\Big]
g^{\alpha \mu}
\nonumber
\\
&&\hspace{-0.35cm}1
+
\Big[
\mathcal{F}_{31} \,
k_1^\mu 
+
\mathcal{F}_{32} \,
k_2^\mu 
\Big]
g^{\alpha \beta}
\\
&&\hspace{-0.35cm}
+
\Big[
\mathcal{F}_{211} \,
k_2^\alpha
+
\mathcal{F}_{311} \,
k_3^\alpha
\Big]
k_1^\beta
k_1^\mu
+
\Big[
\mathcal{F}_{212} \,
k_2^\alpha
+
\mathcal{F}_{312} \,
k_3^\alpha
\Big]
k_1^\beta
k_2^\mu
\nonumber
\\
&&\hspace{-0.35cm}
+
\Big[
\mathcal{F}_{231} \,
k_2^\alpha
+
\mathcal{F}_{331} \,
k_3^\alpha
\Big]
k_3^\beta
k_1^\mu
+
\Big[
\mathcal{F}_{232} \,
k_2^\alpha
+
\mathcal{F}_{332} \,
k_3^\alpha
\Big]
k_3^\beta
k_2^\mu
.
\nonumber
\end{eqnarray}
Due to the fact that have featured 
the on-shell final photon state, 
the gauge-invariant amplitude should 
follow the Ward identity for simplicity 
in our calculations. 
As a result, applying this character 
for photon $\gamma_\mu (k_3)$ is going 
to lead the amplitude vanished partly 
by replacing $\varepsilon^*_\mu (k_3) 
\rightarrow k_3^\mu$, and as long as 
be satisfied by $k_3^\mu \,
\mathcal{M}^{\alpha \beta \mu} = 0$.
Therefore we are likely to obtain 
two useful relations in terms of 
such form factors for photon 
$\gamma_\mu (k_3)$ as detailed below
\begin{eqnarray}
\mathcal{F}_{12}
&=&
-
( k_1 \cdot k_3 )
\,
\mathcal{F}_{231}
-
( k_2 \cdot k_3 )
\,
\mathcal{F}_{232}
,
\\[0.25cm]
\mathcal{F}_{21}
&=&
-
( k_1 \cdot k_3 )
\,
\mathcal{F}_{311}
-
( k_2 \cdot k_3 )
\,
\mathcal{F}_{312}
.
\end{eqnarray}
As a consequence, the Lorentz-invariant 
amplitude $\mathcal{M}^{\alpha \beta \mu}$ 
can be compactly formulated as follows
\begin{eqnarray}
\label{Eq:AmpComp}
\mathcal{M}^{\alpha \beta \mu}
&=&
k_3^\alpha  
\Big[
\mathcal{F}_{13} \,
g^{\beta \mu}
+
\mathcal{F}_{331} \,
k_3^\beta
k_1^\mu
\Big]
+
k_1^\mu 
\Big[
\mathcal{F}_{31} \,
g^{\alpha \beta}
+
\mathcal{F}_{211} \,
k_2^\alpha
k_1^\beta
\Big]
\\[0.125cm]
&&\hspace{-0.35cm}
+
\hspace{+0.125cm}
\mathcal{F}_{311} \,
k_1^\beta
\Big[
k_3^\alpha
k_1^\mu
-
( k_1 \cdot k_3 )
g^{\alpha \mu}
\Big]
+
\mathcal{F}_{312} \,
k_1^\beta
\Big[
k_3^\alpha
k_2^\mu
-
( k_2 \cdot k_3 )
g^{\alpha \mu}
\Big]
\nonumber
\\[0.125cm]
&&\hspace{-0.35cm}
+
\hspace{+0.125cm}
\text{terms} \,
\big\{
k_1 
\leftrightarrow 
k_2,
\alpha 
\leftrightarrow 
\beta
\big\}
,
\nonumber 
\end{eqnarray}
where the ``terms" is also verified 
for these such additional form factors 
by interchanging these Mandelstam 
variables $k_{13} \leftrightarrow 
k_{23}$ in its expressions. 
It is stressed that these all 
form factors, which depend on two 
independent variables among the 
three Lorentz invariant kinematic 
quantities $k_{ij}$ as well as 
the squared masses of the relevant 
particles participating in the loop, 
are completely presented in terms of 
the scalar one-loop Passarino-Veltman 
functions.

As previously known, after writing out 
the corresponding amplitude for 
each Feynman diagram as presented in 
Fig.~\ref{Fig:diags}, the one-loop 
tensor integrals are also reduced to 
scalar PV functions via 
the Passarino-Veltman reduction 
scheme as concentratedly utilized 
in \textsf{FeynCalc} package with 
the analytical aid of \textsf{Package-X} 
in order to clarify totally 
ultraviolet-divergent finiteness 
and to obey renormalization scale 
independence when summing over all 
relevant diagrams, and the total 
amplitude then is properly decomposed 
into these mentioned form factors as 
followed the amplitude structure of 
$\mathcal{M}^{\alpha \beta \mu}$ 
in Eq.~(\ref{Eq:Amp14terms}).
Subsequently, by following the Ward 
identity, these above relations of 
form factors are also verified 
in our analysis.
From now on, the expression for 
such resulting amplitude presented 
in Eq.~(\ref{Eq:AmpComp}) along with 
the identities in the last term 
thus assists as the starting point 
of our research. 
This determines that these six 
independent form factors are 
mandatory to carry out 
the calculation of decay rates 
as well as branching ratios  
in the succeeding investigations.
Therefore the above form factors 
$\mathcal{F}_{ij}$ and $\mathcal{F}_{ijk}$ 
could be totally decomposed into 
two distinguished contributions 
arising from the scalar CP-even 
Higgs poles $\phi = h, H$ diagrams, 
including both triangle and self-energy 
topologies, as well as these one-loop 
box diagrams relevant to 
the decay process.
In particular, the contributions of 
all charged particles $F$ such as SM 
fermions $f = e,\mu,\tau,u,c,t,\ldots$ 
and VLQs as well, the $W^\pm$ gauge boson, 
and the charged scalar boson $H^\pm$ 
are gathering at these $\phi$-pole's 
form factors, whereas box contribution 
only benefits from these charged 
fermions $F$ exchanging in loop.
It should be noted that all box 
diagrams are contributing to 
six form factors in our analysis.

In this work, these mentioned form 
factors are temporarily omitted 
due to the analytically complex 
length of their expressions, and 
we thus only focus on the next 
phenomenological results.
Furthermore, regarding to the detailed 
expressions of total decay widths 
at leading order for the heavy CP-even 
Higgs $H$ and CP-odd Higgs $A$ in 2HDM, 
which are consistently utilized into 
the computing these $\phi$-pole's form 
factors as well as the branching ratios 
of this decay process, respectively, 
are also considered into account.
To totally obtain the complete 
numerical result of partial decay 
rates and branching ratios, 
we need the total decay widths 
of these relevant Higgs bosons, 
which are analytically computed by 
taking into account all feasible 
two-body decay channels as well as 
have already been studied 
in the literature considerably 
of corresponding 
Refs.~\cite{Kanemura:2022NPB} 
and \cite{Aiko:2023NPB}.

Focusing on the subsequent discussions, 
together with all relevant one-loop 
form factors determined, we implement 
to compute the partial decay rates for 
the process.
By employing the earlier derived 
expression for the invariant amplitude
$\mathcal{M}^{\alpha \beta \mu}$ 
corresponding to the decay channel 
$A \, (p) \rightarrow Z_{\alpha} (k_1) \, Z_{\beta} (k_2) \, \gamma_{\mu} (k_3)$, 
a direct summation over the 
polarizations $\varepsilon^*_\nu (k_l)$
of both $Z$ bosons and the photon 
in these final states yields 
the squared amplitude.
Knowing this quantity allows us to 
carry out the decay width via 
the following expressions particularly 
presented in the below Appendix A, 
which are consistently considered 
all cases of polarization 
modes for these final $Z$ bosons 
as well as unpolarized situations.  
Consequently, for an example in 
particular, the occurrence of 
unpolarized decay rate of 
this process is compactly 
formulated for our numerical 
results as detailed below
\begin{eqnarray}
&&
\Gamma_{A \rightarrow Z Z \gamma}
=
\dfrac{1}{2 M_{A}}
\int
d \Phi_3
\,
\big|
\mathcal{M}_{A \rightarrow ZZ\gamma}^{\text{unpol}}
\big|^2
=
\dfrac{1}{256 \pi^3 M_{A}^3}
\int \limits_{k_{12}^\text{min}} ^{k_{12}^\text{max}}
d k_{12}
\int \limits_{k_{13}^\text{min}} ^{k_{13}^\text{max}}
d k_{13}
\,
\big|
\mathcal{M}_{A \rightarrow ZZ\gamma}^{\text{unpol}}
\big|^2
,
\label{DRunpol}
\end{eqnarray}
where the three-body phase space 
$d \Phi_3$, in general, can be 
developed in terms of two Mandelstam 
variables, in which the integration 
of unpolarized square amplitude 
$\big|
\mathcal{M}_{A \rightarrow ZZ\gamma}^{\text{unpol}}
\big|^2$ with respect 
to $k_{13}$ is performed in first, 
followed by $k_{12}$.
In addition, along with the upper 
and lower integration limits are 
generally expressed into 
the allowed kinematic region 
for the decay process 
as follows
\begin{eqnarray}
&&
k_{12}^\text{min}
=
4 M_Z^2
,
\qquad \qquad
k_{12}^\text{max}
=
M_{A}^2
,
\\[0.25cm]
&&
k_{13}^\text{max(min)}
=
\dfrac{1}{2}
\Bigg[
M_{A}^2
+
2 M_Z^2
-
k_{12}
\pm
\big(
M_{A}^2
-
k_{12}
\big)
\sqrt{
	1
	-
	4 M_Z^2 / k_{12}
}
\;
\Bigg]
,
\end{eqnarray}
in which these Lorentz-invariant 
kinematic variables are totally 
interdependent together with 
the corresponding on-shell masses. 
Furthermore, a major advantage of 
this approach is that the evaluated 
process's partial decay rates 
$\Gamma_{A \rightarrow Z Z \gamma}$ 
can be expressed directly through 
these above mentioned six independent 
one-loop form factors $\mathcal{F}_{ij}$ 
and $\mathcal{F}_{ijk}$, which are 
systematically embedded within 
the squared amplitude.
Ultimately, this procedure is 
also done similarly for 
other polarization circumstances.

The last but not the least,
based on all induced one-loop 
contributions from explicit Feynman 
diagram groups in Fig.~\ref{Fig:diags}, 
the allowed kinematic region 
for the decay channel depends on 
the value of on-shell mass for 
the incoming scalar CP-odd Higgs 
boson $A$ as primarily compared 
(greater than or equal to, 
or less than) to the virtual mass 
at poles of both exchanged scalar 
CP-even Higgs bosons $\phi = h, H$. 
We will then analyze their 
characteristics in both scenarios 
for the resulting partial decay 
rates in the following paragraphs.
Note that the allowed parameter space 
of mass $M_{A}$ will be thoroughly 
investigated in the subsequent 
phenomenological results. 

On the one hand, in the kinematic 
circumstance of $M_{A} \geq 
M_{\phi} + M_{Z}$ satisfied, 
which are originated from the 
sub-process $A \rightarrow Z \phi^*$ 
at tree-level, 
indeed the incoming scalar boson $A$ 
is heavy enough to decay both 
on-shell CP-even Higgs bosons $\phi$, 
therefore the Narrow Width Approximation 
(NWA) approach is properly evaluated.
In this scenario, thus, by using 
the Breit--Wigner propagator for 
the Higgs pole of two exchanged 
scalar bosons $\phi$ in the squared 
amplitude, as seen in 
Eq.~(\ref{DRunpol}), can be integrated 
and the decay process 
$A \rightarrow Z Z \gamma$ proceeds 
as the pure two-body decay 
$A \rightarrow Z \phi$, then 
connected by the on-shell decay 
$\phi \rightarrow Z \gamma$ 
as mentioned before, 
can be written as follows
\begin{eqnarray}
\Gamma_{A \rightarrow Z Z \gamma}^{\text{NWA}}
&=&
2 \times
\sum \limits_{\phi}
\Gamma_{A \rightarrow Z \phi}
\times
\text{Br} \,
\{ \phi \rightarrow Z \gamma \}
,
\end{eqnarray}
where the given factor $2$ denoted as 
two identical $Z$-bosons in final state, 
and the tree-level decay rates for 
$\Gamma_{A \rightarrow Z \phi}$ 
could be similarly found as in 
Ref.~\cite{TranDT:2025EPJC} 
for our numerical investigations.
In addition, the branching ratios 
for on-shell channel 
$\phi \rightarrow Z \gamma$ are 
particularly evaluated as $\text{Br} \,
\{\phi \rightarrow Z \gamma\} \equiv 
\Gamma_{\phi \rightarrow Z \gamma} / 
\Gamma_{\phi}^{\text{Total}}$, 
in which these formulas of partial 
decay rate corresponding to 
the one-loop induced process 
$\phi \rightarrow Z \gamma$ 
together with their mass $M_{\phi}^2$
for $\phi \equiv h,H$ are written 
in detail as follows
\begin{eqnarray}
\Gamma_{\phi \rightarrow Z \gamma}
&=&
\dfrac{1}{16 \pi}
\Bigg(
1
-
\dfrac{M_Z^2}{M_{\phi}^2}
\Bigg)
\dfrac{
|\mathcal{M}_{\phi \rightarrow Z \gamma}|^2
}{M_{\phi}}
,
\end{eqnarray}
in which the squared amplitude is given by 
a compact formula, that presented 
in terms of one-loop contributing 
form factors 
$\mathcal{F}^{W^\pm}_{\phi \rightarrow Z \gamma}$, 
$\mathcal{F}^{F}_{\phi \rightarrow Z \gamma}$ 
and 
$\mathcal{F}^{H^\pm}_{\phi \rightarrow Z \gamma}$ within the ’t Hooft--Feynman 
gauge as well.
Furthermore, these form factors 
are also expressed into the tensor 
Passarino-Veltman reduction 
coefficients and primarily denoted 
for these corresponding triangle 
$W^\pm$-boson, all charged fermions $F$ 
and charged scalar Higgs bosons $H^\pm$ 
diagram groups in the specific 2HDM,
then also supplemented by new charged 
states of VLQs such as $T, B, \ldots$ 
depend on each extending scenario.
This squared amplitude is 
expressed as detailed below
\begin{eqnarray}
|\mathcal{M}_{\phi \rightarrow Z \gamma}|^2
&=&
\dfrac{M_{\phi}^4}{8 \pi^4}
\Bigg(
1 
- 
\dfrac{M_Z^2}{M_{\phi}^2}
\Bigg)^2
\Bigg|
\,
g_{\phi \, W^\pm W^\mp}
\times
\mathcal{F}^{W^\pm}_{\phi \rightarrow Z \gamma}
\\[0.125cm]
&&
+
g_{\phi \, H^\pm H^\mp}
\times
\mathcal{F}^{H^\pm}_{\phi \rightarrow Z \gamma}
+
\sum 
\limits_{F \equiv f, T, B, \ldots}
N_C^F
\,
M_F^2
\times
g_{\phi \, F \bar{F}} 
\times
\mathcal{F}^{F}_{\phi \rightarrow Z \gamma}
\,
\Bigg|^2
.
\nonumber
\end{eqnarray}
All the necessary coupling 
constants $g_{\phi \, W^\pm W^\mp}, 
g_{\phi \, H^\pm H^\mp}$ and 
$g_{\phi \, F \bar{F}}$ 
in the above equation, 
as consistently presented in 
Table 1 as well as Appendix A of 
the Ref.~\cite{TranDT:2025EPJC} 
in the context of the 2HDM Type II, 
are denoted as the general couplings 
for a pair of corresponding 
charged particles interacting with 
the scalar CP-even Higgs bosons 
$\phi \equiv h,H$.
It should be noted that 
the general representation 
of all relevant fermion couplings 
for each VLQs situation considered 
as $g_{\phi \, F \bar{F}} = 
\big( e M_{F} / 2 M_{W} s_W \big) 
\times \mathcal{Y}^{\,\phi}_{FF}$ 
is taken into account in their 
contributions, therein the detailed 
expression of $\mathcal{Y}^{\,\phi}_{FF}$
is apparently referred to 
Ref.~\cite{Arhrib:2025EPJC} 
as mentioned in the Sect.~$2$.
Furthermore the factor $N_C^F$ denotes 
the number of colors for all charged 
fermion $F$ propagating in loop, 
acquiring the value of 1 for leptons 
and 3 for quarks, and their 
corresponding mass $M_{F}$.
Subsequently, each type of 
contributing form factor is 
reading as detailed follows
\begin{eqnarray}
\mathcal{F}^{W^\pm}_{\phi \rightarrow Z \gamma}
&=&
-\dfrac{e^2}{2 M_{W}^2 s_{2W}}
\Big\{
\Big[
c_{W}^2 
\big(M_{\phi}^2+10 M_{W}^2\big)
-
s_{W}^2 
\big(M_{\phi}^2+2 M_{W}^2\big)
\Big] 
\big(
C_{2}
+
C_{12}
+
C_{22}
\big)
\\[0.125cm]
&&\hspace{2.75cm}
+
2 M_{W}^2 
(3 c_{W}^2-s_{W}^2) 
C_{0}
\Big\}
(M_{Z}^2,0,M_{\phi}^2,
M_{W}^2,M_{W}^2,M_{W}^2)
\nonumber
\end{eqnarray}
in which the form factor 
$\mathcal{F}^{W^\pm}_
{\phi \rightarrow Z \gamma}$ is 
also verified by 
the ultraviolet-divergent finiteness, 
and these remaining ones for charged 
Higgs bosons $H^\pm$ and fermions $F$ 
then are expressed as below 
\begin{eqnarray}
\mathcal{F}^{H^\pm}_{\phi \rightarrow Z \gamma}
&=&
g_{\gamma H^\pm H^\mp}
\times
g_{Z H^\pm H^\mp}
\times
\big(
C_{2}
+
C_{12}
+
C_{22}
\big)
(M_{Z}^2,0,M_{\phi}^2,
M_{H^\pm}^2,M_{H^\pm}^2,M_{H^\pm}^2),
\\
\mathcal{F}^{F}_{\phi \rightarrow Z \gamma}
&=&
-
g_{\gamma F \bar{F}}
\times
g_{Z F \bar{F}}
\times
\Big[
4 
\big(
C_{2}
+
C_{12}
+
C_{22}
\big)
+
C_{0}
\Big]
(M_{Z}^2,0,M_{\phi}^2,
M_{F}^2,M_{F}^2,M_{F}^2)
\end{eqnarray}
where these general couplings for 
gauge bosons interacting with 
all charged fermion $F$ contributions 
are expressed as $g_{\gamma F \bar{F}} 
= e \times Q_{F}$ as well as 
the vector coupling component of 
the constant $g_{Z F \bar{F}} = 
\big( e / s_{2W} \big) \times 
\big( T^{3}_{F} - 2 s_{W}^2 Q_{F} \big)$ 
for the third component $T^{3}_{F}$ 
of isospin.
Moreover, the related charged Higgs 
couplings in the 2HDM are given by 
$g_{\gamma H^\pm H^\mp} = - e$ and 
$g_{Z H^\pm H^\mp} 
= - e \, c_{2W} / s_{2W}$. 

On the other hand, in the remaining 
scenario, it means that the condition 
$M_{A} < M_{\phi} + M_{Z}$ is strictly 
obeyed, the incoming pseudo-scalar 
boson $A$ will not be heavy in 
their mass enough to produce 
an on-shell Higgs boson $\phi$ 
in addition to the real $Z$ gauge boson.
Accordingly, we will acquire a pure 
three-body decaying process persistently 
induced by both reducible and box 
diagrams, the corresponding decay rate 
formula can be officially implemented 
as described in Eq.~(\ref{DRunpol}).
It should be also noted that 
in the mentioned kinematic region, 
$M_{A} \geq M_{\phi} + M_{Z}$ satisfied, 
we firmly verify that the pure 
three-body process augmented by 
the whole presence of box 
contributions in this work, 
the resulting decay rates 
are carried out as compared to 
the above NWA approach 
$\Gamma_{A \rightarrow Z Z \gamma}^{\text{NWA}}$ for 
completeness of our studying as well.
Nevertheless, the following 
investigating the invariant-mass 
distributions is totally utilized 
from the derived formula of decay rate 
$\Gamma_{A \rightarrow Z Z \gamma}$ 
in the entire kinematic region of 
incoming mass $M_{A}$ for 
the decay channel with 
the aid of these orders of 
integration reversal.
\section{Phenomenological results}
We now turn to the numerical analysis of the
one-loop-induced decay process
$A \rightarrow ZZ\gamma$,
focusing on its phenomenological implications
within the Type-II 2HDM and its extensions
incorporating Vector-Like Quarks (VLQs).
As discussed in Sect.~2, the seven independent
parameters of the 2HDM relevant to our study
are associated primarily with the Higgs sector
and are collected in $\mathcal{P}_{\text{2HDM}}$.
All other SM input parameters are fixed to the
values given in
Refs.~\cite{PDG:2024PRD,KHP:2024PTEP,KHP:2025PTEP}.
To incorporate the additional VLQ scenarios,
the corresponding supplementary parameters
used in our numerical analysis are listed in
Table~\ref{VLQsParameters}.
We then present numerical results for the
partial decay widths and the corresponding
branching ratios at several benchmark points
within the allowed parameter space.
Finally, we investigate the invariant-mass
distributions of the $Z\gamma$ system,
$m_{Z\gamma}$, for a selected final-state
$Z$ boson and the photon.

To begin with, we particularly 
present an overview of 
the allowed values of these 
input parameters as described above, 
according to relevant experimental 
and theoretical constraints of 
the 2HDM situation, in particular 
Type II scenario, which is consequently 
considered for embedded expansion 
of VLQs. 
Actually, from the recent analyses 
of the ATLAS and CMS 
Collaborations~\cite{ATLAS:2022Nature} 
after the Higgs boson discovery, 
it should be noted that 
the observed properties of 
the 125.1 GeV scalar boson found at 
the LHC are highly consistent with 
the SM expectations, consequently 
imposing strong constraints on 
the scalar sector of SM extensions.
In the specific 2HDM, as a result, 
the scalar boson $h$ is usually 
proposed to be the lightest one 
and then is identified with 
the SM Higgs boson, which constrains 
the parameter space of this model 
to a region very close to 
the so-called ``alignment limit'' 
of the 2HDM~\cite{Draper:2021JHEP}. 
It means that the value of 
$s_{\beta - \alpha}$ should be 
strictly taken the limit 
nearly tend to a value of $1$.
Therefore, the viable parameter 
space of the 2HDM is explicitly 
introduced in this work, 
such as enforcing 
$s_{\beta - \alpha} = 0.98$ 
adapted from the consistency with 
the SM alignment limit together with 
the mass of the light CP-even Higgs 
boson $M_{h} = 125.1$ GeV, as well as 
satisfied with the reliability of  
Higgs signal strength measurements 
currently reported by the LHC.
In addition, recent measurements on 
the $h \rightarrow Z \gamma$ decay 
rate~\cite{Aad:2024PRL} still leave 
room for deviations from SM predictions. 
A possibility to capture at once all 
such anomalies can be comprehensively 
addressed by introducing VLQs, 
which naturally affect the one-loop 
induced SM-like Higgs decay channels 
that are otherwise dominated by 
the SM top $t$ quark.
Similarly expanding VLQs in 
the context of the 2HDM Type II for 
both diphoton and one $Z$-boson with 
a photon channels~\cite{Andrei:2016PLB, 
Arhrib:2018PRD} 
is also previously investigated 
in one-loop contributions.  

Additionally, in the 2HDM Type II, 
by constraints from flavour physics 
such as experimental observables 
relevant to decay processes of 
$B$ mesons, a rather stringent 
lower limit applies on the charged 
Higgs mass $M_{H^\pm}$, 
of typical $580$ 
GeV~\cite{Misiak:2017EPJC, 
Benbrik:2024arxiv} 
or so, quite irrespectively of 
the other model parameters.
Furthermore, it should be noted 
that incorporating VLQs singlet 
and doublet scenarios into 
the 2HDM Type-II could relax 
this limit~\cite{Benbrik:2025PRD}, 
which is set to exceed $600$ GeV 
at present, together with the large 
$t_{\beta}$ is not excluded by 
current data therein. 
However, through large mixing angles 
and cancellations in flavour processes, 
electroweak precision observables 
(EWPOs) impose constraints that 
tend to keep its mass near 
the standard 2HDM Type-II limit.
Due to the updated lower bound 
of the charged Higgs boson mass, 
$M_{H^\pm} = 650$ GeV is set up 
as a representative benchmark.
Moreover, since compliance with 
EWPOs in the 2HDM Type-II generally 
restricts the difference of mass 
states within the scalar sector 
to a maximum of approximately 
$150$ GeV, we select that the mass of 
scalar heavy CP-even Higgs $H$ boson
as well as the dimensional parameter 
for the softly-breaking scale of 
the $Z_2$ symmetry defined as 
$m_{12}^2 = M^2 s_{\beta} c_{\beta}$ 
are both fixed at 
$M \equiv M_{H} = 500$ GeV.
Subsequently, for reinforcing these 
choices within the 2HDM Type II 
extending VLQs XT doublet scenario, 
which is numerically considered 
in the next phenomenological 
results, the consistency from updated 
$W^\pm$-boson mass of the CDF 
collaboration~\cite{Abouabid:2024JPG} 
may be taken into account therein.
By investigating the oblique parameters 
$S, T$ and $U$ with the presence of VLQs 
from the CDF-II data, the maximum mass 
splitting between the charged Higgs boson 
$H^\pm$ and the CP-even Higgs $H$ has 
the potential to arrive at 
approximately $200$ GeV by allowing 
only small VLQs mixing angles.

Eventually, in our numerical 
examination, we also consider 
two remaining parameter 
ranges as referred in 
Ref.~\cite{TranDT:2025EPJC} 
for the CP-odd Higgs boson mass 
$M_{A}$ and the mixing parameter 
$t_\beta$ such as $300 \; \text{GeV} 
\leq M_{A} \leq 1000\; \text{GeV}$ 
as well as $2 \leq t_\beta \leq 10$ 
as favored by both flavour physics 
constraints and high-precision 
measurements of EWPOs.
Recall that due to the Higgs sector 
of the MSSM, which is also concerned 
with $M_{A}$ and $t_{\beta}$ 
as totally independent 
inputs~\cite{Djouadi:2013JHEP}, 
is a special incarnation of 
the 2HDM whose all Yukawa 
interactions with Higgs doublets 
is of Type II.
As a consequence, the parameter 
variations emphasizing on both 
$\{ M_{A}, t_{\beta} \}$ will 
be implemented in the 
further discussions.
Aside from that the other detailed 
representation of soft-symmetric 
$Z_2$-breaking parameter, such as
$m_{12}^2 = M_{A}^2 \, t_{\beta}^2 
\, / \, (1 \, + \, t_{\beta}^2)$, 
may be further considered 
when the presence of VLQs in 
our computations, as have already 
been opted for in the literature 
significantly of 
Refs.~\cite{Arhrib:2024EPJC, 
Arhrib:2025EPJC}. 
Regarding to such decaying studies 
from the off-shell sub-process 
$\phi^* \rightarrow Z \gamma$, 
however instead of dependently 
varying in both $M_{A}$ and 
$t_{\beta}$, we intentionally 
choose a fixed value of $m_{12}^2$, 
which is totally attributed into 
all related charged Higgs vertices 
$\phi \, H^\pm H^\mp$ for 
$\phi \equiv h,H$, together with 
their relevant one-loop contributions 
are estimated substantially less 
as compared with these remaining ones. 

In these below Tables~\ref{BR2HDMtB2} 
and \ref{BR2HDMtB6}, we exam the 
partial decay rates from total 
contributions constituted by 
both triangle and box diagrams 
obeying as a pure three-body process, 
and their corresponding branching 
ratios for the considered 
$A \rightarrow Z Z \gamma$ channel
at various points of the allowed 
parameter space within 
the 2HDM Type II.
These results are also specifically 
investigated in the case of 
polarization modes $\lambda_2 = 0,L,R$ 
only for the massive vector boson 
$Z_{\beta} (k_2)$.
In which, the first result row 
denotes for case of unpolarized 
amplitude, whereas two remaining 
lines are depicted in polarization 
modes such as one longitudinal 
$\lambda_2 = 0$ and two transversal 
$\lambda_2 = L,R$ patterns, 
respectively.
In addition, the second column 
shows the total decay widths 
$\Gamma_{A}^{\textrm{Total}}$ of 
CP-odd Higgs boson $A$ consistently 
corresponding to their mass 
in the first one, combined with 
the value of $t_{\beta} = 2, 6$ 
as representative studying points. 
Besides the third column provides 
the partial decay rates, 
the relevant branching ratios 
are also reported in the last one.

Although, in general, the recorded 
partial decay rates of this process 
in the 2HDM Type II for each 
polarization mode as well as the 
unpolarized case have constantly 
an upward tendency across such 
entire scope of the CP-odd Higgs 
boson mass $M_{A} \in 
[400, 800] \, \text{GeV}$ under 
current studying, the resulting 
branching ratios always peak at 
$M_{A} = 600$ GeV, consequently 
decreases slightly to 
$M_{A} = 800$ GeV, due to 
the combination of a completely 
opposite trend for their 
corresponding total decay widths. 
These effects may also be driven 
at both instances of 
the mixing angle $t_{\beta}$ 
in these Tables~\ref{BR2HDMtB2} 
and \ref{BR2HDMtB6}.
Furthermore, these figures of 
the partial decay rate 
straightforwardly proportional 
to their corresponding mass 
$M_{A}$, while remaining show 
only an inconsequential dependence 
on both fixed parameters $t_{\beta}$.
Our documented discoveries, 
in addition, also indicate that 
the resulting branching ratios 
for this decay process are routinely 
of the order $\mathcal{O} (10^{-4})$ 
at various specific points within 
the allowed parameter space.
These figures, however, have dropped 
sharply to $\mathcal{O} (10^{-5})$ 
at totally $M_{A} = 400$ GeV with 
respect to such both $t_{\beta}$, 
whereas the remaining ones could 
reach up to $\mathcal{O} (10^{-3})$ 
consistently at only a distinct 
point $M_{A} = 600$ GeV along with 
$t_{\beta} = 6$ due to the fact 
that a lowest value of their total 
decay width could be resulted in.
Regarding to the contributions 
of each polarization mode, 
the longitudinal situation 
$\lambda_2 = 0$ tends to be greater 
than the case of transversal 
$\lambda_2 = L,R$ mode by 
an approximate factor of $4$ 
at almost all these parameter 
points of $\{ M_{A}, t_{\beta} \}$ 
under our investigation.

It should be noted that Table 3 
of Ref.~\cite{TranDT:2025EPJC}, 
which details the previously studied 
one-loop induced decay channel 
$A \rightarrow Z \gamma \gamma$ 
in unpolarized situation, 
is also considered into account 
for our comparison.
Together with the 2HDM Type II 
concerned and a specific input 
parameter space, 
these current results achieve 
a lower proportion of both partial 
decay rate as well as branching 
ratio weighed against the formers. 
Due to based on all relevant 
Feynman diagrams, a suppressed 
correlation intensely depending on 
such two-body phase spaces 
towards both decay sub-processes 
$\phi^* \rightarrow Z \gamma$ and 
$\phi^* \rightarrow \gamma \gamma$, 
which are consistently connected by 
the coupling $A Z \phi$ before, 
is dominantly presented in terms of 
$\big(1 - M_Z^2 / M_{\phi}^2 \big)^3$. 
That is the reason why there are 
such differences resulted in.
\begin{table}[H]
	\begin{center}
		\footnotesize
		\begin{tabular}{
				l@{\hspace{2cm}}
				l@{\hspace{2cm}}
				l@{\hspace{2cm}}
				l}
			\hline\hline
			\\
			$M_{A}$
			& $\Gamma_{A}^{\textrm{Total}}$ 
			& $\Gamma_{A \rightarrow 
				Z Z \gamma}$
			& Br $A \rightarrow 
			Z Z \gamma$ 
			\\
			\hline\hline 
			\\
			$400$
			&
			$159.828$
			&
			$(1.45498
			\pm \, 0.00146) 
			\times 10^{-3}$
			&
			$0.911 \times 10^{-5}$
			\\
			\\
			&
			$$
			&
			$(0.95655
			\pm \, 0.00096) 
			\times 10^{-3}$
			&
			$0.599 \times 10^{-5}$
			\\
			\\
			&
			$$
			&
			$(0.24897
			\pm \, 0.00025) 
			\times 10^{-3}$
			&
			$0.156 \times 10^{-5}$
			\\[0.075cm]
			\hline
			\\
			$500$
			&
			$22.676$
			&
			$(3.50468
			\pm \, 0.00351) 
			\times 10^{-3}$
			&
			$1.546 \times 10^{-4}$
			\\
			\\
			&
			$$
			&
			$(2.30743
			\pm \, 0.00231) 
			\times 10^{-3}$
			&
			$1.018 \times 10^{-4}$
			\\
			\\
			&
			$$
			&
			$(0.59937
			\pm \, 0.00061) 
			\times 10^{-3}$
			&
			$0.264 \times 10^{-4}$
			\\[0.075cm]  
			\hline
			\\
			$600$
			&
			$9.713$
			&
			$(6.69204
			\pm \, 0.00669) 
			\times 10^{-3}$
			&
			$6.891 \times 10^{-4}$
			\\
			\\
			&
			$$
			&
			$(4.43595
			\pm \, 0.00444) 
			\times 10^{-3}$
			&
			$4.567 \times 10^{-4}$
			\\
			\\
			&
			$$
			&
			$(1.14841
			\pm \, 0.00115) 
			\times 10^{-3}$
			&
			$1.182 \times 10^{-4}$
			\\[0.075cm]  
			\hline
			\\
			$700$
			&
			$21.768$
			&
			$(1.13692
			\pm \, 0.00114) 
			\times 10^{-2}$
			&
			$5.223 \times 10^{-4}$
			\\
			\\
			&
			$$
			&
			$(0.74999
			\pm \, 0.00075) 
			\times 10^{-2}$
			&
			$3.446 \times 10^{-4}$
			\\
			\\
			&
			$$
			&
			$(0.19452
			\pm \, 0.00019) 
			\times 10^{-2}$
			&
			$0.894 \times 10^{-4}$
			\\[0.075cm]  
			\hline
			\\
			$800$
			&
			$55.903$
			&
			$(1.77219
			\pm \, 0.00177) 
			\times 10^{-2}$
			&
			$3.171 \times 10^{-4}$
			\\
			\\
			&
			$$
			&
			$(1.16646
			\pm \, 0.00117) 
			\times 10^{-2}$
			&
			$2.087 \times 10^{-4}$
			\\
			\\
			&
			$$
			&
			$(0.30384
			\pm \, 0.00031) 
			\times 10^{-2}$
			&
			$0.544 \times 10^{-4}$
			\\[0.075cm]
			\hline\hline 
		\end{tabular}
	\end{center}
	\caption{
		\label{BR2HDMtB2}
		Partial decay rates and their
		corresponding branching ratios
		at $t_{\beta} = 2$ within 
		the viable parameter space of 
		the 2HDM Type II. 
		The first resulting row 
		denotes for case of unpolarized 
		amplitude, whereas two remaining 
		lines are depicted in polarization 
		modes such as one longitudinal 
		$\lambda_2 = 0$ and two transversal 
		$\lambda_2 = L,R$ patterns, respectively.
		Partial and total decay rates, 
		and masses are in GeV.
		} 
\end{table}


\begin{table}[H]
	\begin{center}
		\footnotesize
		\begin{tabular}{
				l@{\hspace{2cm}}
				l@{\hspace{2cm}}
				l@{\hspace{2cm}}
				l}
			\hline\hline
			$M_{A}$
			& $\Gamma_{A}^{\textrm{Total}}$ 
			& $\Gamma_{A \rightarrow 
				Z Z \gamma}$
			& Br $A \rightarrow 
			Z Z \gamma$ 
			\\
			\hline\hline 
			\\
			$400$
			&
			$157.663$
			&
			$(1.45243
			\pm \, 0.00145) 
			\times 10^{-3}$
			&
			$0.922 \times 10^{-5}$
			\\
			\\
			&
			$$
			&
			$(0.95427
			\pm \, 0.00095) 
			\times 10^{-3}$
			&
			$0.605 \times 10^{-5}$
			\\
			\\
			&
			$$
			&
			$(0.24849
			\pm \, 0.00025) 
			\times 10^{-3}$
			&
			$0.158 \times 10^{-5}$
			\\[0.075cm]  
			\hline
			\\
			$500$
			&
			$18.529$
			&
			$(3.49843
			\pm \, 0.00351) 
			\times 10^{-3}$
			&
			$1.888 \times 10^{-4}$
			\\
			\\
			&
			$$
			&
			$(2.30963
			\pm \, 0.00231) 
			\times 10^{-3}$
			&
			$1.247 \times 10^{-4}$
			\\
			\\
			&
			$$
			&
			$(0.59875
			\pm \, 0.00061) 
			\times 10^{-3}$
			&
			$0.323 \times 10^{-4}$
			\\[0.075cm]  
			\hline
			\\
			$600$
			&
			$3.991$
			&
			$(6.69127
			\pm \, 0.00669) 
			\times 10^{-3}$
			&
			$1.677 \times 10^{-3}$
			\\
			\\
			&
			$$
			&
			$(4.42462
			\pm \, 0.00443) 
			\times 10^{-3}$
			&
			$1.109 \times 10^{-3}$
			\\
			\\
			&
			$$
			&
			$(1.14784
			\pm \, 0.00115) 
			\times 10^{-3}$
			&
			$0.288 \times 10^{-3}$
			\\[0.075cm]  
			\hline
			\\
			$700$
			&
			$14.612$
			&
			$(1.14412
			\pm \, 0.00114) 
			\times 10^{-2}$
			&
			$7.831 \times 10^{-4}$
			\\
			\\
			&
			$$
			&
			$(0.75161
			\pm \, 0.00075) 
			\times 10^{-2}$
			&
			$5.144 \times 10^{-4}$
			\\
			\\
			&
			$$
			&
			$(0.19617
			\pm \, 0.00019) 
			\times 10^{-2}$
			&
			$1.343 \times 10^{-4}$
			\\[0.075cm]  
			\hline
			\\
			$800$
			&
			$47.387$
			&
			$(1.79863
			\pm \, 0.00179) 
			\times 10^{-2}$
			&
			$3.796 \times 10^{-4}$
			\\
			\\
			&
			$$
			&
			$(1.17896
			\pm \, 0.00118) 
			\times 10^{-2}$
			&
			$2.488 \times 10^{-4}$
			\\
			\\
			&
			$$
			&
			$(0.31042
			\pm \, 0.00031) 
			\times 10^{-2}$
			&
			$0.655 \times 10^{-4}$
			\\[0.075cm]
			\hline\hline 
		\end{tabular}
	\end{center}
	\caption{
		\label{BR2HDMtB6}
		Same as Table~\ref{BR2HDMtB2}, 
		but for the value of 
		$t_{\beta} = 6$.
		} 
\end{table}

In the subsequent 
Tables~\ref{BR2HDMVLQstB2} 
and \ref{BR2HDMVLQstB6}, 
partial decay rates and their
corresponding branching ratios 
within the representative 
parameter space of 2HDM 
Type II supplemented by VLQs 
in the scenario of doublet XT 
are specifically presented 
for both $t_{\beta} = 2,6$ 
and various points of 
$300 \; \text{GeV} \leq M_{A} 
\leq 1000\; \text{GeV}$.
In the context of such model, 
the VLQs $T$ and $X$, whose 
obvious mixing effects is 
temporarily not considered 
in this work, are individually 
implemented into these relevant 
one-loop contributions, including 
both triangle and box topologies
only related to all charged fermion 
$F$ Feynman diagrams as presented 
in Fig.~\ref{Fig:diags}.
However, it should be noted that 
though the coupling $V X \bar{X}$ 
towards $V \equiv Z, \gamma$ 
could be carried out in this 
framework, the VLQ $X$ only 
manifestly has charged current 
interactions with $G^\pm$ and $H^\pm$
in the Yukawa Lagrangian [as seen in 
Eq.~(\ref{Eq:LagYukDoubletXT}) 
for details]. 
As a result, in an obvious way, 
the interaction vertices of VLQ 
$T$ and its mass $m_{T}$ directly 
participate in such augmented 
contributions.

Additionally, as discussed in 
Refs.~\cite{Arhrib:2024EPJC, 
Arhrib:2025EPJC}, by imposing 
stringent restrictions on 
small mixing angles for 
the oblique parameters in order to 
maintain compatibility with 
current LHC bounds on VLQ masses, 
a comprehensive scope of VLQs mass 
starting at $600$ GeV and tending 
to arrive at $2000$ GeV is also 
considered into account therein. 
Due to, in our current investigations, 
there is no participating in of 
the mass $m_{X}$ towards VLQ $X$, 
which is strictly constrained by 
the following mass-dependent 
relation~\cite{Aguilar:2013PRD} 
regarding to the XT doublet scenario, 
is given as a function at tree level 
of such a mixing angle 
$\theta^u_R$, $m_{T}$, as well as 
SM top quark mass $m_{t}$ by,
\begin{eqnarray}
m_{X}^2
&=&
m_{T}^2
\, 
(c^u_R)^2
+
m_{t}^2
\, 
(s^u_R)^2
,
\end{eqnarray}
the key role of $m_{T}$, therefore, 
at present is taken into account as 
a free parameter for our examinations 
when such a value of $s^u_R$ is precisely 
enforced, which is completely obeyed 
in the Table~\ref{VLQsParameters}.
Furthermore, respecting to the XT 
doublet scenario, the leading 
condition $|s^u_R| \leq 0.22$ is 
totally constrained by EWPOs to 
remain small, which has reported in 
Ref.~\cite{Arhrib:2025JPG} 
for details. 

As a consequence, the mass of 
vector-like top partner $T$ is 
chosen at $m_{T} = 800$ GeV 
(these second and third columns) 
and $m_{T} = 1200$ GeV 
(these last two columns) 
in these below 
Tables~\ref{BR2HDMVLQstB2} 
and \ref{BR2HDMVLQstB6},
as well as the mixing angle term 
$s^u_R = 0.2$ is carried out 
as a specific example in 
the allowed constraint region.
These results are also specifically 
investigated in the separated cases 
of polarization mode $\lambda_2 = 0$ 
and $\lambda_2 = L,R$ 
only for the massive 
vector boson $Z_{\beta} (k_2)$ 
corresponding to these resulting 
second and third lines, whereas 
the first result row denotes as 
case of unpolarized amplitude.
Note that due to the on-shell 
condition $M_{A} > 2 m_{T}$ for 
such two-body decay channel 
$A \rightarrow T \bar{T}$
at tree level is totally not 
satisfied, the total decay width 
$\Gamma_{A}^{\textrm{Total}}$ 
has still significantly unchanged 
by additional VLQs top partner $T$.
Aside from that these partial decay 
rates of unpolarized case within 
the only standard 2HDM Type II 
could be similarly found in these 
corresponding Tables~\ref{BR2HDMtB2} 
and \ref{BR2HDMtB6} for 
our comparisons about such 
embedded VLQs contributions.

Overall, together with these VLQs 
contributing to the one-loop exchanging 
of such considered process at almost 
range of mass parameter $M_{A}$, 
our findings indicate that the 
unpolarized partial decay rates and 
their corresponding branching ratios 
at both cases of mass $m_{T}$ contribute 
approximately $5\%$ to these related 
results in the 2HDM Type II.
Such comparable splittings, 
whose total contributions are 
relatively small, may also be driven 
at both situations of 
the mixing angle $t_{\beta}$ 
as well as the whole cases 
of polarization mode.
This is partly because VLQs tend 
to decouple from the decay process 
as their masses increase 
$(m_{T} \gg m_{t})$ and also due to 
the mixing angle term $(s^u_R)^2$ 
appears consistently in three 
heavy-heavy interaction couplings 
$\phi \, T \bar{T}$ and $A \, T \bar{T}$ 
further suppressed by their squares.
However, it is noteworthy that 
the primary, albeit small, contribution 
arises from compatible modifications 
of a factor $(c^u_R)^2$ 
in three light-light interaction 
couplings $\phi \, t \bar{t}$ and 
$A \, t \bar{t}$.

Furthermore, along with 
the inclusion of VLQs in the 
Table~\ref{BR2HDMVLQstB2}, the 
resulting growth in these branching 
ratios prominently peaks at 
$M_{A} = 600$ GeV (reaching up to 
nearly $5.65\%$) and drops to 
its lowest value at $M_{A} = 800$ GeV 
(down to around $4.5\%$).
In addition, increasing 
the mass from $m_{T} = 800$ 
GeV to $m_{T} = 1200$ GeV suppresses 
this enhancement rate across almost all 
$M_{A}$ points by $0.12\%$ to $0.45\%$, 
except for a negligible upward fluctuation 
at $M_{A} = 700$ GeV 
(ascending slightly to  $0.02\%$). 
\begin{table}[H]
	\begin{center}
		\footnotesize
		\begin{tabular}{
				l@{\hspace{1.3cm}}
				l@{\hspace{0.7cm}}
				l@{\hspace{1.3cm}}
				l@{\hspace{0.7cm}}
				l}
			\hline\hline
			\\
			$M_{A}$
			& $\Gamma_{A \rightarrow 
				Z Z \gamma}$ 
			& Br $A \rightarrow 
			Z Z \gamma$ 
			& $\Gamma_{A \rightarrow 
				Z Z \gamma}$ 
			& Br $A \rightarrow 
			Z Z \gamma$ 
			\\
			\hline\hline 
			\\
			$400$
			&
			$(1.53018
			\pm \, 0.00153) 
			\times 10^{-3}$
			&
			$0.958 \times 10^{-5}$
			&
			$(1.52798
			\pm \, 0.00153) 
			\times 10^{-3}$
			&
			$0.956 \times 10^{-5}$
			\\
			\\[0.125cm]
			&
			$(1.00488
			\pm \, 0.00101) 
			\times 10^{-3}$
			&
			$0.629 \times 10^{-5}$
			&
			$(1.00419
			\pm \, 0.00101) 
			\times 10^{-3}$
			&
			$0.628 \times 10^{-5}$
			\\
			\\[0.125cm]
			&
			$(0.26163
			\pm \, 0.00026) 
			\times 10^{-3}$
			&
			$0.164 \times 10^{-5}$
			&
			$(0.26138
			\pm \, 0.00026) 
			\times 10^{-3}$
			&
			$0.164 \times 10^{-5}$
			\\[0.075cm]
			\hline
			\\
			$500$
			&
			$(3.68713
			\pm \, 0.00371) 
			\times 10^{-3}$
			&
			$1.626 \times 10^{-4}$
			&
			$(3.68215
			\pm \, 0.00371) 
			\times 10^{-3}$
			&
			$1.624 \times 10^{-4}$
			\\
			\\[0.125cm]
			&
			$(2.42888
			\pm \, 0.00241) 
			\times 10^{-3}$
			&
			$1.071 \times 10^{-4}$
			&
			$(2.42518
			\pm \, 0.00241) 
			\times 10^{-3}$
			&
			$1.069 \times 10^{-4}$
			\\
			\\[0.125cm]
			&
			$(0.63008
			\pm \, 0.00061) 
			\times 10^{-3}$
			&
			$0.278 \times 10^{-4}$
			&
			$(0.62949
			\pm \, 0.00061) 
			\times 10^{-3}$
			&
			$0.278 \times 10^{-4}$
			\\[0.075cm]  
			\hline
			\\
			$600$
			&
			$(7.06967
			\pm \, 0.00707) 
			\times 10^{-3}$
			&
			$7.279 \times 10^{-4}$
			&
			$(7.03975
			\pm \, 0.00704) 
			\times 10^{-3}$
			&
			$7.248 \times 10^{-4}$
			\\
			\\[0.125cm]
			&
			$(4.66191
			\pm \, 0.00466) 
			\times 10^{-3}$
			&
			$4.801 \times 10^{-4}$
			&
			$(4.65847
			\pm \, 0.00466) 
			\times 10^{-3}$
			&
			$4.796 \times 10^{-4}$
			\\
			\\[0.125cm]
			&
			$(1.20727
			\pm \, 0.00121) 
			\times 10^{-3}$
			&
			$1.239 \times 10^{-4}$
			&
			$(1.20657
			\pm \, 0.00121) 
			\times 10^{-3}$
			&
			$1.242 \times 10^{-4}$
			\\[0.075cm]  
			\hline
			\\
			$700$
			&
			$(1.19508
			\pm \, 0.00119) 
			\times 10^{-2}$
			&
			$5.491 \times 10^{-4}$
			&
			$(1.19557
			\pm \, 0.00119) 
			\times 10^{-2}$
			&
			$5.492 \times 10^{-4}$
			\\
			\\[0.125cm]
			&
			$(0.78732
			\pm \, 0.00079) 
			\times 10^{-2}$
			&
			$3.617 \times 10^{-4}$
			&
			$(0.78765
			\pm \, 0.00079) 
			\times 10^{-2}$
			&
			$3.618 \times 10^{-4}$
			\\
			\\[0.125cm]
			&
			$(0.20418
			\pm \, 0.00021) 
			\times 10^{-2}$
			&
			$0.938 \times 10^{-4}$
			&
			$(0.20436
			\pm \, 0.00021) 
			\times 10^{-2}$
			&
			$0.939 \times 10^{-4}$
			\\[0.075cm]  
			\hline
			\\
			$800$
			&
			$(1.86045
			\pm \, 0.00186) 
			\times 10^{-2}$
			&
			$3.328 \times 10^{-4}$
			&
			$(1.85262
			\pm \, 0.00185) 
			\times 10^{-2}$
			&
			$3.314 \times 10^{-4}$
			\\
			\\[0.125cm]
			&
			$(1.22429
			\pm \, 0.00123) 
			\times 10^{-2}$
			&
			$2.191 \times 10^{-4}$
			&
			$(1.22201
			\pm \, 0.00122) 
			\times 10^{-2}$
			&
			$2.186 \times 10^{-4}$
			\\
			\\[0.125cm]
			&
			$(0.31887
			\pm \, 0.00032) 
			\times 10^{-2}$
			&
			$0.571 \times 10^{-4}$
			&
			$(0.31862
			\pm \, 0.00032) 
			\times 10^{-2}$
			&
			$0.571 \times 10^{-4}$
			\\[0.075cm]
			\hline\hline 
		\end{tabular}
	\end{center}
	\caption{
		\label{BR2HDMVLQstB2}
		Partial decay rates and their
		corresponding branching ratios 
		within the representative 
		parameter space of 2HDM 
		Type II supplemented by VLQs 
		in the scenario of doublet XT 
		are specifically presented 
		for $t_{\beta} = 2$.
		The mass of vector-like top 
		partner $T$ is chosen at 
		$m_{T} = 800$ GeV 
		(these second and third columns) 
		and $m_{T} = 1200$ GeV 
		(these last two columns).
		Partial decay rates and 
		remaining masses 
		are in GeV.
		} 
\end{table}

Subsequently, the shifting of mixing 
angle parameter with $t_{\beta} = 6$ 
in the Table~\ref{BR2HDMVLQstB6} 
noticeably weakens the relative impact 
of vector-like quarks 
on the branching ratios.
Specifically, the boost in these 
resulting ones drops from a maximum of 
roughly $5.65\%$ in the lower $t_{\beta}$ 
region to a lower peak of approximately 
$5.15\%$ as $t_{\beta}$ increases, 
indicating that higher $t_{\beta}$ 
possibly suppresses these VLQs exchanging 
into one-loop contributions. 
Crucially, the decoupling phenomenon 
undergoes a clear inversion at higher 
kinematic thresholds, specifically 
for both values of $M_{A} = 600$ 
and $800$ GeV.
In these two heavy CP-odd Higgs 
mass regions, the heavier scenario 
$m_{T} = 1200$ GeV anomalously yields 
a higher percentage gain than the lighter 
one $m_{T} = 800$ GeV, demonstrating 
that an elevated $t_{\beta}$ combined 
with specific mass scales significantly 
alters the interference patterns 
within the one-loop fermion diagrams. 

\begin{table}[H]
	\begin{center}
		\footnotesize
		\begin{tabular}{
				l@{\hspace{1.3cm}}
				l@{\hspace{0.7cm}}
				l@{\hspace{1.3cm}}
				l@{\hspace{0.7cm}}
				l}
			\hline\hline
			\\
			$M_{A}$
			& $\Gamma_{A \rightarrow 
				Z Z \gamma}$ 
			& Br $A \rightarrow 
			Z Z \gamma$ 
			& $\Gamma_{A \rightarrow 
				Z Z \gamma}$ 
			& Br $A \rightarrow 
			Z Z \gamma$ 
			\\
			\hline\hline 
			\\
			$400$
			&
			$(1.52194
			\pm \, 0.00152) 
			\times 10^{-3}$
			&
			$0.966 \times 10^{-5}$
			&
			$(1.52157
			\pm \, 0.00152) 
			\times 10^{-3}$
			&
			$0.965 \times 10^{-5}$
			\\
			\\[0.125cm]
			&
			$(1.00098
			\pm \, 0.00101) 
			\times 10^{-3}$
			&
			$0.635 \times 10^{-5}$
			&
			$(0.99974
			\pm \, 0.00101) 
			\times 10^{-3}$
			&
			$0.634 \times 10^{-5}$
			\\
			\\[0.125cm]
			&
			$(0.26036
			\pm \, 0.00026) 
			\times 10^{-3}$
			&
			$0.165 \times 10^{-5}$
			&
			$(0.26022
			\pm \, 0.00026) 
			\times 10^{-3}$
			&
			$0.165 \times 10^{-5}$
			\\[0.075cm]  
			\hline
			\\
			$500$
			&
			$(3.66953
			\pm \, 0.00371) 
			\times 10^{-3}$
			&
			$1.981 \times 10^{-4}$
			&
			$(3.66364
			\pm \, 0.00371) 
			\times 10^{-3}$
			&
			$1.977 \times 10^{-4}$
			\\
			\\[0.125cm]
			&
			$(2.41938
			\pm \, 0.00241) 
			\times 10^{-3}$
			&
			$1.306 \times 10^{-4}$
			&
			$(2.41517
			\pm \, 0.00241) 
			\times 10^{-3}$
			&
			$1.304 \times 10^{-4}$
			\\
			\\[0.125cm]
			&
			$(0.62731
			\pm \, 0.00061) 
			\times 10^{-3}$
			&
			$0.339 \times 10^{-4}$
			&
			$(0.62669
			\pm \, 0.00061) 
			\times 10^{-3}$
			&
			$0.338 \times 10^{-4}$
			\\[0.075cm]  
			\hline
			\\
			$600$
			&
			$(7.02961
			\pm \, 0.00703) 
			\times 10^{-3}$
			&
			$1.762 \times 10^{-3}$
			&
			$(7.03621
			\pm \, 0.00704) 
			\times 10^{-3}$
			&
			$1.763 \times 10^{-3}$
			\\
			\\[0.125cm]
			&
			$(4.63584
			\pm \, 0.00464) 
			\times 10^{-3}$
			&
			$1.162 \times 10^{-3}$
			&
			$(4.64294
			\pm \, 0.00464) 
			\times 10^{-3}$
			&
			$1.163 \times 10^{-3}$
			\\
			\\[0.125cm]
			&
			$(1.20212
			\pm \, 0.00121) 
			\times 10^{-3}$
			&
			$0.301 \times 10^{-3}$
			&
			$(1.20121
			\pm \, 0.00121) 
			\times 10^{-3}$
			&
			$0.301 \times 10^{-3}$
			\\[0.075cm]  
			\hline
			\\
			$700$
			&
			$(1.19838
			\pm \, 0.00121) 
			\times 10^{-2}$
			&
			$8.202 \times 10^{-4}$
			&
			$(1.19733
			\pm \, 0.00121) 
			\times 10^{-2}$
			&
			$8.194 \times 10^{-4}$
			\\
			\\[0.125cm]
			&
			$(0.78803
			\pm \, 0.00079) 
			\times 10^{-2}$
			&
			$5.393 \times 10^{-4}$
			&
			$(0.78862
			\pm \, 0.00079) 
			\times 10^{-2}$
			&
			$5.397 \times 10^{-4}$
			\\
			\\[0.125cm]
			&
			$(0.20512
			\pm \, 0.00021) 
			\times 10^{-2}$
			&
			$1.404 \times 10^{-4}$
			&
			$(0.20527
			\pm \, 0.00021) 
			\times 10^{-2}$
			&
			$1.405 \times 10^{-4}$
			\\[0.075cm]  
			\hline
			\\
			$800$
			&
			$(1.87945
			\pm \, 0.00188) 
			\times 10^{-2}$
			&
			$3.966 \times 10^{-4}$
			&
			$(1.88059
			\pm \, 0.00188) 
			\times 10^{-2}$
			&
			$3.969 \times 10^{-4}$
			\\
			\\[0.125cm]
			&
			$(1.23449
			\pm \, 0.00124) 
			\times 10^{-2}$
			&
			$2.605 \times 10^{-4}$
			&
			$(1.23378
			\pm \, 0.00124) 
			\times 10^{-2}$
			&
			$2.604 \times 10^{-4}$
			\\
			\\[0.125cm]
			&
			$(0.32487
			\pm \, 0.00033) 
			\times 10^{-2}$
			&
			$0.686 \times 10^{-4}$
			&
			$(0.32459
			\pm \, 0.00033) 
			\times 10^{-2}$
			&
			$0.685 \times 10^{-4}$
			\\[0.075cm]
			\hline\hline 
		\end{tabular}
	\end{center}
	\caption{
		\label{BR2HDMVLQstB6}
		Same as Table~\ref{BR2HDMVLQstB2}, 
		but for the value of 
		$t_{\beta} = 6$.
		} 
\end{table}


In the next Table~\ref{BR2HDMPolaZ}, 
these resulting partial decay rates 
and their corresponding branching 
ratios for the considered 
$A \rightarrow Z Z \gamma$ process 
at four representative input points 
$M_{A} = 400$ and $800$ GeV together 
with $t_{\beta} = 2, 6$ in the 
2HDM Type II are specifically 
analyzed in distinct circumstances 
of polarization modes 
$\lambda_1,\lambda_2 = 0,L,R$ for 
both polarized $Z$-bosons 
in the final state.
Further, the other input parameters 
consisting of total decay widths 
as well as the corresponding 
results of polarization sum 
served for our subsequent 
discussions could be found in 
these both Tables~\ref{BR2HDMtB2} 
and \ref{BR2HDMtB6}.

In general, these obtained results 
of branching ratio exhibit an absolute 
dominance of the whole longitudinal 
polarization configuration, whereas 
being profoundly modified by opposing 
effects from the mixing angle 
parameter $t_{\beta}$. 
Specifically, these figures for purely 
longitudinal state $\lambda_1 = 
\lambda_2 = 0$ consistently yields 
the highest values, acquiring 
$\mathcal{O} (10^{-5})$ at 
$M_{A} = 400$ GeV, and then reaching 
up to $\mathcal{O} (10^{-4})$ 
at $M_{A} = 800$ GeV. 
Meanwhile the mixing states 
(one longitudinal, one transversal) 
remain roughly identical and describe 
approximately half the value of 
such solely longitudinal state, 
the exclusively transversal modes 
$\lambda_1 = \lambda_2 = L,R$ or 
mixing states $\lambda_1 = L \, (R), \, 
\lambda_2 = R \, (L)$ are excessively 
suppressed, falling below 
the longitudinal-containing modes 
by $7$ to $9$ orders of magnitude, 
ranging only from $\mathcal{O} (10^{-14})$ 
to $\mathcal{O} (10^{-11})$.

\begin{table}[H]
	\begin{center}
		\footnotesize
		\begin{tabular}{
				l@{\hspace{0.75cm}} 
				l@{\hspace{1.0cm}}   
				l@{\hspace{1.25cm}} 
				l}
			\hline\hline
			\\
			\big(
			$M_{A} \,, t_{\beta}$
			\big)
			&
			Polarized Modes
			& $\Gamma_{A \rightarrow 
				Z Z \gamma}$
			& Br $A \rightarrow 
			Z Z \gamma$ 
			\\
			\hline\hline 
			\\[-0.25cm]
			$$
			&
			$\lambda_1 = 
			\lambda_2 = 0$
			&
			$(0.45985
			\pm \, 0.00046) 
			\times 10^{-3}$
			&
			$0.288 \times 10^{-5}$
			\\[-0.125cm]
			\\
			$$
			&
			$\lambda_1 = 0, \;
			\lambda_2 = L,R$
			&
			$(0.24885
			\pm \, 0.00025) 
			\times 10^{-3}$
			&
			$0.156 \times 10^{-5}$
			\\[-0.125cm]
			\\
			$(400, 2)$
			&
			$\lambda_2 = 0, \;
			\lambda_1 = L,R$
			&
			$(0.24808
			\pm \, 0.00025) 
			\times 10^{-3}$
			&
			$0.155 \times 10^{-5}$
			\\[-0.125cm]
			\\
			$$
			&
			$\lambda_1 = 
			\lambda_2 = L,R$
			&
			$(1.25327
			\pm \, 0.03672) 
			\times 10^{-11}$
			&
			$7.841 \times 10^{-14}$
			\\[-0.125cm]
			\\
			$$
			&
			$\lambda_1 = L \, (R), \;
			\lambda_2 = R \, (L)$
			&
			$(1.74992
			\pm \, 0.00659) 
			\times 10^{-10}$
			&
			$1.095 \times 10^{-12}$
			\\[0.25cm]
			\hline
			\\
			$$
			&
			$\lambda_1 = 
			\lambda_2 = 0$
			&
			$(0.45895
			\pm \, 0.00046) 
			\times 10^{-3}$
			&
			$0.291 \times 10^{-5}$
			\\[-0.125cm]
			\\
			$$
			&
			$\lambda_1 = 0, \;
			\lambda_2 = L,R$
			&
			$(0.24863
			\pm \, 0.00025) 
			\times 10^{-3}$
			&
			$0.158 \times 10^{-5}$
			\\[-0.125cm]
			\\
			$(400, 6)$
			&
			$\lambda_2 = 0, \;
			\lambda_1 = L,R$
			&
			$(0.24845
			\pm \, 0.00025) 
			\times 10^{-3}$
			&
			$0.158 \times 10^{-5}$
			\\[-0.125cm]
			\\
			$$
			&
			$\lambda_1 = 
			\lambda_2 = L,R$
			&
			$(1.39304
			\pm \, 0.04082) 
			\times 10^{-12}$
			&
			$0.884 \times 10^{-14}$
			\\[-0.125cm]
			\\
			$$
			&
			$\lambda_1 = L \, (R), \;
			\lambda_2 = R \, (L)$
			&
			$(1.94507
			\pm \, 0.05593) 
			\times 10^{-11}$
			&
			$0.124 \times 10^{-12}$
			\\[0.25cm]
			\hline
			\\
			$$
			&
			$\lambda_1 = 
			\lambda_2 = 0$
			&
			$(0.55983
			\pm \, 0.00056) 
			\times 10^{-2}$
			&
			$1.002 \times 10^{-4}$
			\\[-0.125cm]
			\\
			$$
			&
			$\lambda_1 = 0, \;
			\lambda_2 = L,R$
			&
			$(0.30395
			\pm \, 0.00031) 
			\times 10^{-2}$
			&
			$0.544 \times 10^{-4}$
			\\[-0.125cm]
			\\
			$(800, 2)$
			&
			$\lambda_2 = 0, \;
			\lambda_1 = L,R$
			&
			$(0.30337
			\pm \, 0.00031) 
			\times 10^{-2}$
			&
			$0.543 \times 10^{-4}$
			\\[-0.125cm]
			\\
			$$
			&
			$\lambda_1 = 
			\lambda_2 = L,R$
			&
			$(1.08498
			\pm \, 0.00107) 
			\times 10^{-9}$
			&
			$1.941 \times 10^{-11}$
			\\[-0.125cm]
			\\
			$$
			&
			$\lambda_1 = L \, (R), \;
			\lambda_2 = R \, (L)$
			&
			$(1.27351
			\pm \, 0.00127) 
			\times 10^{-9}$
			&
			$2.278 \times 10^{-11}$
			\\[0.25cm]
			\hline
			\\
			$$
			&
			$\lambda_1 = 
			\lambda_2 = 0$
			&
			$(0.55944
			\pm \, 0.00056) 
			\times 10^{-2}$
			&
			$1.181 \times 10^{-4}$
			\\[-0.125cm]
			\\
			$$
			&
			$\lambda_1 = 0, \;
			\lambda_2 = L,R$
			&
			$(0.31076
			\pm \, 0.00031) 
			\times 10^{-2}$
			&
			$0.656 \times 10^{-4}$
			\\[-0.125cm]
			\\
			$(800, 6)$
			&
			$\lambda_2 = 0, \;
			\lambda_1 = L,R$
			&
			$(0.31045
			\pm \, 0.00031) 
			\times 10^{-2}$
			&
			$0.655 \times 10^{-4}$
			\\[-0.125cm]
			\\
			$$
			&
			$\lambda_1 = 
			\lambda_2 = L,R$
			&
			$(1.20615
			\pm \, 0.00955) 
			\times 10^{-10}$
			&
			$0.255 \times 10^{-11}$
			\\[-0.125cm]
			\\
			$$
			&
			$\lambda_1 = L \, (R), \;
			\lambda_2 = R \, (L)$
			&
			$(1.41456
			\pm \, 0.00854) 
			\times 10^{-10}$
			&
			$0.299 \times 10^{-11}$
			\\[-0.25cm]
			\\
			\hline\hline 
		\end{tabular}
	\end{center}
	\caption{
		\label{BR2HDMPolaZ}
		Partial decay rates and their corresponding branching ratios 
		for the considered 
		$A \rightarrow Z Z \gamma$ process 
		at four representative input-points 
		$M_{A} = 400, 800 \; \text{GeV}$ 
		and $t_{\beta} = 2, 6$ in 
		the 2HDM Type II are considered
		in terms of polarization modes 
		$\lambda_1,\lambda_2 = 0,L,R$ 
		for both polarized $Z$-bosons 
		in the final state.
		Partial decay rates and masses 
		are in GeV.
		} 
\end{table}

Furthermore, varying between 
$t_{\beta} = 2$ and $t_{\beta} = 6$ 
introduces a distinctive impact 
depending strictly on the polarization 
modes in the provided 
Table~\ref{BR2HDMPolaZ}. 
For the purely longitudinal state, 
an elevated $t_{\beta}$ acts as 
an enhancement factor, causing 
these figures to increase marginally 
by roughly $1\%$ at the lower mass 
threshold $M_{A} = 400$ GeV and surge 
significantly by approximately $17.85\%$ 
at the higher mass threshold 
$M_{A} = 800$ GeV.
In totally contrast, a larger 
$t_{\beta}$ severely suppresses 
the whole transversal states.
In details, most notably at 
$M_{A} = 400$ GeV, the 
$\lambda_1 = \lambda_2 = L,R$ mode is 
diminished by nearly a factor of $9$, 
with a comparable suppression rate 
of more than $7.5$ times distinctly 
observed at $M_{A} = 800$ GeV. 

\begin{table}[H]
	\begin{center}
		\footnotesize
		\begin{tabular}{
				l@{\hspace{0.75cm}} 
				l@{\hspace{1.0cm}}   
				l@{\hspace{1.25cm}} 
				l}
			\hline\hline
			\\
			\big(
			$M_{A} \,, t_{\beta}$
			\big)
			&
			Polarized Modes
			& $\Gamma_{A \rightarrow 
				Z Z \gamma}$
			& Br $A \rightarrow 
			Z Z \gamma$ 
			\\
			\hline\hline 
			\\[-0.25cm]
			$$
			&
			$\lambda_1 = 
			\lambda_2 = 0$
			&
			$(0.48319
			\pm \, 0.00048) 
			\times 10^{-3}$
			&
			$0.302 \times 10^{-5}$
			\\[-0.125cm]
			\\
			$$
			&
			$\lambda_1 = 0, \;
			\lambda_2 = L,R$
			&
			$(0.26161
			\pm \, 0.00026) 
			\times 10^{-3}$
			&
			$0.164 \times 10^{-5}$
			\\[-0.125cm]
			\\
			$(400, 2)$
			&
			$\lambda_2 = 0, \;
			\lambda_1 = L,R$
			&
			$(0.26156
			\pm \, 0.00026) 
			\times 10^{-3}$
			&
			$0.164 \times 10^{-5}$
			\\[-0.125cm]
			\\
			$$
			&
			$\lambda_1 = 
			\lambda_2 = L,R$
			&
			$(1.14841
			\pm \, 0.03365) 
			\times 10^{-11}$
			&
			$7.185 \times 10^{-14}$
			\\[-0.125cm]
			\\
			$$
			&
			$\lambda_1 = L \, (R), \;
			\lambda_2 = R \, (L)$
			&
			$(1.60361
			\pm \, 0.00604) 
			\times 10^{-10}$
			&
			$1.004 \times 10^{-12}$
			\\[0.25cm]
			\hline
			\\
			$$
			&
			$\lambda_1 = 
			\lambda_2 = 0$
			&
			$(0.48082
			\pm \, 0.00048) 
			\times 10^{-3}$
			&
			$0.305 \times 10^{-5}$
			\\[-0.125cm]
			\\
			$$
			&
			$\lambda_1 = 0, \;
			\lambda_2 = L,R$
			&
			$(0.26036
			\pm \, 0.00026) 
			\times 10^{-3}$
			&
			$0.165 \times 10^{-5}$
			\\[-0.125cm]
			\\
			$(400, 6)$
			&
			$\lambda_2 = 0, \;
			\lambda_1 = L,R$
			&
			$(0.26043
			\pm \, 0.00026) 
			\times 10^{-3}$
			&
			$0.165 \times 10^{-5}$
			\\[-0.125cm]
			\\
			$$
			&
			$\lambda_1 = 
			\lambda_2 = L,R$
			&
			$(1.27647
			\pm \, 0.03741) 
			\times 10^{-12}$
			&
			$0.811 \times 10^{-14}$
			\\[-0.125cm]
			\\
			$$
			&
			$\lambda_1 = L \, (R), \;
			\lambda_2 = R \, (L)$
			&
			$(1.78241
			\pm \, 0.05125) 
			\times 10^{-11}$
			&
			$0.113 \times 10^{-12}$
			\\[0.25cm]
			\hline
			\\
			$$
			&
			$\lambda_1 = 
			\lambda_2 = 0$
			&
			$(0.58806
			\pm \, 0.00059) 
			\times 10^{-2}$
			&
			$1.052 \times 10^{-4}$
			\\[-0.125cm]
			\\
			$$
			&
			$\lambda_1 = 0, \;
			\lambda_2 = L,R$
			&
			$(0.31852
			\pm \, 0.00032) 
			\times 10^{-2}$
			&
			$0.571 \times 10^{-4}$
			\\[-0.125cm]
			\\
			$(800, 2)$
			&
			$\lambda_2 = 0, \;
			\lambda_1 = L,R$
			&
			$(0.31878
			\pm \, 0.00032) 
			\times 10^{-2}$
			&
			$0.571 \times 10^{-4}$
			\\[-0.125cm]
			\\
			$$
			&
			$\lambda_1 = 
			\lambda_2 = L,R$
			&
			$(9.94902
			\pm \, 0.00981) 
			\times 10^{-10}$
			&
			$1.779 \times 10^{-11}$
			\\[-0.125cm]
			\\
			$$
			&
			$\lambda_1 = L \, (R), \;
			\lambda_2 = R \, (L)$
			&
			$(1.16748
			\pm \, 0.00117) 
			\times 10^{-9}$
			&
			$2.088 \times 10^{-11}$
			\\[0.25cm]
			\hline
			\\
			$$
			&
			$\lambda_1 = 
			\lambda_2 = 0$
			&
			$(0.58641
			\pm \, 0.00059) 
			\times 10^{-2}$
			&
			$1.238 \times 10^{-4}$
			\\[-0.125cm]
			\\
			$$
			&
			$\lambda_1 = 0, \;
			\lambda_2 = L,R$
			&
			$(0.32484
			\pm \, 0.00033) 
			\times 10^{-2}$
			&
			$0.686 \times 10^{-4}$
			\\[-0.125cm]
			\\
			$(800, 6)$
			&
			$\lambda_2 = 0, \;
			\lambda_1 = L,R$
			&
			$(0.32451
			\pm \, 0.00033) 
			\times 10^{-2}$
			&
			$0.685 \times 10^{-4}$
			\\[-0.125cm]
			\\
			$$
			&
			$\lambda_1 = 
			\lambda_2 = L,R$
			&
			$(1.10771
			\pm \, 0.00981) 
			\times 10^{-10}$
			&
			$0.234 \times 10^{-11}$
			\\[-0.125cm]
			\\
			$$
			&
			$\lambda_1 = L \, (R), \;
			\lambda_2 = R \, (L)$
			&
			$(1.29623
			\pm \, 0.00934) 
			\times 10^{-10}$
			&
			$0.274 \times 10^{-11}$
			\\[-0.25cm]
			\\
			\hline\hline 
		\end{tabular}
	\end{center}
	\caption{
		\label{BR2HDMVLQsPolaZmT800}
		Same as Table~\ref{BR2HDMPolaZ}, 
		but for the 2HDM Type II 
		embedded by VLQs 
		in the scenario of doublet XT.
		In which, the VLQs $T$ mass 
		$m_{T} = 800$ GeV and the 
		mixing angle term $s^u_R = 0.2$ 
		are distinctly chosen for 
		our examination.
		} 
\end{table}

In these ultimate 
Tables~\ref{BR2HDMVLQsPolaZmT800} 
and \ref{BR2HDMVLQsPolaZmT1200}, 
the augmenting of vector-like quarks 
in the XT doublet scenario systematically 
modulates such branching ratio enlargements 
(also reaching up approximately $5\%$ 
in the similar investigation) 
without changing the essential behaviors 
observed in the standard 2HDM situation 
[as discussed in Table~\ref{BR2HDMPolaZ}], 
where the purely longitudinal polarization 
mode retains its complete dominance, and 
is nearly twice as large as the values 
for each mixing state, which consists of 
one longitudinal and one transversal modes 
such as $\lambda_1 = 0, \, \lambda_2 = L,R$ 
as well as 
$\lambda_2 = 0, \, \lambda_1 = L,R$. 

In addition, the presence of VLQs 
in both Tables~\ref{BR2HDMVLQsPolaZmT800} 
and \ref{BR2HDMVLQsPolaZmT1200} induces 
an over $4.65\%$ to nearly $5\%$ 
enhancement in the dominant purely 
longitudinal situation, whereas it 
further exhausts the highly suppressed 
solely transversal states by roughly 
$8.4\%$ to approximately $6.5\%$. 
\begin{table}[H]
	\begin{center}
		\footnotesize
		\begin{tabular}{
				l@{\hspace{0.75cm}} 
				l@{\hspace{1.0cm}}   
				l@{\hspace{1.25cm}} 
				l}
			\hline\hline
			\\
			\big(
			$M_{A} \,, t_{\beta}$
			\big)
			&
			Polarized Modes
			& $\Gamma_{A \rightarrow 
				Z Z \gamma}$
			& Br $A \rightarrow 
			Z Z \gamma$ 
			\\
			\hline\hline 
			\\
			$$
			&
			$\lambda_1 = 
			\lambda_2 = 0$
			&
			$(0.48241
			\pm \, 0.00048) 
			\times 10^{-3}$
			&
			$0.302 \times 10^{-5}$
			\\[-0.125cm]
			\\
			$$
			&
			$\lambda_1 = 0, \;
			\lambda_2 = L,R$
			&
			$(0.26142
			\pm \, 0.00026) 
			\times 10^{-3}$
			&
			$0.164 \times 10^{-5}$
			\\[-0.125cm]
			\\
			$(400, 2)$
			&
			$\lambda_2 = 0, \;
			\lambda_1 = L,R$
			&
			$(0.26139
			\pm \, 0.00026) 
			\times 10^{-3}$
			&
			$0.164 \times 10^{-5}$
			\\[-0.125cm]
			\\
			$$
			&
			$\lambda_1 = 
			\lambda_2 = L,R$
			&
			$(1.17187
			\pm \, 0.03433) 
			\times 10^{-11}$
			&
			$7.332 \times 10^{-14}$
			\\[-0.125cm]
			\\
			$$
			&
			$\lambda_1 = L \, (R), \;
			\lambda_2 = R \, (L)$
			&
			$(1.63636
			\pm \, 0.00616) 
			\times 10^{-10}$
			&
			$1.024 \times 10^{-12}$
			\\[0.25cm]
			\hline
			\\
			$$
			&
			$\lambda_1 = 
			\lambda_2 = 0$
			&
			$(0.48044
			\pm \, 0.00048) 
			\times 10^{-3}$
			&
			$0.305 \times 10^{-5}$
			\\[-0.125cm]
			\\
			$$
			&
			$\lambda_1 = 0, \;
			\lambda_2 = L,R$
			&
			$(0.26029
			\pm \, 0.00026) 
			\times 10^{-3}$
			&
			$0.165 \times 10^{-5}$
			\\[-0.125cm]
			\\
			$(400, 6)$
			&
			$\lambda_2 = 0, \;
			\lambda_1 = L,R$
			&
			$(0.26027
			\pm \, 0.00026) 
			\times 10^{-3}$
			&
			$0.165 \times 10^{-5}$
			\\[-0.125cm]
			\\
			$$
			&
			$\lambda_1 = 
			\lambda_2 = L,R$
			&
			$(1.30255
			\pm \, 0.03817) 
			\times 10^{-12}$
			&
			$0.826 \times 10^{-14}$
			\\[-0.125cm]
			\\
			$$
			&
			$\lambda_1 = L \, (R), \;
			\lambda_2 = R \, (L)$
			&
			$(1.81882
			\pm \, 0.05231) 
			\times 10^{-11}$
			&
			$0.115 \times 10^{-12}$
			\\[0.25cm]
			\hline
			\\
			$$
			&
			$\lambda_1 = 
			\lambda_2 = 0$
			&
			$(0.58805
			\pm \, 0.00059) 
			\times 10^{-2}$
			&
			$1.052 \times 10^{-4}$
			\\[-0.125cm]
			\\
			$$
			&
			$\lambda_1 = 0, \;
			\lambda_2 = L,R$
			&
			$(0.31833
			\pm \, 0.00032) 
			\times 10^{-2}$
			&
			$0.569 \times 10^{-4}$
			\\[-0.125cm]
			\\
			$(800, 2)$
			&
			$\lambda_2 = 0, \;
			\lambda_1 = L,R$
			&
			$(0.31835
			\pm \, 0.00032) 
			\times 10^{-2}$
			&
			$0.569 \times 10^{-4}$
			\\[-0.125cm]
			\\
			$$
			&
			$\lambda_1 = 
			\lambda_2 = L,R$
			&
			$(1.01506
			\pm \, 0.00101) 
			\times 10^{-9}$
			&
			$1.816 \times 10^{-11}$
			\\[-0.125cm]
			\\
			$$
			&
			$\lambda_1 = L \, (R), \;
			\lambda_2 = R \, (L)$
			&
			$(1.19111
			\pm \, 0.00119) 
			\times 10^{-9}$
			&
			$2.131 \times 10^{-11}$
			\\[0.25cm]
			\hline
			\\
			$$
			&
			$\lambda_1 = 
			\lambda_2 = 0$
			&
			$(0.58563
			\pm \, 0.00059) 
			\times 10^{-2}$
			&
			$1.236 \times 10^{-4}$
			\\[-0.125cm]
			\\
			$$
			&
			$\lambda_1 = 0, \;
			\lambda_2 = L,R$
			&
			$(0.32445
			\pm \, 0.00033) 
			\times 10^{-2}$
			&
			$0.685 \times 10^{-4}$
			\\[-0.125cm]
			\\
			$(800, 6)$
			&
			$\lambda_2 = 0, \;
			\lambda_1 = L,R$
			&
			$(0.32485
			\pm \, 0.00033) 
			\times 10^{-2}$
			&
			$0.686 \times 10^{-4}$
			\\[-0.125cm]
			\\
			$$
			&
			$\lambda_1 = 
			\lambda_2 = L,R$
			&
			$(1.13013
			\pm \, 0.00999) 
			\times 10^{-10}$
			&
			$0.239 \times 10^{-11}$
			\\[-0.125cm]
			\\
			$$
			&
			$\lambda_1 = L \, (R), \;
			\lambda_2 = R \, (L)$
			&
			$(1.32246
			\pm \, 0.00953) 
			\times 10^{-10}$
			&
			$0.279 \times 10^{-11}$
			\\[-0.25cm]
			\\
			\hline\hline 
		\end{tabular}
	\end{center}
	\caption{
		\label{BR2HDMVLQsPolaZmT1200}
		Same as Table~\ref{BR2HDMVLQsPolaZmT800}, 
		but for the vector-like quark 
		mass $m_{T} = 1200$ GeV.
		} 
\end{table}

What's more, when examining the mass 
$m_{T}$ dependence within the VLQ 
configuration, the dominant longitudinal 
states (include these mixing ones with them) 
exhibit virtually identical branching ratio 
values with deviations below $0.2\%$, 
demonstrating that the heavy VLQ mass 
$m_{T}$ exerts a negligible impact on 
such primary decaying modes.
Conversely, increasing $m_{T}$ from $800$ 
GeV to $1200$ GeV slightly elevates 
the suppressed transversal branching ratios 
by $2\%$ to $3\%$ across all examined 
kinematic thresholds of 
$\{ M_{A}, t_{\beta} \}$ in these both 
mentioned Tables~\ref{BR2HDMVLQsPolaZmT800} 
and \ref{BR2HDMVLQsPolaZmT1200}.

In the following paragraphs, 
we will analyze the investigation 
of distributions for the differential 
decay rates with respect to the 
mass-invariant $m_{Z\gamma}$ relevant 
to one selected $Z$ boson and a photon 
in the final state, explicitly across 
the entire scope of 
$M_{Z} \leq m_{Z\gamma} \leq M_{A}-M_{Z}$, 
in which we opt for the Mandelstam 
variable $k_{13} \equiv m_{Z\gamma}^2$ 
is consistently presented in our 
numerical computations.
As a representative example, 
the studied configuration of input 
parameters such as the CP-odd Higgs 
mass $M_{A} = 800$ GeV 
specifically for both $t_{\beta} = 2,6$
is taken into account in the 2HDM 
Type II, and then supplemented by VLQs 
in the scenario of doublet XT along with 
$m_{T} = 800$ GeV as well as a mixing 
angle term $s^u_R = 0.2$.

\begin{figure}[H]
\centering
\includegraphics[width=15cm,height=7cm]
{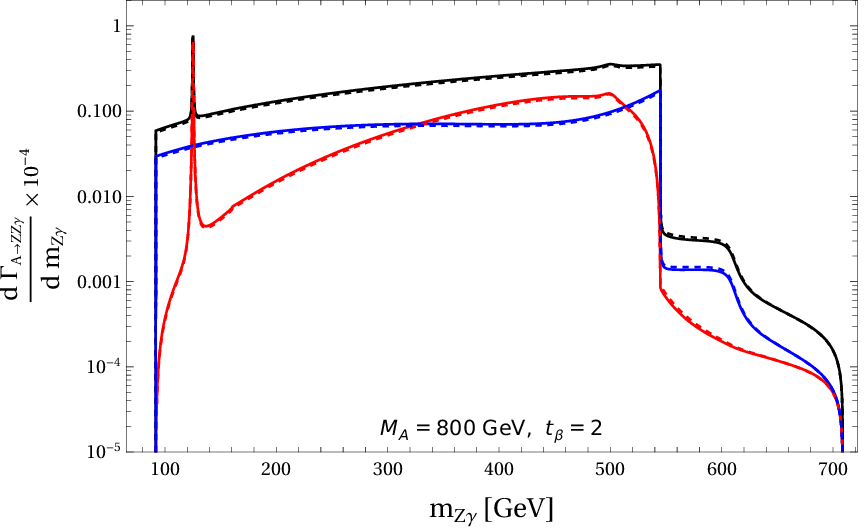}
\\[0.25cm]
\hfill
\\[0.25cm]
\includegraphics[width=15cm,height=7cm]
{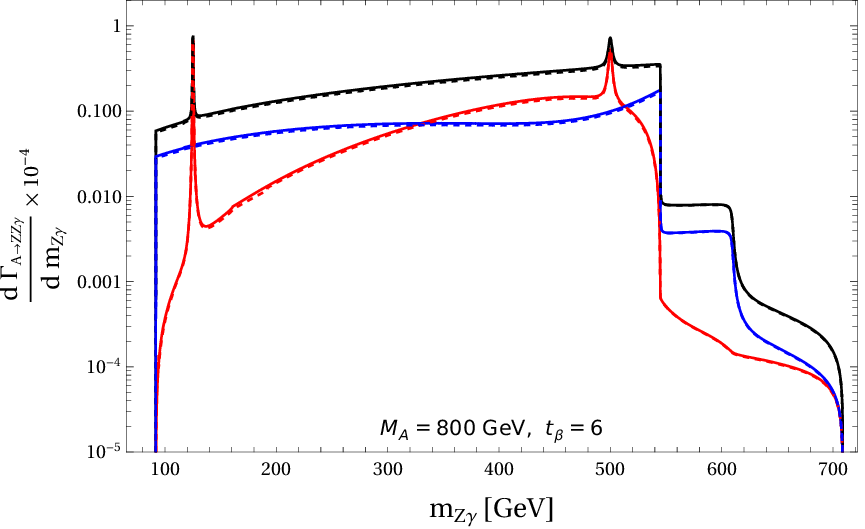}
\caption{
\label{Fig:mZApol1Z}
The differential decay rates as functions
of the invariant mass $m_{Z\gamma}$ of the
$Z\gamma$ system are investigated for two
characteristic polarization states of the
vector boson $Z_{\beta}(k_2)$, as well as
for the unpolarized case, within the
Type-II 2HDM extended with VLQs in the
doublet-$XT$ scenario.
We emphasize that the vertical axis is
labeled as
$d\Gamma_{A\rightarrow ZZ\gamma}/dm_{Z\gamma}
\times 10^{-4}$.
Accordingly, the numerical values shown
on the vertical axis should be multiplied
by a factor of $10^{-4}$.}
\end{figure}

As depicted in the Fig.~\ref{Fig:mZApol1Z}, 
we focus on two distinct plots of 
the differential decay rates with 
respect to the mass-invariant 
quality $m_{Z\gamma}$ corresponding 
to $t_{\beta} = 2$ and $t_{\beta} = 6$ 
at the same mass of $M_{A} = 800$ GeV,
which are simultaneously examined 
for the 2HDM Type II in two 
characteristic polarization modes 
of vector boson $Z_{\beta} (k_2)$ 
such as one longitudinal 
$\lambda_2 = 0$ (red dashed line) 
and two transversal $\lambda_2 = L,R$ 
(blue dashed line) patterns 
as well as the unpolarized situation 
(black dashed line).
Whereas the embedded contributions 
of vector-like quark are described 
as the solid and corresponding color 
curves for the subsequent discussions 
among them. 

Overall, our explicit findings 
in the 2HDM Type II indicate 
that the invariant mass $m_{Z\gamma}$ 
distribution exhibits distinct geometric 
behaviors across different polarization modes, 
driven by the remarkable SM-like Higgs 
boson $h$ pole at $M_{h} = 125.1$ GeV 
as well as the heavy CP-even Higgs $H$ 
resonance at pole 
$m_{Z\gamma} \simeq M_{H} = 500$ GeV. 
In the low-mass range of $m_{Z\gamma}$ 
at both circumstances of $t_{\beta}$, 
a striking disparity is observed as 
the longitudinal mode $\lambda_2 = 0$ 
undergoes a vertical surge to establish 
a narrow and sharp peak before plunging 
drastically, whereas the transverse mode 
$\lambda_2 = L,R$ maintains a delicate 
fluctuation that dominates this region. 
Synthesizing these features, 
the unpolarized distribution effectively 
captures both the edged peak from 
the longitudinal state and 
the stable contribution from 
the transverse situation.

Furthermore, a profound reversal in 
tendency occurs beyond such 
$m_{Z\gamma} \simeq 300$ GeV threshold, 
where the longitudinal mode climbs 
steeply and undergoes a crossover 
to surpass the transverse mode, 
subsequently attributing 
the high-energy regime of 
$m_{Z\gamma}$.
Approaching the heavy Higgs pole 
at about $500$ GeV, as a consequence, 
the behavior of these curves 
splits into two completely opposite 
patterns depending on the mixing 
parameter $t_{\beta}$. 
Under the $t_{\beta} = 2$ scenario, 
the longitudinal as well as unpolarized 
channels merely reach up to a broad 
and smooth peak, whereas shifting to 
$t_{\beta} = 6$ triggers a distinguish 
resonance effect that enforces both 
dashed lines to shoot up into 
corresponding peaks.
Immediately after this kinematic 
$m_{Z\gamma}$ threshold, all curves 
experience a precipitous and 
multi-order-of-magnitude drop. 

Eventually, the inclusion of VLQs 
via one-loop contributions modifies 
the magnitude rather than 
these distributed configurations.
All solid lines replicate the precise 
shape of the standard 2HDM Type II 
distribution, but are consistently 
shifted upward across the nearly 
entire energy range of $m_{Z\gamma}$. 
Moreover, this supplement not only 
enhances these overall differential 
decay rates, but also mitigates 
these slopes in the less contributed 
scope of $m_{Z\gamma} > 540$ GeV 
within the case of $t_{\beta} = 2$, 
establishing a visibly more gradual 
decay tendency at the high-energy region, 
which totally disappears in 
$t_{\beta} = 6$. 

\begin{figure}[H]
\centering
\includegraphics[width=15cm,height=7cm]
{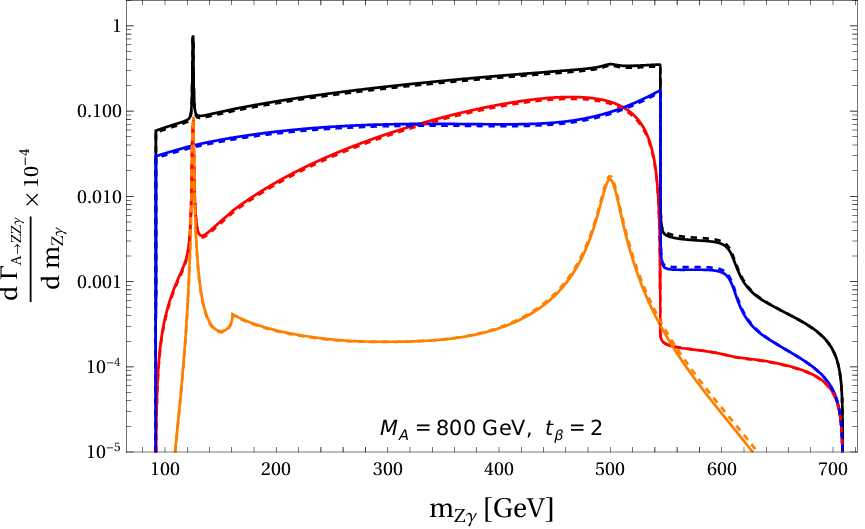}
\\[0.25cm]
\hfill
\\[0.25cm]
\includegraphics[width=15cm,height=7cm]
{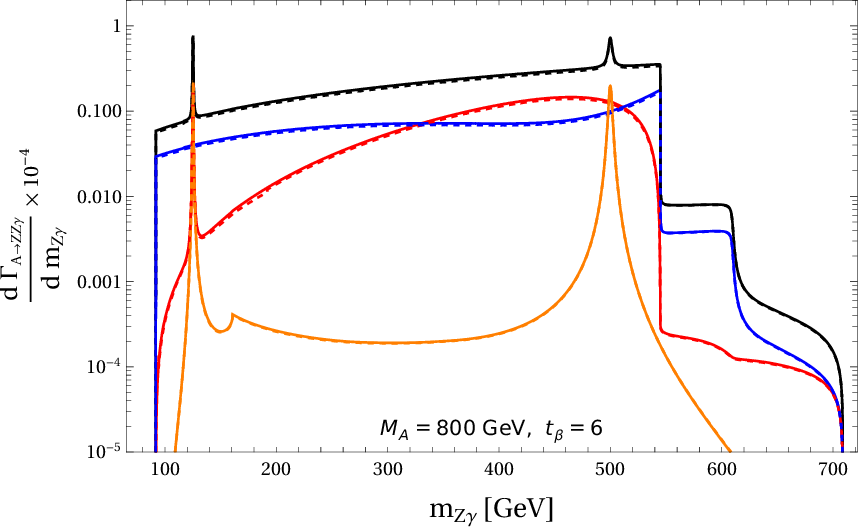}
\caption{
\label{Fig:mZApol2Z}
Same as Figure~\ref{Fig:mZApol1Z},
but for all representative
polarization modes of both
$Z$-bosons in the final state.
}
\end{figure}

In the shown Fig.~\ref{Fig:mZApol2Z}, 
we similarly emphasize that 
the differential decay rates 
with respect to the mass-invariant 
$m_{Z\gamma}$ are also fully considered 
into account in all representative 
polarization modes of both Z-bosons 
in the final state. 
In this case, we only investigate all 
polarization states serving as such 
significant contributions in 
the 2HDM Type II as discussed 
in these above Tables, such as 
one purely longitudinal state 
$\lambda_1 = \lambda_2 = 0$ 
(red dashed line) and two mixing 
polarized situations: (i)
$\lambda_1 = 0, \, \lambda_2 = L,R$ 
(blue dashed line) as well as (ii) 
$\lambda_2 = 0, \, \lambda_1 = L,R$ 
(orange dashed line), and then the 
unpolarized case (black dashed line), 
in general, is also appended to 
our phenomenological observations.
Analogously, the solid and 
corresponding color-coded lines 
illustrate the augmented VLQs 
contributions for the subsequent 
comparative discussions.

On the whole, these both dashed 
and solid blue lines denote for 
the mixed polarization 
state (i) perfectly replicates 
the slight fluctuation and 
a downward-sloping inclination at 
high-energy region of $m_{Z\gamma}$, 
as well as acquiring an equivalent 
magnitude consistently observed 
in the transverse mode 
$\lambda_2 = L,R$ [as discussed in 
Fig.~\ref{Fig:mZApol1Z}]. 
This is due to relying on a relation 
specifically presented in 
Eq.~(\ref{Eq:1Z2Zrelation}),
these remaining contributions 
stemming from four polarization 
modes $\lambda_1 = L \, (R), 
\, \lambda_2 = L,R$ 
are totally negligible, 
as numerically examined in 
these above Tables.
Consequently, the investigative 
focus at present shifts entirely 
to the behavior comparison of such 
remaining components in 
the 2HDM Type II, including 
a purely longitudinal state and 
the situation (ii), whose sum, 
in analogy, are consistently 
constitute of the longitudinal 
mode $\lambda_2 = 0$ as depicted 
in Fig.~\ref{Fig:mZApol1Z}. 

Initially, within the low-mass range 
around $m_{Z\gamma} \simeq 125.1$ GeV, 
these both curves display an 
impressive synchronicity, 
characterized by coincident and 
sharp vertical peaks precisely at 
this light CP-even Higgs pole. 
Moving beyond this resonance, 
however, a profound discrepancy in 
their behaviors occurs. 
Whereas the polarization state 
$\lambda_1 = \lambda_2 = 0$ 
undergoes a slight adjustment 
before climbing steadily, the mode 
$\lambda_2 = 0, \, \lambda_1 = L,R$ 
plunges drastically to develop 
a deep region around the band of 
$m_{Z\gamma} \simeq 150$ GeV. 
Thereafter exceeding the approximate 
$300$ GeV threshold, the solely 
longitudinal state experiences 
a massive surge with an accelerating 
slope to absolutely assert dominance 
in magnitude, whereas the situation 
(ii) ascends at a more gradual pace 
when it recovers toward the high-energy 
region of $m_{Z\gamma}$. 

In further, by approaching the heavy 
CP-even Higgs $H$ pole at 
$m_{Z\gamma} \simeq 500$ GeV, 
the red dashed line for case of 
$\lambda_1 = \lambda_2 = 0$ progresses 
monotonically and exhibits a non-peaking 
behavior under both circumstances as in 
Fig.~\ref{Fig:mZApol2Z}, whereas 
the topology of such dashed orange curve 
varies dramatically depending on 
the mixing angle parameter $t_{\beta}$. 
Under the $t_{\beta} = 2$ scenario, 
the curve of polarization mode 
$\lambda_2 = 0, \, \lambda_1 = L,R$ 
merely culminates in a depressed 
and broad peak, remaining entirely 
below the dominant red dashed line. 
In sharp contrast to this, shifting 
to $t_{\beta} = 6$ triggers a monumental 
resonance impact that forces this orange 
dashed line to shoot up vertically into 
a narrow and sharp resonance vertex, 
effectively reversing the hierarchy to 
surpass the curve of solely longitudinal 
state $\lambda_1 = \lambda_2 = 0$ 
at this pole. 
Notably, within this $500$ GeV range, 
the one-loop corrections from extra 
VLQs yield highly contrasting effects 
on the spectrum of situation (ii). 
In details, they exert a suppressing 
impact at $t_{\beta} = 2$, where 
the orange dashed line possesses visibly 
higher than the corresponding solid line, 
but cause these two lines to become 
virtually indistinguishable 
at $t_{\beta} = 6$. 

In addition, immediately following 
the kinematic threshold of
$m_{Z\gamma} > 540$ GeV, all curves 
suffer a perpendicular and massive 
decrease in magnitude. 
Within this high-mass boundary, 
the orange lines of polarization state 
$\lambda_2 = 0, \, \lambda_1 = L,R$ 
collapse rapidly, and the virtual 
impact of vector-like quarks on 
this decay channel diminishes 
significantly as compared to lower 
energy regions. 
Concurrently, the VLQs contributions 
to the red solid line become entirely 
negligible as the spectrum continues 
its steep descent toward the high-energy 
regime of $m_{Z\gamma}$. 

To sum up, despite localized shifts 
in amplitude, the solid lines consistently 
map out within the majorly shifted upward 
tendency as well as reinforce the fundamental 
distributed configurations established by 
the dashed curves in the standard 
2HDM Type II. 
\section{Conclusions}
In summary, we presented a detailed one-loop analysis of the decay channel $A \rightarrow ZZ\gamma$ in both the 2HDM and the
2HDM-VLQ. We calculated the partial decay widths and branching ratios for both unpolarized and polarized $Z$ bosons and quantified the effects of Vector-Like Quarks in the loop. The branching ratios were found to vary from approximately $10^{-5}$ to $10^{-4}$ for selected benchmark points within the allowed parameter space of the model investigated.
Moreover, the presence of Vector-Like Quarks produced non-negligible corrections to the total decay rates, reaching approximately $\sim 5\%$. These findings demonstrated that the $A \rightarrow ZZ\gamma$ channel could provide a complementary probe of extended Higgs sectors and Vector-Like Quark effects. In particular, the high luminosity expected at future multi-TeV muon colliders could offer promising opportunities to test these effects experimentally.
\section*{Appendix A: 
One-loop induced Feynman diagrams 
of $A \rightarrow \phi^* \gamma$
with $\phi^* \rightarrow ZZ$}
\begin{figure}[H]
	\centering
	\includegraphics[width=14cm,height=9cm]
	{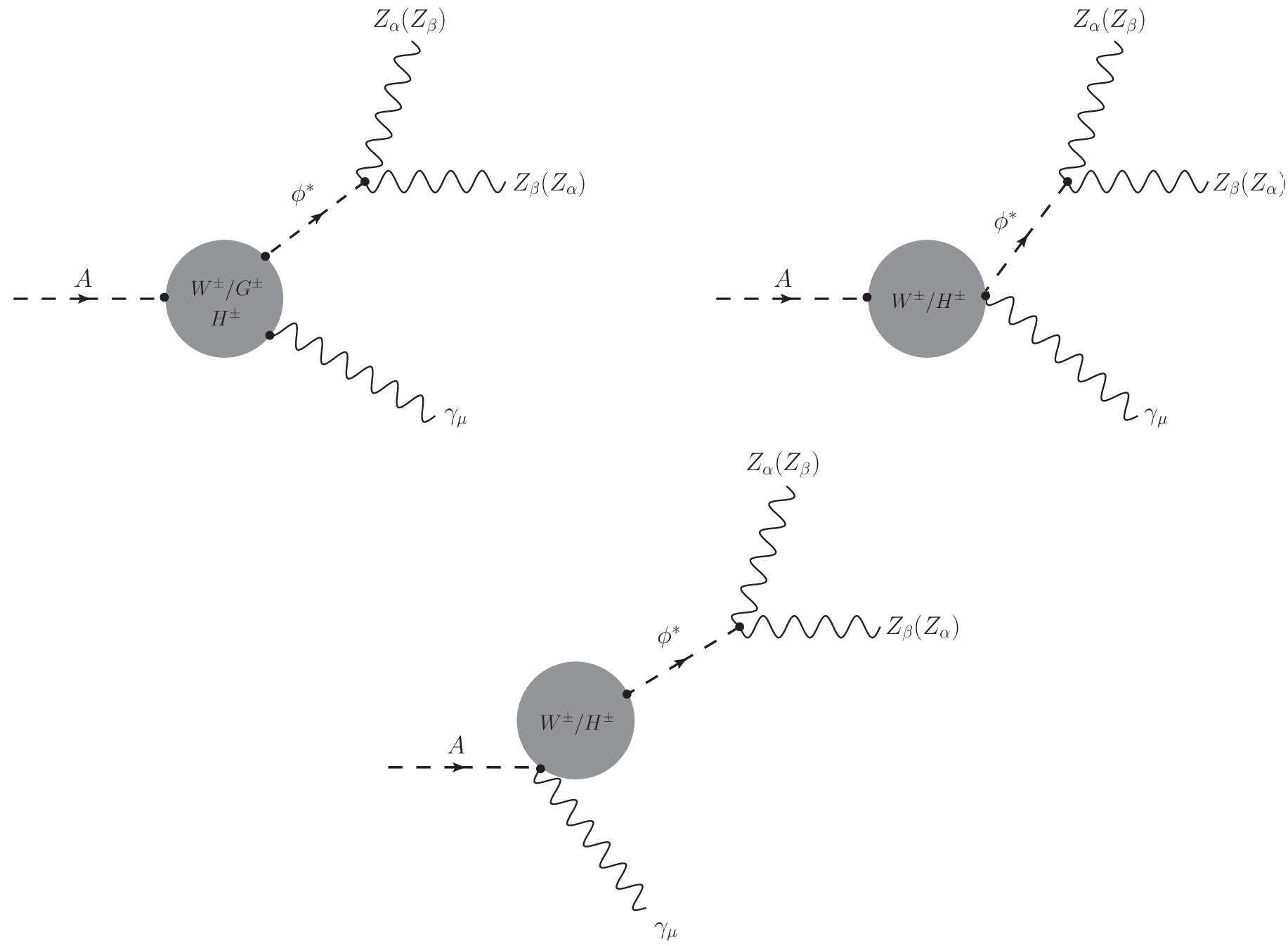}
	\caption{
\label{Fig:TrigsdiagsnoCB}
		The other distinctive triangle
		topology relevant to the one-loop
		decaying induced process
		$A \rightarrow Z Z \gamma$
		is also described in
		the general $R_{\xi}$ gauge.
		There are such additional diagrams
		that are unanimously obtained
		by exchanging the $Z$ bosons
		in the final state.
	}
\end{figure}

Aside from the main one-loop
contributions as described above
in Fig.~\ref{Fig:diags},
there are also two distinguished
topologies relevant to the one-loop
decaying induced process
$A \rightarrow Z Z \gamma$, which
primarily give a zero contribution
when appropriately summing over
all diagrams is carried out.
First of all, the set of one-loop
Feynman diagrams totally consists of
reducible triangle ones,
particularly depicted in the
Fig.~\ref{Fig:TrigsdiagsnoCB}
for our convenience.
These diagrams simultaneously arise
from the off-shell one-loop decay
processes $A \rightarrow \phi^* \gamma$,
connected by the $\phi ZZ$ coupling
for each corresponding pole
$\phi \equiv h,H$, where all combinations
of charged particles $W^\pm, G^\pm$
and $H^\pm$ are obviously propagated
in loop corrections.
Furthermore, the first diagram only
include all charged fermions $F$
contributing into these current loops,
because the all quartic couplings
related to scalar-vector bosons and
a fermion pair such as
$\gamma \phi F \bar{F}$ and
$\gamma A F \bar{F}$ do not
exist in the 2HDM as well as
the corresponding VLQs embedded model.
In this situation, however, it should
be emphasized that the contributions of
these charged fermions in two opposite
loop diagrams cancel out to zero in terms
of featured Levi-Civita tensors when
their momentum directions are reversed
and summed up.
Accompany with these previous triangle
contributions, the self-energy topology
stems from the decay processes of
the both off-shell final states
$A \rightarrow \phi^* Z^*$
at the tree-level, subsequently
followed by the $\phi^* ZZ$
couplings, as well as the one-loop
$Z-\gamma$ mixing diagrams connected
through a virtual massive vector
boson $Z^*$ as elucidated in the following
Fig.~\ref{Fig:SelfdiagsnoCB}.
In which all relevant charged particles
such as only $W^\pm, G^\pm$ bosons
and relevant ghosts $u^\pm$ contributing
into loops are consistently considered
for all their cancellations.
Moreover, it should be noted that
the contribution of charged Higgs
$H^\pm$ in such self-energy topology
is also separately taken into account
in our subsequent discussions.
\begin{figure}[H]
	\centering
	\includegraphics[width=7cm,
	height=4cm]
	{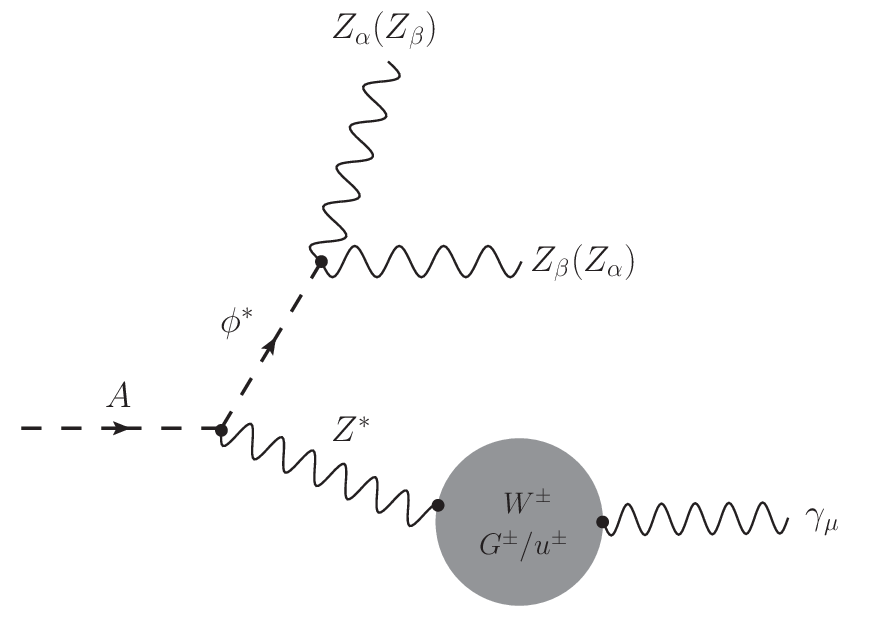}
	\caption{
		\label{Fig:SelfdiagsnoCB}
		Same as
		Figure~\ref{Fig:TrigsdiagsnoCB},
		but for the corresponding
		self-energy topology of
		one-loop $Z - \gamma$ mixing.
		It is stressed that
		there are also such supplementary
		diagrams that are consistently
		acquired by interchanging
		the $Z$ bosons in the final state.
	}
\end{figure}

Following the gauge-invariant Lorentz
structure of Eq.~(\ref{Eq:Amplitude}),
in general, the one-loop amplitude for
these aforementioned distinct topologies
of the off-shell decay channel
$A \rightarrow \phi^* \gamma$
followed by the tree-level process
$\phi^* \rightarrow ZZ$
is given particularly in only terms of
two form factors $\mathcal{T}_{3i}$ for
$i = 1,2$, as detailed below
\begin{eqnarray}
\mathcal{M}_{A \rightarrow \phi^* \gamma}
&=&
g^{\alpha \beta}
\Big[
\mathcal{T}_{31} \,
k_1^\mu
+
\mathcal{T}_{32} \,
k_2^\mu
\Big]
\varepsilon^*_\alpha (k_1) \,
\varepsilon^*_\beta (k_2) \,
\varepsilon^*_\mu (k_3)
,
\end{eqnarray}
in which we also analytically clarify
that the form factor $\mathcal{T}_{32} =
\mathcal{T}_{31} \{k_{13} \leftrightarrow
k_{23}\}$ is acquired as well.
However, the pole $\phi \equiv h,H$
contributions for these coupling
$\phi^* ZZ$ is consistently expressed
in terms of Mandelstam variable
$k_{12} = M_{A}^2 + 2 M_Z^2 - k_{13} -
k_{23}$, therefore we get that
$\mathcal{T}_{31} = \mathcal{T}_{32}
\equiv \mathcal{T}$.
Subsequently, the general form factor
$\mathcal{T}$ is decomposed into two
individual contributions at each pole
$\phi$ such as triangle and self-energy
ones corresponding to
Figs.~\ref{Fig:TrigsdiagsnoCB}
and~\ref{Fig:SelfdiagsnoCB},
\begin{eqnarray}
\mathcal{T}
&=&
\sum \limits_{\phi \equiv h,H}
g_{\phi ZZ} \,
\Big(
\mathcal{T}_{\text{Trig}}^{\phi}
+
g_{A Z \phi} \,
\mathcal{T}_{\text{Self}}^{\phi}
\Big)
.
\end{eqnarray}
where these general couplings $g_{\phi ZZ}
\equiv i e M_{W} / s_W c_W^2 \times
s_{\beta - \alpha} (c_{\beta - \alpha})$
as well as $g_{A Z \phi} \equiv e / s_{2W}
\times c_{\beta - \alpha}
(- s_{\beta - \alpha})$
are denoted for the each specific pole
$\phi \equiv h \, (H)$, respectively.
The self-energy form factor
$\mathcal{T}_{\text{Self}}^{\phi}$
then has a form of
\begin{eqnarray}
\mathcal{T}_{\text{Self}}^{\phi}
&=&
\dfrac{e^2}{
(6 \pi^2) M_{Z}^2 s_{2W}}
\dfrac{1}{\big( k_{12}-M_{\phi}^2 \big)}
\times
\\[0.25cm]
&&\hspace{-0.25cm} \times
\Bigg\{
2 M_{W}^2
\big(4 c_W^2+s_W^2\big)
B_{0}(0,M_{W}^2,M_{W}^2)
+
\big(5 c_W^2-s_W^2\big)
\Big[
M_{W}^2
-
A_{0}(M_{W}^2)
\Big]
\Bigg\}
.
\nonumber
\end{eqnarray}
Furthermore, afterwards the
Passarino-Veltman reduction scheme
was made use of in details, these
contributing expressions of
the triangle form factor
$\mathcal{T}_{\text{Trig}}^{\phi}$
for the pole $\phi = h,H$
are analytically derived as follows
\begin{eqnarray}
\mathcal{T}_{\text{Trig}}^{h}
&=&
-
\dfrac{e^3}{(32 \pi^2)  s_W^2}
\, c_{\beta - \alpha}
\,
\Bigg[
\dfrac{4}{\big( k_{12}-M_h^2 \big)}
B_{0}(0,M_{W}^2,M_{W}^2)
-
2 C_{0}
(0,k_{12},M_A^2,M_{W}^2,M_{W}^2,M_{H^\pm}^2)
\nonumber
\\[0.05cm]
&&\hspace{+3.25cm}
-
\dfrac{1}{\big( k_{12}-M_A^2 \big)}
\Big[
B_{0}(k_{12},M_{H^\pm}^2,M_{W}^2)
-
B_{0}(M_A^2,M_{H^\pm}^2,M_{W}^2)
\Big]
\Bigg]
,
\\[0.5cm]
\mathcal{T}_{\text{Trig}}^{H}
&=&
\mathcal{T}_{\text{Trig}}^{h}
\,
\Big\{
M_h^2
\rightarrow
M_H^2
\, , \,
c_{\beta - \alpha}
\rightarrow
- s_{\beta - \alpha}
\,
\Big\}
.
\end{eqnarray}
Because of these last two terms
expressed in the square bracket of
such form factors
$\mathcal{T}_{\text{Trig}}^{\phi}$
are transparently independent on
the scalar CP-even Higgs boson masses
$M_{\phi}^2$ with $\phi \equiv h,H$,
our resulting findings in the general
form factor $\mathcal{T}$ are partly
acquired as detailed below
\begin{eqnarray}
\sum \limits_{\phi \equiv h,H}
g_{\phi ZZ} \,
\mathcal{T}_{\text{Trig}}^{\phi}
&=&
-
\dfrac{i e^4}{(24 \pi^2)}
\dfrac{
s_{\beta - \alpha} \,
c_{\beta - \alpha} \,
\big( M_h^2 - M_H^2 \big)
}{\big( k_{12}-M_h^2 \big)
\big( k_{12}-M_H^2 \big)}
\times
\dfrac{3 M_{W}}{s_W^3 c_W^2}
B_{0}(0,M_{W}^2,M_{W}^2)
,
\label{Eq:Gtrig}
\end{eqnarray}
whereas we also have the remaining one
of form factor $\mathcal{T}$ is simplified
presented into a compact form as follows,
\begin{eqnarray}
\label{Eq:Gself}
\sum \limits_{\phi \equiv h,H}
g_{\phi ZZ} \,
g_{A Z \phi} \,
\mathcal{T}_{\text{Self}}^{\phi}
&=&
\dfrac{i e^4}{
(24 \pi^2)}
\dfrac{
	s_{\beta - \alpha} \,
	c_{\beta - \alpha} \,
	\big( M_h^2 - M_H^2 \big)
}{\big( k_{12}-M_h^2 \big)
	\big( k_{12}-M_H^2 \big)}
\times
\dfrac{M_{W}}{M_{Z}^2 \, s_W^3 c_W^4}
\\[0.25cm]
&&\hspace{-0.5cm} \times
\Bigg\{
2 M_{W}^2
\big(4 c_W^2+s_W^2\big)
B_{0}(0,M_{W}^2,M_{W}^2)
+
\big(5 c_W^2-s_W^2\big)
\Big[
M_{W}^2
-
A_{0}(M_{W}^2)
\Big]
\Bigg\}
.
\nonumber
\end{eqnarray}
As a consequence, once the PV reduction
scheme was explicitly carried out for
the above triangle contributions,
the sum of these two aforementioned
distinctive topologies in such
Figs.~\ref{Fig:TrigsdiagsnoCB}
and~\ref{Fig:SelfdiagsnoCB} with
both Higgs poles $\phi$ simultaneously
is canceled out by the analytical aid
of a well-known relation for such scalar
PV functions $A_0$ and $B_0$ at zero
external momentum, as detailed below
\begin{eqnarray}
B_0 (0,M_W^2,M_W^2)
&=&
A_0 (M_W^2) / M_W^2 - 1
\end{eqnarray}
as could be similarly found in the
Appendix A of Ref.~\cite{Marciano:2012PRD}
for the $W^\pm$-boson mass such that
$x \equiv M_W^2$ at accurately
the dimension $D = 4$.
Furthermore, the tree-level relation
defined as $s_{W}^2 = 1 - M_W^2 / M_Z^2$,
which is consistently accomplished to
an all-order relation in the on-shell
renormalization scheme for the Weinberg
angle $s_{W}$, is also instrumental
for the vanishing of these above
contributing results in total.
To conclude, by summing over these
Eqs.~(\ref{Eq:Gtrig}) and~(\ref{Eq:Gself}),
we obtain that the general form factor
$\mathcal{T} = 0$ is reading
in our analysis.
Last but not least, the contribution of
charged Higgs $H^\pm$ for the
self-energy topology as described in
the Fig.~\ref{Fig:SelfdiagsnoCB} is
also gained the value of zero,
due to the similar relation like
$B_0 (0,M_{H^\pm}^2,M_{H^\pm}^2) =
A_0 (M_{H^\pm}^2) / M_{H^\pm}^2 - 1$
fully factored out of their
total amplitude.
\section*{Appendix B: 
Squared amplitudes for
polarized $Z$-bosons}
In this work, we consider all polarization
modes of each $Z$ boson in the final state.
First, the polarization states of the massive
vector boson $Z_{\beta}(k_2)$ are treated
explicitly, while the other $Z$ boson and the
photon are summed over their polarization
states in the squared amplitude. Specifically,
we have
\begin{alignat}{2}
\varepsilon^{*}_\alpha (k_1)
\,
\varepsilon_\rho (k_1)
&=
-
g^{\alpha \rho}
+
k_1^{\alpha}
k_1^{\rho}
/ M_Z^2
\qquad & \qquad
\text{for vector boson}
\;\,
Z_{\alpha} (k_1),
\\[0.25cm]
\varepsilon^{*}_\mu (k_3)
\,
\varepsilon_\nu (k_3)
&=
-
g^{\mu \nu}
\qquad & \qquad
\text{for photon}
\;\,
\gamma_{\mu} (k_3).
\end{alignat}
Next, both $Z$ bosons are treated
similarly in our analysis.

Working in the rest frame of the CP-odd
scalar Higgs boson $A$, the four-momenta
of the particles involved satisfy the
momentum-conservation relation
$p^\mu = \sum\limits_{i=1}^{3} k_i^\mu$,
and their components are expressed as follows:
\begin{eqnarray}
p^\mu 
&\equiv&
\big(
M_{A}
,
\,
\vec{0}
\,
\big),\\
k_{1}^\mu 
&\equiv&
\big(
E_1
,
\,
\vec{k_1}
\,
\big)
=
\big(
E_1
,
\,
0,
\,
0,
\,
|P_1|
\,
\big),
\label{Eq:k1mu}
\\
k_{2}^\mu 
&\equiv&
\big(
E_2
,
\,
\vec{k_2}
\,
\big)
=
\big(
E_2
,
\,
|P_2| \, s_{\theta} \, c_{\phi},
\,
|P_2| \, s_{\theta} \, s_{\phi},
\,
|P_2| \, c_{\theta}
\,
\big),
\\
k_{3}^\mu 
&\equiv&
\big(
E_{\gamma}
,
\,
\vec{k_{\gamma}}
\,
\big)
=
\big(
E_{\gamma}
,
\,
-
\vec{k_1}
-
\vec{k_2}
\,
\big)
\end{eqnarray}
in which we denote that $s_{x} (c_{x}) 
\equiv \sin x \, (\cos x)$ for the second
on-shell vector boson $Z$ propagates 
along the $(\theta, \phi)$ axis in range 
of $\theta \in [0, \pi]$ and
$\phi \in [0, 2\pi]$.
By applying three-momentum conservation,
expressed as $|\vec{k}_\gamma|^2
= |\vec{k}_1|^2 + |\vec{k}_2|^2
+ 2\,\vec{k}_1 \cdot \vec{k}_2$,
we obtain an expression for the polar
angle $\theta$ in this
frame, which is given by
\begin{eqnarray}
c_{\theta}
&=&
\dfrac{1}{
2
|P_1|
|P_2|}
\Big(
E_{\gamma}^2
-
|P_1|^2
-
|P_2|^2
\Big).
\end{eqnarray}
The formula is expressed in terms of 
the magnitude of momentum three-vectors 
$|P_i|$ as well as these on-shell 
particle energies like that 
$E_{i}^2 = |P_i|^2 + M_Z^2$ for 
$i = 1,2$ and then $|\vec{k_{\gamma}}|^2 
\equiv E_{\gamma}^2 = 
\big(M_{A} - E_1 - E_2 \big)^2$ 
acquired from the massless photon.
In particular, these explicit formulas are 
presented in terms of invariant masses and 
Mandelstam variables with the help of 
$k_{ij} \equiv (k_i + k_j)^2 
= (p - k_l)^2$ for $l \neq i,j = 1,2,3$ 
satisfied with the momentum four-vector 
conservation as detailed below
\begin{eqnarray}
E_{\gamma}
&=&
\dfrac{1}{
2 M_{A}}
\Lambda^{1/2}
(M_{A}^2, 0, k_{12})
,
\\
|P_1|
&=&
\dfrac{1}{
2 M_{A}}
\Lambda^{1/2}
(M_{A}^2, M_Z^2, k_{23})
,
\\
|P_2|
&=&
\dfrac{1}{
2 M_{A}}
\Lambda^{1/2}
(M_{A}^2, M_Z^2, k_{13})
,
\end{eqnarray}
in which the Källén function for these 
above kinematical factors gets well-defined 
form of $\Lambda(x, y, z) = (x - y - z)^2 
- 4yz$.
From the above Higgs rest frame, we 
properly choose that these full 
polarization four-vectors include 
longitudinal (index $``0"$) 
and transversal (indices $``L", ``R"$) 
modes~\cite{Kadeer:2008EPJC, Maina:2021PLB} 
for only vector boson $Z_{\beta} (k_2)$ 
traveling along the $(\theta, \phi)$ axis
in final state are given in 
the representative form as follows:
\begin{eqnarray}
\varepsilon^{0}_\beta (k_2)
&=&
\dfrac{1}{M_Z}
\big(
|P_2|
,
\,
E_2 \, s_{\theta} \, c_{\phi},
\,
E_2 \, s_{\theta} \, s_{\phi},
\,
E_2 \, c_{\theta}
\,
\big),
\\
\varepsilon^{L,R}_\beta (k_2)
&=&
\dfrac{1}{\sqrt{2}}
\big(
0
,
\,
\pm \, c_{\theta} \, c_{\phi}
+
i \, s_{\phi},
\,
\pm \, c_{\theta} \, s_{\phi}
-
i \, c_{\phi},
\,
\mp \, s_{\theta}
\,
\big).
\end{eqnarray}
Additionally we also analytically clarify 
three helpful relations of polarized 
vectors $\varepsilon^{\lambda_2}_\beta (k_2)$ 
corresponding to $Z_{\beta} (k_2)$ such as 
transversal and orthonormal condition,
completeness relation, respectively,
as detailed below
\begin{eqnarray}
k_{2}^\beta \,
\varepsilon^{\lambda_2}_\beta (k_2)
\, \equiv \,
k_{2}^\beta \,
{\varepsilon^{\lambda_2}_\beta}^{*} (k_2)
&=&
0,
\label{Eq:Transversal}
\\
\varepsilon^{\lambda_2}_\beta (k_2)
\,
{\varepsilon^{\lambda_2'}_\beta}^{*} (k_2)
&=&
- \delta_{\lambda_2 \lambda_2'},
\label{Eq:Orthonormal}
\\
\sum \limits_{\lambda_2}
{\varepsilon^{\lambda_2}_\beta}^{*} (k_2)
\,
\varepsilon^{\lambda_2}_\tau (k_2)
&=&
-
g^{\beta \tau}
+
k_2^{\beta}
k_2^{\tau}
/ M_Z^2
\label{Eq:Completeness}
\end{eqnarray}
in which each polarization mode 
$\lambda_2, \lambda'_2 \equiv 0,L,R$ is 
individually considered.
As a result, the one-loop induced amplitude
is rewritten from Eq.~(\ref{Eq:Amplitude}) 
in terms of two identically 
polarized modes when the condition 
$\lambda_2 \equiv \lambda'_2$ is evaluating 
to sum over polarizations for only 
second $Z$-boson as follows
\begin{eqnarray}
\mathcal{M}_{A \rightarrow ZZ\gamma}^{(\lambda_2)}
&=&
\mathcal{M}^{\alpha \beta \mu} \,
\varepsilon^{*}_\alpha (k_1) \,
{\varepsilon^{\lambda_2}_\beta}^{*} (k_2) 
\,
\varepsilon^*_\mu (k_3)
.
\end{eqnarray}
Thus unpolarized square amplitude, 
in general, is computed as follows 
\begin{eqnarray}
\big|
\mathcal{M}_{A \rightarrow ZZ\gamma}^{\text{unpol}}
\big|^2
&=&
\sum
\limits_{\lambda_2 \; \equiv \; \lambda'_2}
\mathcal{M}_{A \rightarrow ZZ\gamma}^{(\lambda_2)}
\times
\Big[
\mathcal{M}_{A \rightarrow ZZ\gamma}^{(\lambda'_2)}
\Big]^{*}
\; \equiv \;
\sum
\limits_{\lambda_2}
\big|
\mathcal{M}_{A \rightarrow ZZ\gamma}^{(\lambda_2)}
\big|^2
\end{eqnarray}
in which conjugation of the above 
amplitude has form of 
\begin{eqnarray}
\Big[
\mathcal{M}_{A \rightarrow ZZ\gamma}^{(\lambda'_2)}
\Big]^{*}
&=&
\mathcal{M}^{\rho \tau \nu} \,
\varepsilon_\rho (k_1) \,
\varepsilon^{\lambda'_2}_\tau (k_2) 
\,
\varepsilon_\nu (k_3).
\end{eqnarray}
Furthermore, subsequent to 
the squared amplitude 
$\big|
\mathcal{M}_{A \rightarrow ZZ\gamma}^{(\lambda_2)}
\big|^2$ particularly considered for 
each polarization mode $\lambda_2$ equals 
to $\lambda'_2$, which gives non-zero contributions, is consistently constituted 
of the unpolarized square amplitude and completely expressed in terms of 
invariant masses and Mandelstam variables.
Eventually, an analytical identification 
that we firmly clarify as follows
\begin{eqnarray}
\big|
\mathcal{M}_{A \rightarrow ZZ\gamma}^{(L)}
\big|^2
&=&
\big|
\mathcal{M}_{A \rightarrow ZZ\gamma}^{(R)}
\big|^2.
\end{eqnarray}
Last but not the least, we will also enlarge 
our calculations by extending more these 
polarization modes for the remaining 
massive vector boson $Z_{\alpha} (k_1)$, 
which moves along $z$-axis as in 
Eq~(\ref{Eq:k1mu}).
The analog polarization four-vectors for 
case of $z$-axis have form of
\begin{eqnarray}
\varepsilon^{0}_\alpha (k_1)
&=&
\dfrac{1}{M_Z}
\big(
|P_1|
,
\,
0,
\,
0,
\,
E_1
\,
\big),
\\
\varepsilon^{L,R}_\alpha (k_1)
&=&
\dfrac{1}{\sqrt{2}}
\big(
0
,
\,
\pm 1,
\,
- i,
\,
0
\,
\big)
\end{eqnarray}
which are also satisfied by these conditions 
and relation in Eqs.~(\ref{Eq:Transversal}), 
(\ref{Eq:Orthonormal}) and (\ref{Eq:Completeness}) for 
polarization modes 
$\lambda_1, \lambda'_1 \equiv 0,L,R$.
In a similar way, the one-loop induced amplitude in Eq.~(\ref{Eq:Amplitude}) 
is also presented by expand both modes 
as detailed below
\begin{eqnarray}
\mathcal{M}_{A \rightarrow ZZ\gamma}^{(\lambda_1, \lambda_2)}
&=&
\mathcal{M}^{\alpha \beta \mu} \,
{\varepsilon^{\lambda_1}_\alpha}^{*} (k_1) \,
{\varepsilon^{\lambda_2}_\beta}^{*} (k_2) 
\,
\varepsilon^*_\mu (k_3)
,
\end{eqnarray}
therefore we could confirm that nine modes 
$(\lambda_1, \lambda_2)$ are totally 
functions of invariant masses and 
Mandelstam variables, and perfectly 
contributed to the unpolarized 
square amplitude as follows
\begin{eqnarray}
\big|
\mathcal{M}_{A \rightarrow ZZ\gamma}^{\text{unpol}}
\big|^2
&\equiv&
\sum
\limits_{\lambda_1, \lambda_2}
\big|
\mathcal{M}_{A \rightarrow ZZ\gamma}^{(\lambda_1, \lambda_2)}
\big|^2
.
\end{eqnarray}
Subsequently, we analytically clarified that
some convenient relations are listed as  
\begin{eqnarray}
\big|
\mathcal{M}_{A \rightarrow ZZ\gamma}^{(0, L)}
\big|^2
&=&
\big|
\mathcal{M}_{A \rightarrow ZZ\gamma}^{(0, R)}
\big|^2,
\\
\big|
\mathcal{M}_{A \rightarrow ZZ\gamma}^{(L,0)}
\big|^2
&=&
\big|
\mathcal{M}_{A \rightarrow ZZ\gamma}^{(R,0)}
\big|^2,
\\
\big|
\mathcal{M}_{A \rightarrow ZZ\gamma}^{(L,R)}
\big|^2
&=&
\big|
\mathcal{M}_{A \rightarrow ZZ\gamma}^{(R,L)}
\big|^2,
\\
\big|
\mathcal{M}_{A \rightarrow ZZ\gamma}^{(L,L)}
\big|^2
&=&
\big|
\mathcal{M}_{A \rightarrow ZZ\gamma}^{(R,R)}
\big|^2
\end{eqnarray}
and furthermore others related to these 
squared amplitudes $\big|
\mathcal{M}_{A \rightarrow ZZ\gamma}^{(\lambda_2)}
\big|^2$ for only extending 
$Z_{\beta} (k_2)$ polarizations 
are also evaluated here, 
\begin{eqnarray}
\label{Eq:1Z2Zrelation}
\big|
\mathcal{M}_{A \rightarrow ZZ\gamma}^{(\lambda_2)}
\big|^2
&=&
\sum
\limits_{\lambda_1 \; = \; 0,L,R}
\big|
\mathcal{M}_{A \rightarrow ZZ\gamma}^{(\lambda_1, \lambda_2)}
\big|^2
,
\end{eqnarray}
thus there is also a perfect agreement 
of the completeness relation in 
Eq.~(\ref{Eq:Completeness}) for the 
polarization mode $\lambda_1$.

\begin{thebibliography}{200}
\bibitem{Dermisek:2009fd}
R.~Dermisek and J.~F.~Gunion,
Phys. Rev. D \textbf{81} (2010), 055001
doi:10.1103/PhysRevD.81.055001
[arXiv:0911.2460 [hep-ph]].
\bibitem{ATLAS:2015kpj}
G.~Aad \textit{et al.} [ATLAS],
Phys. Lett. B \textbf{744} (2015), 163-183
doi:10.1016/j.physletb.2015.03.054
[arXiv:1502.04478 [hep-ex]].


\bibitem{CMS:2020ffa}
A.~M.~Sirunyan \textit{et al.} [CMS],
JHEP \textbf{08} (2020), 139
doi:10.1007/JHEP08(2020)139
[arXiv:2005.08694 [hep-ex]].


\bibitem{ATLAS:2024rzd}
G.~Aad \textit{et al.} [ATLAS],
JHEP \textbf{12} (2024), 126
doi:10.1007/JHEP12(2024)126
[arXiv:2409.20381 [hep-ex]].

\bibitem{CMS:2012fgd}
S.~Chatrchyan \textit{et al.} [CMS],
Phys. Rev. Lett. \textbf{109} (2012), 121801
doi:10.1103/PhysRevLett.109.121801
[arXiv:1206.6326 [hep-ex]].


\bibitem{ATLAS:2023zkt}
G.~Aad \textit{et al.} [ATLAS],
JHEP \textbf{02} (2024), 197
doi:10.1007/JHEP02(2024)197
[arXiv:2311.04033 [hep-ex]].

\bibitem{Aiko:2022gmz}
M.~Aiko, S.~Kanemura and K.~Sakurai,
Nucl. Phys. B \textbf{986} (2023), 116047
doi:10.1016/j.nuclphysb.2022.116047
[arXiv:2207.01032 [hep-ph]].

\bibitem{Kanemura:2017gbi}
S.~Kanemura, M.~Kikuchi, K.~Sakurai and K.~Yagyu,
Comput. Phys. Commun. \textbf{233} (2018), 134-144
doi:10.1016/j.cpc.2018.06.012
[arXiv:1710.04603 [hep-ph]].

\bibitem{Kanemura:2019slf}
S.~Kanemura, M.~Kikuchi, K.~Mawatari, K.~Sakurai and K.~Yagyu,
Comput. Phys. Commun. \textbf{257} (2020), 107512
doi:10.1016/j.cpc.2020.107512
[arXiv:1910.12769 [hep-ph]].

\bibitem{Aiko:2023xui}
M.~Aiko, S.~Kanemura, M.~Kikuchi, K.~Sakurai and K.~Yagyu,
Comput. Phys. Commun. \textbf{301} (2024), 109231
doi:10.1016/j.cpc.2024.109231
[arXiv:2311.15892 [hep-ph]].

\bibitem{Akeroyd:2023kek}
A.~G.~Akeroyd, S.~Alanazi and S.~Moretti,
J. Phys. G \textbf{50} (2023) no.9, 095001
doi:10.1088/1361-6471/ace3e1
[arXiv:2301.00728 [hep-ph]].

\bibitem{Weber:2003tw}
C.~Weber, H.~Eberl and W.~Majerotto,
Phys. Rev. D \textbf{68} (2003), 093011
doi:10.1103/PhysRevD.68.093011
[arXiv:hep-ph/0308146 [hep-ph]].


\bibitem{Chishtie:1999dd}
F.~A.~Chishtie, V.~Elias and T.~G.~Steele,
J. Phys. G \textbf{26} (2000), 93-98
doi:10.1088/0954-3899/26/1/309
[arXiv:hep-ph/9909374 [hep-ph]].


\bibitem{Arhrib:2018pdi}
A.~Arhrib, R.~Benbrik, J.~El Falaki, M.~Sampaio and R.~Santos,
Phys. Rev. D \textbf{99} (2019) no.3, 035043
doi:10.1103/PhysRevD.99.035043
[arXiv:1809.04805 [hep-ph]].


\bibitem{Arco:2025ydq}
F.~Arco, T.~Biek{\"o}tter, P.~Stylianou and G.~Weiglein,
JHEP \textbf{06} (2025), 170
doi:10.1007/JHEP06(2025)170
[arXiv:2502.03443 [hep-ph]].


\bibitem{Banik:2024ugs}
S.~Banik, G.~Coloretti, A.~Crivellin and H.~E.~Haber,
Phys. Rev. D \textbf{111} (2025) no.7, 075021
doi:10.1103/PhysRevD.111.075021
[arXiv:2412.00523 [hep-ph]].


\bibitem{Phan:2024zus}
K.~H.~Phan, D.~T.~Tran and T.~H.~Nguyen,
PTEP \textbf{2024} (2024) no.8, 083B02
doi:10.1093/ptep/ptae103
[arXiv:2404.02417 [hep-ph]].


\bibitem{Sanchez-Velez:2018xdj}
R.~S{\'a}nchez-V{\'e}lez and G.~Tavares-Velasco,
Phys. Rev. D \textbf{97} (2018) no.9, 095038
doi:10.1103/PhysRevD.97.095038
[arXiv:1802.01222 [hep-ph]].

\bibitem{Yin:2002sq}
J.~Yin, W.~G.~Ma, R.~Y.~Zhang and H.~S.~Hou,
Phys. Rev. D \textbf{66} (2002), 095008.

\bibitem{Akeroyd:1999gu}
A.~G.~Akeroyd, A.~Arhrib and M.~Capdequi Peyranere,
Mod. Phys. Lett. A \textbf{14} (1999), 2093-2108
[erratum: Mod. Phys. Lett. A \textbf{17} (2002), 373]
doi:10.1142/S0217732399002157
[arXiv:hep-ph/9907542 [hep-ph]].

\bibitem{Akeroyd:2001aka}
A.~G.~Akeroyd, A.~Arhrib and M.~Capdequi Peyranere,
Phys. Rev. D \textbf{64} (2001), 075007
[erratum: Phys. Rev. D \textbf{65} (2002), 099903]
doi:10.1103/PhysRevD.65.099903
[arXiv:hep-ph/0104243 [hep-ph]].

\bibitem{Sasaki:2017fvk}
K.~Sasaki and T.~Uematsu,
Phys. Lett. B \textbf{781} (2018), 290-294
doi:10.1016/j.physletb.2018.04.005
[arXiv:1712.00197 [hep-ph]].

\bibitem{Farris:2003pn}
T.~Farris, J.~F.~Gunion, H.~E.~Logan and S.~f.~Su,
Phys. Rev. D \textbf{68} (2003), 075006
doi:10.1103/PhysRevD.68.075006
[arXiv:hep-ph/0302266 [hep-ph]].


\bibitem{Arhrib:2002ti}
A.~Arhrib,
Phys. Rev. D \textbf{67} (2003), 015003
doi:10.1103/PhysRevD.67.015003
[arXiv:hep-ph/0207330 [hep-ph]].
\bibitem{Branco:2011iw}
G.~C.~Branco, P.~M.~Ferreira, L.~Lavoura, M.~N.~Rebelo, M.~Sher and J.~P.~Silva,
Phys. Rept. \textbf{516} (2012), 1-102
doi:10.1016/j.physrep.2012.02.002
[arXiv:1106.0034 [hep-ph]].
\bibitem{Aoki:2009PRD}
M.~Aoki, S.~Kanemura, K.~Tsumura and K.~Yagyu,
Phys. Rev. D \textbf{80} (2009), 015017
doi:10.1103/PhysRevD.80.015017
[arXiv:0902.4665 [hep-ph]].
\bibitem{Arhrib:2024EPJC}
A.~Arhrib, R.~Benbrik, M.~Boukidi and S.~Moretti,
Eur. Phys. J. C \textbf{84} (2024) no.10, 1008
doi:10.1140/epjc/s10052-024-13390-5
[arXiv:2403.13021 [hep-ph]].
\bibitem{Arhrib:2025EPJC}
A.~Arhrib, R.~Benbrik, M.~Boukidi, B.~Manaut and S.~Moretti,
Eur. Phys. J. C \textbf{85} (2025) no.1, 2
doi:10.1140/epjc/s10052-024-13692-8
[arXiv:2401.16219 [hep-ph]].
\bibitem{Aguilar:2003PRD}
J.~A.~Aguilar-Saavedra,
Phys. Rev. D \textbf{67} (2003), 035003
[erratum: Phys. Rev. D \textbf{69} (2004), 099901]
doi:10.1103/PhysRevD.69.099901
[arXiv:hep-ph/0210112 [hep-ph]].
\bibitem{Aguilar:2013PRD}
J.~A.~Aguilar-Saavedra, R.~Benbrik, S.~Heinemeyer and M.~P{\'e}rez-Victoria,
Phys. Rev. D \textbf{88} (2013) no.9, 094010
doi:10.1103/PhysRevD.88.094010
[arXiv:1306.0572 [hep-ph]].
\bibitem{Benbrik:2025PRD}
R.~Benbrik, M.~Boukidi and S.~Moretti,
Phys. Rev. D \textbf{109} (2024) no.5, 055016
doi:10.1103/PhysRevD.109.055016
[arXiv:2211.07259 [hep-ph]].
\bibitem{TranDT:2025EPJC}
D.~T.~Tran, T.~H.~Nguyen and K.~H.~Phan,
Eur. Phys. J. C \textbf{85} (2025) no.10, 1123
doi:10.1140/epjc/s10052-025-14852-0
[arXiv:2503.22974 [hep-ph]].
\bibitem{Aoki:1982PTPS}
K. Aoki, Z. Hioki, R. Kawabe, M. Konuma, and T. Muta,
Prog. Theor. Phys. Suppl. \textbf{73}, 1 (1982)
doi:10.1143/PTPS.73.1.
\bibitem{Kanemura:2017PRD}
S.~Kanemura, M.~Kikuchi, K.~Sakurai and K.~Yagyu,
Phys. Rev. D \textbf{96} (2017) no.3, 035014
doi:10.1103/PhysRevD.96.035014
[arXiv:1705.05399 [hep-ph]].
\bibitem{Hahn:1999CPC}
T. Hahn, M. Pérez-Victoria,
T.~Hahn and M.~Perez-Victoria,
Comput. Phys. Commun. \textbf{118} (1999), 153-165
doi:10.1016/S0010-4655(98)00173-8
[arXiv:hep-ph/9807565 [hep-ph]].
\bibitem{Denner:1993FP}
A.~Denner,
Fortsch. Phys. \textbf{41} (1993), 307-420
doi:10.1002/prop.2190410402
[arXiv:0709.1075 [hep-ph]].
\bibitem{Hooft:1972NPB}
G.~'t Hooft and M.~J.~G.~Veltman,
Nucl. Phys. B \textbf{44} (1972), 189-213
doi:10.1016/0550-3213(72)90279-9.
\bibitem{Hahn:2001CPC}
T.~Hahn,
Comput. Phys. Commun. \textbf{140} (2001), 418-431
doi:10.1016/S0010-4655(01)00290-9
[arXiv:hep-ph/0012260 [hep-ph]].
\bibitem{Mertig:1991CPC}
R.~Mertig, M.~Bohm and A.~Denner,
Comput. Phys. Commun. \textbf{64} (1991), 345-359
doi:10.1016/0010-4655(91)90130-D.
\bibitem{Mertig:2016CPC}
V.~Shtabovenko, R.~Mertig and F.~Orellana,
Comput. Phys. Commun. \textbf{207} (2016), 432-444
doi:10.1016/j.cpc.2016.06.008
[arXiv:1601.01167 [hep-ph]].
\bibitem{Patel:2015CPC}
H.~H.~Patel,
Comput. Phys. Commun. \textbf{197} (2015), 276-290
doi:10.1016/j.cpc.2015.08.017
[arXiv:1503.01469 [hep-ph]].
\bibitem{Vladyslav:2017CPC}
V.~Shtabovenko,
Comput. Phys. Commun. \textbf{218} (2017), 48-65
doi:10.1016/j.cpc.2017.04.014
[arXiv:1611.06793 [physics.comp-ph]].
\bibitem{Denner:2006NPB}
A.~Denner and S.~Dittmaier,
Nucl. Phys. B \textbf{734} (2006), 62-115
doi:10.1016/j.nuclphysb.2005.11.007
[arXiv:hep-ph/0509141 [hep-ph]].
\bibitem{Denner:2011NPB}
A.~Denner and S.~Dittmaier,
Nucl. Phys. B \textbf{844} (2011), 199-242
doi:10.1016/j.nuclphysb.2010.11.002
[arXiv:1005.2076 [hep-ph]].
\bibitem{Denner:2017CPC}
A.~Denner, S.~Dittmaier and L.~Hofer,
Comput. Phys. Commun. \textbf{212} (2017), 220-238
doi:10.1016/j.cpc.2016.10.013
[arXiv:1604.06792 [hep-ph]].
\bibitem{Peter:1978JCP}
G.~P.~Lepage,
J. Comput. Phys. \textbf{27} (1978), 192
doi:10.1016/0021-9991(78)90004-9.
\bibitem{Hahn:2005CPC}
T.~Hahn,
Comput. Phys. Commun. \textbf{168} (2005), 78-95
doi:10.1016/j.cpc.2005.01.010
[arXiv:hep-ph/0404043 [hep-ph]].
\bibitem{Kanemura:2022NPB}
S.~Kanemura, M.~Kikuchi and K.~Yagyu,
Nucl. Phys. B \textbf{983} (2022), 115906
doi:10.1016/j.nuclphysb.2022.115906
[arXiv:2203.08337 [hep-ph]].
\bibitem{Aiko:2023NPB}
M.~Aiko, S.~Kanemura and K.~Sakurai,
Nucl. Phys. B \textbf{986} (2023), 116047
doi:10.1016/j.nuclphysb.2022.116047
[arXiv:2207.01032 [hep-ph]].
\bibitem{PDG:2024PRD}
S.~Navas \textit{et al.} [Particle Data Group],
Phys. Rev. D \textbf{110} (2024) no.3, 030001
doi:10.1103/PhysRevD.110.030001.
\bibitem{KHP:2024PTEP}
K.~H.~Phan, D.~T.~Tran and T.~H.~Nguyen,
PTEP \textbf{2024} (2024) no.8, 083B02
doi:10.1093/ptep/ptae103
[arXiv:2404.02417 [hep-ph]].
\bibitem{KHP:2025PTEP}
K.~H.~Phan, D.~T.~Tran and T.~H.~Nguyen,
PTEP \textbf{2025} (2025) no.2, 023B05
doi:10.1093/ptep/ptaf012
[arXiv:2406.15749 [hep-ph]].
\bibitem{ATLAS:2022Nature}
G.~Aad \textit{et al.} [ATLAS],
Nature \textbf{607} (2022) no.7917, 52-59
[erratum: Nature \textbf{612} (2022) no.7941, E24]
doi:10.1038/s41586-022-04893-w
[arXiv:2207.00092 [hep-ex]].
\bibitem{Draper:2021JHEP}
P.~Draper, A.~Ekstedt and H.~E.~Haber,
JHEP \textbf{05} (2021), 235
doi:10.1007/JHEP05(2021)235
[arXiv:2011.13159 [hep-ph]].
\bibitem{Aad:2024PRL}
G.~Aad \textit{et al.} [ATLAS and CMS],
Phys. Rev. Lett. \textbf{132} (2024) no.2, 021803
doi:10.1103/PhysRevLett.132.021803
[arXiv:2309.03501 [hep-ex]].
\bibitem{Andrei:2016PLB}
A.~Angelescu, A.~Djouadi and G.~Moreau,
Phys. Lett. B \textbf{756} (2016), 126-132
doi:10.1016/j.physletb.2016.02.064
[arXiv:1512.04921 [hep-ph]].
\bibitem{Arhrib:2018PRD}
A.~Arhrib, R.~Benbrik, S.~J.~D.~King, B.~Manaut, S.~Moretti and C.~S.~Un,
Phys. Rev. D \textbf{97} (2018), 095015
doi:10.1103/PhysRevD.97.095015
[arXiv:1607.08517 [hep-ph]].
\bibitem{Misiak:2017EPJC}
M.~Misiak and M.~Steinhauser,
Eur. Phys. J. C \textbf{77} (2017) no.3, 201
doi:10.1140/epjc/s10052-017-4776-y
[arXiv:1702.04571 [hep-ph]].
\bibitem{Benbrik:2024arxiv}
R.~Benbrik, M.~Boukidi and S.~Moretti,
PoS \textbf{ICHEP2024} (2025), 083
doi:10.22323/1.476.0083
[arXiv:2409.16054 [hep-ph]].
\bibitem{Abouabid:2024JPG}
H.~Abouabid, A.~Arhrib, R.~Benbrik, M.~Boukidi and J.~E.~Falaki,
J. Phys. G \textbf{51} (2024) no.7, 075001
doi:10.1088/1361-6471/ad3f34
[arXiv:2302.07149 [hep-ph]].
\bibitem{Djouadi:2013JHEP}
A.~Djouadi and J.~Quevillon,
JHEP \textbf{10} (2013), 028
doi:10.1007/JHEP10(2013)028
[arXiv:1304.1787 [hep-ph]].
\bibitem{Arhrib:2025JPG}
A.~Arhrib, R.~Benbrik, M.~Boukidi and S.~Moretti,
J. Phys. G \textbf{52} (2025) no.10, 105002
doi:10.1088/1361-6471/ae09be
[arXiv:2409.20104 [hep-ph]].
\bibitem{Marciano:2012PRD}
W.~J.~Marciano, C.~Zhang and S.~Willenbrock,
Phys. Rev. D \textbf{85} (2012), 013002
doi:10.1103/PhysRevD.85.013002
[arXiv:1109.5304 [hep-ph]].
\bibitem{Kadeer:2008EPJC}
A.~Kadeer, J.~G.~Korner and U.~Moosbrugger,
Eur. Phys. J. C \textbf{59} (2009), 27-47
doi:10.1140/epjc/s10052-008-0801-5
[arXiv:hep-ph/0511019 [hep-ph]].
\bibitem{Maina:2021PLB}
E.~Maina,
Phys. Lett. B \textbf{818} (2021), 136360
doi:10.1016/j.physletb.2021.136360
[arXiv:2007.12080 [hep-ph]].
\end{thebibliography}
\end{document}